\documentclass[11pt]{article}
\usepackage{amsmath,amsfonts,amssymb,epsfig,color,psfrag}
\usepackage{rotating}
\begin{document}
\title{Geometric Complexity Theory VIII: On canonical bases for
the nonstandard quantum groups \\ {\small (extended abstract)}}
\author{ 
Dedicated to Sri Ramakrishna \\ \\
Ketan D. Mulmuley 
 \\
The University of Chicago
\\  \\
(Technical Report TR-2007-15\\
Computer Science Department \\
The University of Chicago \\
September 2007) \\ Revised version  \\
http://ramakrishnadas.cs.uchicago.edu 
}

\maketitle

\newtheorem{prop}{Proposition}[section]
\newtheorem{claim}[prop]{Claim}
\newtheorem{goal}[prop]{Goal}
\newtheorem{theorem}[prop]{Theorem}
\newtheorem{metathesis}[prop]{Metathesis}
\newtheorem{mainpoint}{Main Point}
\newtheorem{hypo}[prop]{Hypothesis}
\newtheorem{guess}[prop]{Guess}
\newtheorem{problem}[prop]{Problem}
\newtheorem{axiom}[prop]{Axiom}
\newtheorem{question}[prop]{Question}
\newtheorem{remark}[prop]{Remark}
\newtheorem{lemma}[prop]{Lemma}
\newtheorem{claimedlemma}[prop]{Claimed Lemma}
\newtheorem{claimedtheorem}[prop]{Claimed Theorem}
\newtheorem{cor}[prop]{Corollary}
\newtheorem{defn}[prop]{Definition}
\newtheorem{ex}[prop]{Example}
\newtheorem{conj}[prop]{Conjecture}
\newtheorem{obs}[prop]{Observation}
\newtheorem{phyp}[prop]{Positivity Hypothesis}
\newcommand{\bitlength}[1]{\langle #1 \rangle}
\newcommand{\ca}[1]{{\cal #1}}
\newcommand{\floor}[1]{{\lfloor #1 \rfloor}}
\newcommand{\ceil}[1]{{\lceil #1 \rceil}}
\newcommand{\gt}[1]{{\langle  #1 |}}
\newcommand{\C}{\mathbb{C}}
\newcommand{\N}{\mathbb{N}}
\newcommand{\R}{\mathbb{R}}
\newcommand{\Z}{\mathbb{Z}}
\newcommand{\frcgc}[5]{\left(\begin{array}{ll} #1 &  \\ #2 & | #4 \\ #3 & | #5
\end{array}\right)}

\newcommand{\cgc}[6]{\left(\begin{array}{ll} #1 ;& \quad #3\\ #2 ; & \quad #4
\end{array}\right| \left. \begin{array}{l} #5 \\ #6 \end{array} \right)}

\newcommand{\wigner}[6]
{\left(\begin{array}{ll} #1 ;& \quad #3\\ #2 ; & \quad #4
\end{array}\right| \left. \begin{array}{l} #5 \\ #6 \end{array} \right)}

\newcommand{\rcgc}[9]{\left(\begin{array}{ll} #1 & \quad #4\\ #2  & \quad #5
\\ #3 &\quad #6
\end{array}\right| \left. \begin{array}{l} #7 \\ #8 \\#9 \end{array} \right)}

\newcommand{\srcgc}[4]{\left(\begin{array}{ll} #1 & \\ #2 & | #4  \\ #3 & |
\end{array}\right)}

\newcommand{\arr}[2]{\left(\begin{array}{l} #1 \\ #2   \end{array} \right)}
\newcommand{\unshuffle}[1]{\langle #1 \rangle}
\newcommand{\ignore}[1]{}
\newcommand{\f}[2]{{\frac {#1} {#2}}}
\newcommand{\tableau}[5]{
\begin{array}{ccc} 
#1 & #2  &#3 \\
#4 & #5 
\end{array}}
\newcommand{\embed}[1]{{#1}^\phi}
\newcommand{\stab}{{\mbox {stab}}}
\newcommand{\perm}{{\mbox {perm}}}
\newcommand{\trace}{{\mbox {trace}}}
\newcommand{\polylog}{{\mbox {polylog}}}
\newcommand{\sign}{{\mbox {sign}}}
\newcommand{\limit}{{\mbox {Lim}}}
\newcommand{\proj}{{\mbox {Proj}}}
\newcommand{\poly}{{\mbox {poly}}}
\newcommand{\std}{{\mbox {std}}}
\newcommand{\m}{\mbox}
\newcommand{\formula}{{\mbox {Formula}}}
\newcommand{\circuit}{{\mbox {Circuit}}}
\newcommand{\sgn}{{\mbox {sgn}}}
\newcommand{\core}{{\mbox {core}}}
\newcommand{\orbit}{{\mbox {orbit}}}
\newcommand{\cycle}{{\mbox {cycle}}}
\newcommand{\ideal}{{\mbox {ideal}}}
\newcommand{\qed}{{\mbox {Q.E.D.}}}
\newcommand{\proof}{\noindent {\em Proof: }}
\newcommand{\weight}{{\mbox {wt}}}
\newcommand{\tab}{{\mbox {Tab}}}
\newcommand{\level}{{\mbox {level}}}
\newcommand{\vol}{{\mbox {vol}}}
\newcommand{\vect}{{\mbox {Vect}}}
\newcommand{\val}{{\mbox {wt}}}
\newcommand{\sym}{{\mbox {Sym}}}
\newcommand{\convex}{{\mbox {convex}}}
\newcommand{\spec}{{\mbox {spec}}}
\newcommand{\strong}{{\mbox {strong}}}
\newcommand{\adm}{{\mbox {Adm}}}
\newcommand{\eval}{{\mbox {eval}}}
\newcommand{\for}{{\quad \mbox {for}\ }}
\newcommand{\Q}{Q}
\newcommand{\mand}{{\quad \mbox {and}\ }}
\newcommand{\invlim}{{\mbox {lim}_\leftarrow}}
\newcommand{\directlim}{{\mbox {lim}_\rightarrow}}
\newcommand{\sformal}{{\cal S}^{\mbox f}}
\newcommand{\vformal}{{\cal V}^{\mbox f}}
\newcommand{\crystal}{\mbox{crystal}}
\newcommand{\conje}{\mbox{\bf Conj}}
\newcommand{\graph}{\mbox{graph}}
\newcommand{\ind}{\mbox{index}}

\newcommand{\rank}{\mbox{rank}}
\newcommand{\id}{\mbox{id}}
\newcommand{\str}{\mbox{string}}
\newcommand{\RSK}{\mbox{RSK}}
\newcommand{\wt}{\mbox{wt}}
\setlength{\unitlength}{.75in}

\begin{abstract} 
This article  gives conjecturally correct algorithms
to  construct   canonical bases of the irreducible polynomial
representations and the matrix coordinate rings of the nonstandard 
quantum groups in GCT4 and GCT7, 
and   canonical bases of the dually paired  nonstandard deformations
of the symmetric group algebra therein. 
These  are generalizations of 
the canonical bases of the irreducible polynomial representations 
and the matrix coordinate ring of 
the standard 
quantum group, as  constructed by Kashiwara and Lusztig, and 
the Kazhdan-Lusztig basis  of  the Hecke algebra.
A positive ($\#P$-) formula for the well-known plethysm constants follows
from their conjectural properties and the duality and reciprocity 
conjectures in \cite{GCT7}.
\end{abstract}

\section{Introduction} \label{sintro}
Let $H$ be a complex connected classical reductive group, 
$X=V_\mu(H)$ its irreducible polynomial representation with highest weight
$\mu$, $G=GL(X)$, and 
$\rho: H\rightarrow G$  the representation map. 
Given a highest weight $\pi$ of $H$ and $\lambda$ of $G$, the plethysm
constant  $a_{\lambda,\mu}^\pi$ is defined to be 
the multiplicity of $V_\pi(H)$ in $V_\lambda(G)$, 
considered  an $H$-module via $\rho$. 
A fundamental problem in representation theory 
is to find a positive ($\#P$-) formula (rule) for the plethysm constant 
\cite{GCT7,stanley} akin to the Littlewood-Richardson rule.
Motivated by this problem, the article \cite{GCT7} constructs a 
quantization $\rho_q$ of the homomorphism $\rho$ in the form
\begin{equation} \label{eqhomoq}  \rho_q: H_q \rightarrow G_q^H,
\end{equation}
where $H_q$ is the standard (Drinfeld-Jimbo) quantum group
\cite{drinfeld,jimbo,rtf} associated with $H$ and $G_q^H$ is the 
new (possibly singular)
quantum group, called {\em the nonstandard quantum group} 
associated with $\rho$.
In the standard case, i.e., when $H=G$, this specializes to
 the standard quantum group, and in the Kronecker case, i.e., when
$H=GL(V)\times GL(W)$, $X=V\otimes W$ with the natural $H$ action, this
specializes to the nonstandard quantum group in \cite{GCT4}. 
Also constructed in \cite{GCT7} 
is a nonstandard quantization ${\cal B}_r^H(q)$ 
of the group algebra $\C[S_r]$, $S_r$ the symmetric group, whose
relationship with $G_q^H$ is conjecturally similar to that of the Hecke
algebra with the standard quantum group.

This article  gives conjecturally correct algorithms
for constructing   canonical bases of the irreducible polynomial
representations and the matrix coordinate ring of $G_q^H$ 
(Section~\ref{scanonical})
and  a canonical basis of ${\cal B}_r^H(q)$ (Section~\ref{scanonicalB}).
We call these 
{\em nonstandard canonical bases}.
They  are generalizations of 
the canonical bases of the irreducible polynomial representations 
and the matrix coordinate ring of 
the standard 
quantum group, as  constructed by Kashiwara and Lusztig
 \cite{kashiwaracrystal,kashiwaraglobal,lusztigcanonical,lusztigbook}, and 
the Kazhdan-Lusztig basis \cite{kazhdan} 
 of  the Hecke algebra.
A positive ($\#P$-) formula for the plethysm constant 
follows from their conjectural properties (Sections~\ref{sconjp} and
\ref{sconjb}), which are akin to those of the standard canonical basis, 
and the  conjectural  duality and reciprocity
between $G_q^H$ and ${\cal B}_r^H(q)$; cf. \cite{GCT7}.

Experimental evidence (Section~\ref{sexpevid}) suggests that these algorithms  should be correct.
But we can not prove this formally,
nor the required properties of the nonstandard canonical
bases. Mainly because we are unable to deal with   the  complexity 
of the minors of the nonstandard quantum group.
Specifically, in contrast to the elementary formula
for the Laplace expansion of
a minor   of the 
standard quantum group--which is
akin to the classical Laplace expansion  at $q=1$--the Laplace expansion 
of a minor of a nonstandard quantum group  is highly 
nonelementary; cf. \cite{GCT7}.
Its coefficients   depend on the
multiplicative structural constants of a canonical basis akin to 
the canonical basis of the coordinate ring of the 
standard quantum group as per Kashiwara and Lusztig.
In the Kronecker case, these  constants are 
conjecturally polynomials in $\Q[q,q^{-1}]$ with nonnegative coefficients,
and in general, polynomials 
with   a conjectural relaxed form of this property.
To prove this and  to get explicit formulae for the  minors in the
nonstandard setting, one  needs explicit interpretations 
for these structural constants  
in the spirit of the interpretation based on perverse sheaves
for the Kazhdan-Lusztig polynomials \cite{kazhdan1}
and the multiplicative structural constants of the canonical basis 
of the Drinfeld-Jimbo enveloping algebra \cite{lusztigbook}.
Thus  even to get explicit formulae 
for the minors of
the nonstandard quantum group a (nonstandard) extension of the theory of
perverse sheaves \cite{beilinson},
and  the underlying Riemann hypothesis over finite
fields \cite{weil2}  seems necessary.

Minors of the standard quantum group 
are in a sense the simplest  (basic) canonical basis elements 
in its matrix  coordinate ring.
That the simplest canonical basis elements
for the nonstandard quantum group--namely, its minors--are
already so nonelementary in contrast
to the standard case  indicates 
the possible difficulties that may be encountered in proving 
correctness of the algorithms  given here for
constructing nonstandard canonical bases.

\noindent {\bf Acknowledgement:} The author is grateful to David 
Kazhdan for helpful discussions and comments,
and to Milind Sohoni for
the help in explicit computations  in MATLAB.

\noindent {\bf Notation:} We use
the symbols $\pi$ and $\mu$ 
to denote labels of irreducible representations of the standard quantum
group and  the symbols $\alpha,\alpha_0,\alpha_1,\ldots$
to denote labels of irreducible representations of the nonstandard quantum
group. Thus   objects with subscripts 
$\pi$ and $\mu$ are standard
and the objects with subscripts $\alpha,\alpha_0,\ldots$ are nonstandard.

\section{Nonstandard canonical basis for $G_q^H$} 
\label{scanonical}
In this section we describe a conjecturally correct 
algorithm  for constructing 
the canonical basis of the matrix coordinate ring  
of the nonstandard quantum group 
$G_q^H$. 
We follow the same terminology as in \cite{klimyk} for the basic quantum 
group notions.

For the sake of simplicity, let us assume that $H=GL(V)$.
Let $H_q=GL_q(V)$ denote its standard (Drinfeld-Jimbo) quantization
\cite{drinfeld,jimbo,rtf},
and $M_q(V)$ the standard quantization of the matrix space $M(V)$.
Let ${\cal O}(M_q(V))$ be the coordinate ring of $M_q(V)$. We 
 call it {\em the matrix coordinate ring} of $GL_q(V)$. The coordinate ring
${\cal O}(GL_q(V))$ of $GL_q(V)$  is obtained by 
localizing ${\cal O}(M_q(V))$ at the quantum determinant of $GL_q(V)$.
Let ${\cal H}$ denote the Lie algebra of $H$, and $U_q({\cal H})$ the
Drinfeld-Jimbo universal enveloping algebra of $H_q=GL_q(V)$. 

To quantize the homomorphism $\rho: H \rightarrow G=GL(X)$ 
as in (\ref{eqhomoq}), 
the article \cite{GCT7} constructs 
a {\em nonstandard matrix coordinate ring}
${\cal O}(M_q^H(X))$ of a (virtual)
{\em nonstandard  matrix space}  $M_q^H(X)$, 
and then defines the {\em nonstandard quantized universal enveloping algebra} 
$U_q^H({\cal G})$  by dualization. The {\em nonstandard 
quantum group} $G_q^H$ is  the virtual object whose universal enveloping
algebra is $U_q^H({\cal G})$. 
The construction also yields natural bialgebra homomorphisms 
from $U_q({\cal H})$ to $U_q^H({\cal G})$ and from ${\cal O}(M_q^H(X))$ 
to ${\cal O}(M_q(V)$, thereby giving the desired quantizations of
the homomorphisms $U({\cal H}) \rightarrow U({\cal G})$ and
${\cal O}(M(X)) \rightarrow {\cal O}(M(V))$. This is what is meant 
by the quantization (\ref{eqhomoq}) of the representation map $\rho$.
The determinant of $G_q^H$ may vanish, and hence, we cannot, in general,
define its coordinate ring ${\cal O}(G_q^H)$ by localizing
${\cal O}(M_q^H(X))$. Fortunately, this does not matter since
 ${\cal O}(M_q^H(X))$ still has 
properties similar to that  of
the standard matrix coordinate ring ${\cal O}(M_q(V))$. Specifically,
it is cosemisimple.
This means all (finite dimensional)
{\em polynomial representations} of $G_q^H$, by which we mean
corepresentations of ${\cal O}(M_q^H(V))$,
are completely reducible. A nonstandard quantum analogue of
the Peter-Weyl theorem holds: i.e.,
\begin{equation} \label{eqpeternonstd}
{\cal O}(M_q^H(X))= \bigoplus_\alpha W_{q,\alpha}^* \otimes W_{q,\alpha},
\end{equation}
where $W_{q,\alpha}$ runs over all irreducible corepresentations of
${\cal O}(M_q^H(X))$. 
Furthermore, the nonstandard enveloping algebra $U_q^H({\cal G})$ is 
a bialgebra with a compact real form ($*$-structure).

The goal is  to construct a canonical basis for the
matrix coordinate ring ${\cal O}(M_q^H(X))$ akin to the 
canonical basis of the standard matrix coordinate ring 
${\cal O}(M_q(V))$ as per Kashiwara and Lusztig 
\cite{kashiwaracrystal,kashiwaraglobal,lusztigcanonical,lusztigbook}.

\subsection{The standard setting} \label{sstandard}
We begin by reviewing the basic scheme of Kashiwara and Lusztig
for constructing a canonical basis of the matrix coordinate ring 
${\cal O}(M_q(V))$. The canonical basis of the coordinate ring 
${\cal O}(GL_q(V))$ is obtained by localizing at the determinant.

Following Kashiwara, we first define a balanced triple.
Let $A$ and $\bar A$ be the ring of rational functions in $q$ regular at $q=0$
and $q=\infty$, respectively. Let $V$ be a $\Q(q)$-vector space, $L_0$ a 
sub-$A$-module ($A$-lattice) of $V$, $L_\infty$ a sub-$\bar A$-module 
($\bar A$-lattice)  of $V$, and $V_{\Q}$ 
a sub-$\Q[q,q^{-1}]$-module of $V$ such that
\[ V\cong \Q(q) \otimes_{\Q[q,q^{-1}]} V_{\Q}\cong \Q(q)\otimes_A L_0 
\cong \Q(q) \otimes_{\bar A} L_\infty.\] 
We say that $(V_\Q,L_0,L_\infty)$ is a {\em balanced triple} 
if any of the
following three equivalent conditions hold: 

\noindent (a) $E=V_{\Q}\cap L_0 \cap L_\infty \rightarrow L_0/q L_0$ is 
an isomorphism. 

\noindent (b) $E \rightarrow L_\infty/q^{-1} L_\infty$ is 
an isomorphism. 

\noindent (c) $\Q[q,q^{-1}] \otimes_\Q E \rightarrow V_\Q$,
$A\otimes_\Q E \rightarrow L_0$,  
$\bar A\otimes_\Q E \rightarrow L_\infty$ are isomorphisms.

Let $R={\cal O}(M_q(V))$.
Kashiwara constructs an $A$-submodule (lattice)
$L=L(R)\subset R$, an involution $-$ of $R$,
a $\Q[q,q^{-1}]$-submodule $R_\Q \subset R$, and a basis $B$ of
 $L / q L$ such that $(R_\Q, L,\bar L)$  is a balanced triple
and, letting $G$ denote the inverse of the isomorphism $ R_\Q \cap L \cap \bar
L \rightarrow L/ q L$, $\{ G(b) \ | \ b \in B\}$ is the canonical basis
of $R$. The pair  $(L,B)$, called the (upper)  crystal
base of $R$,
has the following form. By the $q$-analogue of the Peter-Weyl theorem 
for the standard quantum group, 
\begin{equation}\label{eqstdpeterweyl}
R=\oplus_\pi V_{q,\pi}^* \otimes V_{q,\pi}
\end{equation}
as a bi-$GL_q(V)$-module, where $V_{q,\pi}=V_{q,\pi}(V)$
is the irreducible polynomial
representation of $GL_q(V)$ with highest weight $\pi$. 
Let $(L_\pi, B_\pi)$ denote the upper crystal base of 
$V_{q,\pi}$. Then 
\begin{equation}
(L,B)=\oplus_\pi (L_\pi^*, B_\pi^*)
\otimes (L_\pi,B_\pi) 
\end{equation} 
 with appropriate normalization.

We  now describe  a construction of 
the upper crystal base  $(L_\pi,B_\pi)$ that can  be generalized to the
nonstandard setting.

Let $r$ be the size of the partition $\pi$. Choose any embedding 
$\rho: V_{q,\pi} \rightarrow V^{\otimes r}$ such that 
the highest weight vector of the image 
$V^\rho_{q,\pi}=\rho(V_{q,\pi})$ belongs to
the $A$-lattice $L(V)^{\otimes r}$ of $V^{\otimes r}$, where $L(V)$ 
denotes the lattice of $V$ generated by its standard basis $\{v_i\}$. 
We also assume that the highest weight vector does not belong to
$q L(V)^{\otimes r}$. 
Choose a Hermitian form on $V^{\otimes r}$ so that its  monomial
basis $\{v_{i_1}\otimes \cdots v_{i_r}\}$  is 
orthonormal. Let $V^{\rho,\bot}_{q,\pi}$ denote the orthogonal
complement of $V^{\rho}_{q,\pi}$. Since $GL_q(V)$ has a 
compact real form $U_q(V)$--i.e., the unitary compact subgroup in the
sense of Woronowicz \cite{wor1}--it follows that $V^{\rho,\bot}_{q,\pi}$
is  a $GL_q(V)$-module. Thus
$V^{\otimes r}= V^{\rho}_{q,\pi} \oplus V^{\rho,\bot}_{q,\pi}$
as a $GL_q(V)$-module. Let 
\[ L^{\rho}_\pi=L(V)^{\otimes r} \cap V^{\rho}_{q,\pi} \quad \mbox{and}
\quad L^{\rho,\bot}_\pi=L(V)^{\otimes r} \cap V^{\rho,\bot}_{q,\pi}.\] 
It follows from Kashiwara's work \cite{kashiwaracrystal} that 
$L(V)^{\otimes r} = L^{\rho}_\pi \oplus L^{\rho,\bot}_\pi$. 

Let $B(V)=\{b_i=\psi(v_i)\}$ denote the basis of $L(V)/q L(V)$, where $\psi: L(V)
\rightarrow L(V)/q L(V)$ is the natural projection. Let 
$B(V)^{\otimes r}=
\{b_{i_1}\otimes \cdots \otimes b_{i_r}\}$ denote the monomial basis 
of $B(V)^{\otimes r}$. Given $b \in B(V)^{\otimes r}$, 
let 
\[ b= \sum_{i_1,\ldots, i_r} f(b;i_1,\ldots,i_r) b_{i_1}\otimes \cdots \otimes 
b_{i_r},\]
be its expansion in the monomial basis. The set of monomials 
$b_{i_1}\otimes \cdots \otimes b_{i_r}$ such that the coefficients 
$f(b;i_1,\ldots,i_r)$ are nonzero is  called the {\em monomial support}
 of $b$. 

It  follows from the works of Kashiwara 
\cite{kashiwaracrystal} and Date et al \cite{date}  that 
$L^{\rho}_\pi/ q L^\rho_\pi$ has a unique basis $B^\rho_\pi$
 (up to scaling by constant multiples) such that the monomial supports of  its
elements are disjoint--in fact, one can choose $\rho$ so that the
monomial support of each basis element consists of just one distinct monomial.
This basis 
can be made completely unique by appropriate normalization.
Then $(L^\rho_\pi,B^{\rho}_\pi)$ coincides with the upper crystal base 
of $V_{q,\pi}$ as constructed by Kashiwara.
Furthermore, this crystal base does not depend 
on the embedding $\rho$ (up to isomorphism). Hence,
we let $(L_\pi,B_\pi)=(L^\rho_\pi,B^\rho_\pi)$ for 
any $\rho$ as above. Kashiwara \cite{kashiwaracrystal}
also shows that $L_\pi$ and 
$B_\pi \cup \{0\}$ are 
invariant under certain crystal operators $\tilde e_i$ and $\tilde f_i$
corresponding to the simple roots of $H_q$.

This  scheme for constructing the upper crystal base $(L_\pi,B_\pi)$
crucially depends on the existence of a compact real form 
$U_q(V) \subseteq GL_q(V)=H_q$.  Even the existence of a compact 
real form of the standard Drinfeld-Jimbo enveloping algebra 
$U_q({\cal H})$ suffices here.

\subsection{Nonstandard  triple} \label{striplen}
We  now generalize the preceding scheme to the nonstandard setting using 
the compact real form of  $U_q^H({\cal G})$,
whose existence is
proved in GCT7. 
The goal is to construct an analogous 
triple for the matrix coordinate ring
 $S={\cal O}(M_q^H(X))$ of $G_q^H$. It will turn out that this 
triple need not be balanced as in the standard case. 
We shall describe in Section~\ref{sglobalgl}  how a canonical basis can be
constructed from such a triple despite the lack of balance.

We begin by recalling  that
the $q$-analogue of the Peter-Weyl theorem  (\ref{eqpeternonstd}) in the
nonstandard setting need not hold over $\Q(q)$ unlike in the 
standard setting. It  holds only over
an appropriate algebraic extension $K$ of $\Q(q)$ \cite{GCT7}--thinking
of $q$ as a transcendental. 
It will be convenient to assume in what follows that $K$ is actually
an algebraic extension $\tilde \Q(q)$, where $\tilde \Q$ is the
algebraic closure of $\Q$.
We let $A_K$ and $\bar A_K$ be the subrings of algebraic functions in 
$K$ that are regular at $q=0$ and $q=\infty$, respectively. 
Let $K_\Q$ be the integral closure of $\Q[q,q^{-1}]$ in $K$.
Clearly, $K_\Q \cap A_K \cap \bar A_K = \hat \Q$, where
$\hat Q$ denotes the integral closure of $Q$ in $\tilde Q$.
In what follows,
we let $\hat Q, A_K,\bar A_K$ 
and $K_\Q$ play the role of $\Q, A,\bar A$ and $\Q[q,q^{-1}]$ in
Section~\ref{sstandard}. Thus by a lattice at $q=0$ we mean an $A_K$-lattice,
by a lattice at $q=\infty$ a $\bar A_K$-lattice, by a $\Q$-form,
a $K_\Q$-form (module). 
Similarly, instead of $\Q$-modules and 
$\Q(q)$-modules, we will be considering $\hat Q$-modules and $K$-modules.
That is,  in what follows 
we shall assuming that the underlying base field is $K$,
instead of $\Q(q)$.

Now let us  describe the construction of the upper crystal base of
an irreducible polynomial representation $W_{q,\alpha}$ of $G_q^H$. 
Let $X_q=V_{q,\mu}$ denote the quantization of $X=V_\mu$; i.e, 
the $H_q$-module with highest weight $\mu$. Since  the
underlying field is $K$,  $X_q$ is a $K$-module.
Let $\{z_i\}$ denote its standard $q$-orthonormal Gelfand-Tsetlin basis.
Let $\{x_i\}$ denote the rescaled version of Gelfand-Tsetlin basis 
(as described in Section 7.3.3  of \cite{klimyk}) so that 
there are no square roots in the explicit formulae for the action 
of the generators of the Drinfeld-Jimbo algebra $U_q{\cal H}$ on this basis.
Alternatively, we can  let $\{x_i\}$ be the standard (upper)
canonical basis \cite{kashiwaracrystal,lusztigbook}
of $X_q$ as an $H_q$-module. We assume that 
$\mbox{lim}_{q\rightarrow 0} (z_i -x_i)=0$; this can always be arranged.
In what follows, we 
sometimes denote $X_q$ by $X$. What is meant should be clear from
the context.
It is shown in \cite{GCT7} that 
$W_{q,\alpha}$ can be embedded in $X^{\otimes r}$ for an appropriate
$r$. We choose  an embedding 
$\rho: W_{q,\alpha} \rightarrow X^{\otimes r}$ as follows.

If $r=1$, $W_{q,\alpha}=X$, so this is trivial.
Otherwise, choose any $\beta$ of degree $r-1$ so that
$W_{q,\alpha}$ occurs as a $G^H_q$-submodule of $W_{q,\beta} \otimes X$. 
By semisimplicity, the latter is completely reducible as a $G^H_q$-module:
\begin{equation} \label{eqpieri1}
W_{q,\beta} \otimes X = \oplus_{\beta_j}  W_{q,\beta_j},
\end{equation}
where we assume that this is also a decomposition as a $K$-module.
Choose any $j$ such that $W_{q,\beta_j}\cong W_{q,\alpha}$. This fixes 
an embedding of $W_{q,\alpha}$ in $W_{q,\beta} \otimes X$.
(It is a plausible conjecture that the decomposition (~\ref{eqpieri1}) is
 multiplicity free. This would be  a conjectural
analogue of Pieri's rule in the nonstandard setting. It would imply
that the  embedding of $W_{q,\alpha}$ in $W_{q,\beta} \otimes X$
is unique. But this is not required here.). 
By induction, we have fixed an embedding of $W_{q,\beta}$ in
$X^{\otimes r-1}$. This fixes an embedding $\rho$ of $W_{q,\alpha}$ in
$X^{\otimes r}$ (among many possible choices).
Let $W^\rho_{q,\alpha}=\rho(W_{q,\alpha})$  be its image.

Choose a Hermitian form on $X^{\otimes r}$ so that its  Gelfand-Tsetlin basis
$\{z_{i_1}\otimes \cdots  \otimes z_{i_r}\}$ is 
orthonormal. Let $W^{\rho,\bot}_{q,\alpha}$ denote the orthogonal
complement of $W^{\rho}_{q,\alpha}$. Since $U_q^H({\cal G})$  has a 
compact real form \cite{GCT7} 
such that $X_q^{\otimes r}$ is its unitary representation with respect to 
this  Hermitian form,
it follows that $W^{\rho,\bot}_{q,\alpha}$
is  a $G_q^H$-module. Thus
$X^{\otimes r}= W^{\rho}_{q,\alpha} \oplus W^{\rho,\bot}_{q,\alpha}$
as a $G_q^H$-module. Let 
\[ L^{\rho}_\alpha=
L(X)^{\otimes r} \cap W^{\rho}_{q,\alpha} \quad \mbox{and}
\quad L^{\rho,\bot}_\alpha=L(X)^{\otimes r} \cap W^{\rho,\bot}_{q,\alpha}.\] 
Then, in analogy with  Kashiwara's work mentioned above:

\begin{prop}
$L(X)^{\otimes r} = L^{\rho}_\alpha \oplus L^{\rho,\bot}_\alpha$. 
\end{prop}
\proof
The r.h.s. is clearly contained in the l.h.s. 
To show the converse it suffices to show that 
$L^{\rho}_\alpha$ and $L^{\rho,\bot}_\alpha$ are projections of the
lattice $L(X)^{\otimes r}$ onto 
$W^{\rho}_{q,\alpha}$ and  $W^{\rho,\bot}_{q,\alpha}$ respectively.
Let us show this for $L^{\rho}_\alpha$, the other case being similar.
Clearly, the projection of $L(X)^{\otimes r}$ onto  $W^{\rho}_{q,\alpha}$
contains $L^{\rho}_\alpha$. We only have to show that
the projection $\hat y$  of any $y\in L(X)^{\otimes r}$ onto
$W^{\rho}_{q,\alpha}$ also belongs to the lattice $L(X)^{\otimes r}$,
and hence to $L^{\rho}_\alpha$. 
Since $y\in L(X)^{\otimes r}$, its length $|y|$ w.r.t. the preceding
Hermitian form tends to  a well defined nonnegative real number
as $q\rightarrow 0$. 
Since, the projection $y \rightarrow \hat y$ is orthonormal, 
the length $|\hat y|$ of $\hat y$ is at most $|y|$, and
hence also tends to a well defined nonnegative
real number 
as $q\rightarrow 0$. This means $\hat y$ is regular at $q=0$ and
hence belongs to  $L(X)^{\otimes r}$.
\qed

Let $B(X)=\{b_i=\phi(x_i)\}$
denote the basis of the $\tilde Q$-module $L(X)/q L(X)$,
where $\phi: L(X)
\rightarrow L(X)/q L(X)$ is the natural projection (the $b_i$'s in this
section are different from the $b_i$'s in Section~\ref{sstandard}). Let 
$B(X)^{\otimes r}=
\{b_{i_1}\otimes \cdots \otimes b_{i_r}\}$ denote the monomial basis 
of $L(X)^{\otimes r}$. Given $b \in B(X)^{\otimes r}$, 
let 
\[ b= 
\sum_{i_1,\ldots, i_r} g(b;i_1,\ldots,i_r) b_{i_1}\otimes \cdots \otimes 
b_{i_r},\]
be its expansion in the monomial basis. The set of monomials 
$b_{i_1}\otimes \cdots \otimes b_{i_r}$ such that the coefficients 
$g(b;i_1,\ldots,i_r)$ are nonzero is
called the {\em monomial support} of $b$. 

In analogy with the  work of Kashiwara  and Date et al   mentioned 
above, it may be conjectured that:

\begin{conj} {\bf (Existence of (local) crystal basis)} \label{clocalcrystalb}

The $\tilde Q$-module
$L^{\rho}_\alpha/ q L^\rho_\alpha$ has a unique basis $B^\rho_\alpha$
(up to scaling by constant multiples, and which
can be made completely unique by appropriate normalization)
such that:
\begin{enumerate} 
\item The monomial supports of its
elements are disjoint, 
\item  $L^{\rho}_\alpha$ and $B^{\rho}_\alpha \cup \{0\}$ are 
invariant under Kashiwara's crystal operators $\tilde e_i$ and $\tilde f_i$
for $H_q$ 
(which are well defined since 
$U_q({\cal H})$ is a subalgebra of $U_q^H({\cal G})$), and
$(L^{\rho}_\alpha,B^{\rho}_\alpha)$ is a local crystal basis,
in the sense of Kashiwara, of
$W_{q,\alpha}$ as an $H_q$-module 
\end{enumerate} 
Furthermore, this crystal base does not depend 
on the embedding $\rho$ (up to isomorphism). 
\end{conj} 
This  conjecture has
been supported by experimental evidence; cf. Section~\ref{expcrystalevi}. 

Assuming this, 
we  let $(L_\alpha,B_\alpha)=(L^\rho_\alpha,B^\rho_\alpha)$ for 
any $\rho$ as above. It is  called the {\em upper crystal base} 
of $W_{q,\alpha}$. 

In standard setting, the embedding $\rho: V_{q,\pi} \rightarrow
V^{\otimes r}$ in Section~\ref{sstandard} can be chosen so that the support of
each basis element in $B_\pi^\rho$ consists of just one monomial. That is,
so that each $b \in B_\pi^\rho$ is a monomial in $B(V)^{\otimes r}$. 
In the nonstandard setting, it is  not always  possible to choose 
an embedding $\rho: W_{q,\alpha} \rightarrow X^{\otimes r}$ so that
the support of each basis element in $B^\rho_\alpha$ in
Conjecture~\ref{sstandard} consists of just one monomial;
 cf. Section~\ref{expcrystalevi} for a
counterexample.

In view of the nonstandard 
$q$-analogue of the Peter-Weyl theorem (\ref{eqpeternonstd}), 
$S={\cal O}(M_q^H(X))$
has a natural upper crystal base 
\begin{equation} \label{eqlocalpeter}
(L(S),B(S))=
\bigoplus_\alpha (L_\alpha^*,B_\alpha^*) \otimes (L_\alpha,B_\alpha)
\end{equation}
at $q=0$ (with appropriate normalization). 
Let $S_\Q$   be the $K_\Q$-forms ($K_\Q$-subring)
of $S$ generated by 
the entries $u^i_j$ of the generic nonstandard quantum matrix ${\bf u}$;
recall (cf. \cite{GCT7})
that $S$ is the quotient of $\C\langle {\bf u} \rangle$
modulo appropriate quadratic relations over the entries $u^i_j$'s of
${\bf u}$.
We  define an involution $-$ over $S$ by a natural generalization 
of its definition in the standard setting. Specifically, $S$ is a
${\cal O}(M_q^H(X))$-bi-comodule. Via the homomorphism from
${\cal O}(M_q^H(X))$ to ${\cal O}(M_q(V))$,
$S$ is also ${\cal O}(M_q(V))$-bi-comodule; i.e., an $H_q$-bi-module.
In the spirit of \cite{kashiwaraglobal},
for any  $u$ and $v$ in $S$ with  $H_q$-bi-weights $(\lambda_r,\lambda_l)$
and $(\mu_r,\mu_l)$, let $\overline{u v} = q^{(\lambda_r,\mu_r)-(\lambda_l,
\mu_l)} \bar v \bar u$, where $(\ ,\ )$ denotes the usual inner product 
in the $H_q$-weight space. We let $\bar u^i_j=u^i_j$, and $\bar q=q^{-1}$.
This defines $-$ on $S$ completely.

Applying $-$ to $(L(S), B(S))$, we get an upper crystal base 
$(\bar L(S), \bar B(S))$ at $q=\infty$. In analogy with the standard
setting, we can now ask:

\begin{question} \label{cbalance}
Is the triple $(S_\Q,L(S),\bar L(S))$  balanced?
In other words, is the map 
$\psi: E=S_\Q \cup  L(S) \cup \bar L(S) \rightarrow L(S)/q L(S)$ 
of $\hat Q$-modules an
isomorphism.
\end{question} 
If it were, we could have defined the canonical basis of $S$ by 
a globalization procedure very much
as in the standard setting, i.e., as $\psi^{-1}(B(S))$.
But, as it turns out, this need not be so; cf. Section~\ref{sexphsl2gsl4}. 
Specifically, for a given $b \in B(S)$, the fibre $\psi^{-1}(b)$ 
need not be singleton. 
This is the major difference from  the standard 
setting that
makes construction of the  canonical basis of $S$ in the nonstandard
setting much more complex. 
We turn to this in the next section.

\subsection{Nonstandard globalization via minimization of degree complexity}
\label{sglobalgl}
We  now give a conjectural
procedure for choosing an unambiguously defined
canonical element $y_b \in \psi^{-1}(b)$, $b \in B(S)$.
The set  $\{y_b \ | b \in B(S) \}$ will then be the {\em canonical basis}
of $S$.

Fix $r$. Let $S_r$ denote the degree $r$ component of $S$.
Let ${\cal A}={\cal A}^H_r={\cal A}_r^H(q)$ 
be the nonstandard $q$-Schur algebra 
\cite{GCT7}, which is the dual $S_r^*=\mbox{Hom}(S_r,K)$ of 
$S_r$. A polynomial  irreducible $G^H_q$-module $W_{q,\alpha}$ of 
degree $r$ is an irreducible ${\cal A}$-module, and conversely
every irreducible ${\cal A}$-module is of this form.
Furthermore, the $q$-analogue of the Peter-Weyl theorem also holds for
${\cal A}$:

\begin{equation} \label{eqpetercalA}
{\cal A}=\oplus_{\alpha} W^*_{q,\alpha} \otimes W_{q,\alpha}.
\end{equation}

For reasons given later (cf. Remark 1 below),
it will be more convenient to
construct the  canonical basis 
of  ${\cal A}^H_r$ first. The canonical basis of $S_r$ will then
be  defined to be its dual.

Let 
\[ \langle \ \rangle: {\cal A} \otimes S_r \rightarrow K \] be
the natural pairing.
The lattice $L({\cal A})$ is defined to be the dual lattice of $L(S_r)$: 
\[ 
L({\cal A})=\{a \in {\cal A} \ | \ \langle a, L(S_r)\rangle \subseteq A_K\}.
\] 
The automorphism $-$ of ${\cal A}$ is defined by:
\[ \langle \bar a, s\rangle = \langle a, \bar s\rangle ^-. \]
We define  $\bar L({\cal A})$ by applying $-$ to $L({\cal A})$. 
We define the $\Q$-form (i.e. $K_\Q$-form) ${\cal A}_\Q$ of ${\cal A}$ by:

\[ {\cal A}_\Q= \{ a \in {\cal A} \ | \ \langle a, S_{r,\Q} \rangle 
\subseteq K_\Q\}, \]
where $S_{r,\Q} = S_r \cap S_\Q$. 
We define the basis $B({\cal A})$ of $L({\cal A})/q L({\cal A})$ to 
be the dual of $B(S)$. Thus $(L({\cal A}),B({\cal A}))$ is the
local crystal basis of ${\cal A}$ as per Conjecture~\ref{clocalcrystalb}
 and we have
an analogue of (\ref{eqlocalpeter}):
\begin{equation} \label{eqlocalpetercalA}
(L({\cal A}),B({\cal A}))=
\bigoplus_\alpha (L_\alpha^*,B_\alpha^*) \otimes (L_\alpha,B_\alpha)
\end{equation}

This defines the triple $({\cal A}_\Q, L({\cal A}), \bar L({\cal A}))$ 
for ${\cal A}$. It
need  not be balanced in the standard sense, just like
the triple $(S_\Q,L(S),\bar L(S))$ above.
Now we describe  the conjectural construction of a canonical basis of
${\cal A}$.

Each component $B_\alpha^* \otimes B_\alpha$ of $B({\cal A})$ 
has the left and right action of Kashiwara's crystal operators for
$H_q$ (cf. Conjecture~\ref{clocalcrystalb}). Thus, by \cite{kashiwara1},
 we get a crystal
graph on $B_\alpha^* \otimes B_\alpha$, whose each connected component
intutively corresponds to an irreducible $H_q$-bi-submodule
$V_{q,\mu_1}^* \otimes V_{q,\mu_2}$  of 
the component $W_{q,\alpha}^* \otimes W_{q,\alpha}$ in the
Peter-Weyl decomposition of ${\cal A}$ (\ref{eqpetercalA}). With 
each element $b\in B({\cal A})$ that occurs in such a connected 
component, we associate the triple $T(b)=(\alpha,\mu_1,\mu_2)$. 
We call it the {\em type} of $b$. The types $T(b)$'s 
can be partially ordered as follows.

First, put a partial  order $\le$ on 
the  labels $\alpha$  of
polynomial irreducible ${\cal A}$-modules $W_{q,\alpha}$ as follows. 
Consider $W_{q,\alpha}$  as an $H_q$-module. 
Let $\mu(\alpha)$ denote the highest weight
in  $W_{q,\alpha}$ as an $H_q$-module.  There may be several highest 
weight vectors in $W_{q,\alpha}$, since $W_{q,\alpha}$ need not
be irreducible as an $H_q$-module. That is fine. 
Let $\le$ denote the usual 
partial order on the highest weights of $H_q$-modules:
$\mu_1 \le \mu_2$ iff $\mu_2-\mu_1 \in \sum_\tau \N \tau$ 
where $\tau$ ranges over the simple positive roots of $H$. 
We say $\alpha \le \alpha'$ iff $\mu(\alpha) \le \mu(\alpha')$.

Now observe that each
component $W_{q,\alpha}^* \otimes W_{q,\alpha}$ in the
Peter-Weyl decomposition of ${\cal A}$ is an $H_q$-bimodule; i.e.,
has a left and right action of $H_q$. 
With each irreducible $H_q$-bimodule in this component 
isomorphic to $V_{q,\mu_1} \otimes V_{q,\mu_2}$, where 
$\mu_1$ and $\mu_2$ are highest left and right weights of $H_q$, we associate
the  type $T=(\alpha,\mu_1,\mu_2)$.
Put a partial order, which we shall again denote by $\le$,
on the types $T$   as per the partial
order $\le$ on the individual components. The type  $T(b)$  associated with
each $b \in B({\cal A})$ above is  similar to this type.
So this also
puts a partial order on the types $T(b)$'s.

Next fix a $b \in B({\cal A})$. Let $T=T(b)$ be its type.
We shall  associate a 
canonical basis element $y_b$ with each such  $b$ by induction on its type 
using the preceding partial order.
The set $\{y_b\}$ will then  be the required canonical basis of ${\cal A}$.

Let ${\cal A}^{\le T}$ denote the span of all $H_q$-bimodules in
${\cal A}$ of types less than or equal to $T$ as per $\le$.
Let 
\[ L({\cal A}^{\le T})= L({\cal A}) \cap {\cal A}^{\le T }.\]
We define $\bar L({\cal A})^{\le T}$, and ${\cal A}^{\le T }_\Q$ similarly.
Consider the natural projection 
\[ \psi_T: {\cal A}^{\le T }_\Q \cap 
 L({\cal A})^{\le T} \cap \bar L({\cal A})^{\le T} \rightarrow 
 L({\cal A})^{\le T} / q  L({\cal A})^{\le T}.\] 
Let $\psi_T^{-1}(b)$ be the fibre of $b$. If this fibre were to contain
a unique element, then we can simply let $y_b$ be this unique element. 
But this need not be so, because the triple 
$({\cal A}^{\le T }_\Q,
 L({\cal A})^{\le T}, \bar L({\cal A})^{\le T})$ need not be
balanced. 
So we have to resolve the ambiguity in some canonical way.
Towards this end, we shall associate with each element 
in ${\cal A}_Q \cap L({\cal A})$ a  complexity measure,
called its {\em degree complexity}. We shall then define 
$y_b$ to be the element in $\psi_T^{-1}(b)$ of minimum degree 
complexity--it would be conjecturally unique; cf. 
Conjecture~\ref{cuniquemindegleft} below.
This scheme is in the  spirit of
\cite{kazhdan}
 where each element of the  Kazhdan-Lusztig basis of the Hecke algebra
is defined to be an element of minimum degree in a certain sense.

So let us define the degree complexity 
of an element $y \in {\cal A}_Q \cap L({\cal A})$. Since 
$X_q^{\otimes r}$ is a represention of ${\cal A}={\cal A}^H_r(q)$,
we have the injection 
\[ \eta: {\cal A} \hookrightarrow Z=\mbox{End}(X_q^{\otimes r})= 
(X_q^{\otimes r})^* \otimes X_q^{\otimes r}.\] 
Let $L(Z)=L(X_q^{\otimes r})^* \otimes L(X_q^{\otimes r})$ be the
lattice associated with $Z$. The $\Q$-form (or rather $K_\Q$-form)
$Z_\Q$
is defined similarly. Then 
\begin{prop} \label{platticeinjection}
The embedding $\eta$ injects the $\Q$-form ${\cal A}_\Q$ into
$Z_\Q$. Furthermore,
assuming Conjecture~\ref{clocalcrystalb}, $\eta$ also injects the lattice
$L({\cal A})$ into $L(Z)$.
\end{prop} 
The proof is easy. (To be filled in).

Fix the upper canonical basis $\{x_i\}$ of $X_q$ as an $H_q$-module.
Let $\{x_i^*\}$ be the dual canonical basis of $X_q^*$. This
fixes the upper canonical basis $CB(Z)$ of $Z$, whose each element 
is of the form 
\[z_{i_1,\ldots,i_r;j_1,\ldots,j_r}=
x^*_{i_1} \otimes \cdots \otimes x^*_{i_r} \otimes
x_{j_1} \otimes \cdots \otimes x_{j_r}.\]
It is also a basis of the $\Q$-form $Z_\Q$ and the lattice
$L(Z)$. Now given any $y \in {\cal A}$, let $w=\eta(y)$. 
Express $w$ in  the canonical basis of $Z$: 
\begin{equation} \label{eqyincanZ}
 w = \eta(y)=\sum_{z} 
a(y,z) z,
\end{equation}
where $z$ ranges of the basis elements in $CB(Z)$.
Since $w \in L(Z) \cap Z_\Q$, each
$a(y,z) \in A_K \cap K_\Q$. 
This means it is integral over $\Q[q]$ and hence has 
a well defined  degree $d(y,z)$ at $q=0$ (the same as the order
of its pole at $q=\infty$); if $a(y,z)=0$, we define $d(y,z)=-\infty$. 
We define {\em  the degree complexity} $d(y)$ of $y$ to be the tuple
$\langle\ldots,  d(y,z), \ldots  \rangle$
of these degrees.
We put a partial order on the degree complexity as follows.

Let $U=U_q(H)$ be the Drinfeld-Jimbo enveloping algebra of $H_q$,
$U^-$ the subalgebra generated by its generators $F_i$'s. For
any string $\nu=\nu_1,\nu_2,\ldots$ of positive integers, let
$U_\nu^{-}$ be the subspace of $U^-$ spanned by the words in $F_i$'s in which
each $F_i$ occurs occurs $\nu_i$ times. 
Given a canonical basis element $x$ of $X_q$, we define its
length to be  $|x|=\sum_i \nu_i$, 
where  $x \in U^-_\nu x_0$, and  $x_0$ is the highest weight vector of
$X_q$. Order the canonical basis elements of $X_q$ as per the reverse order
on their lengths; so that $x_0$ is the highest element in this order.
Put a similar order on $X_q^*$. This puts an induced
partial order 
\ignore{
Since $Z$ is an $H_q$-bi-module, each $z \in CB(Z)$
has an $H_q$-bi-weight. The partial order $\le$ on $H_q$-weights also
puts a partial order on $H_q$-bi-weights. This also puts a
partial order $\le $}
on the elements of the canonical basis $CB(Z)$ of $Z$.
We let $<$ denote the strict less than relation
as per this partial order.
Given $y$ and $y'$, and letting 
$w=\eta(y), w'=\eta(y')$, we say that $d(y) \le d(y')$ if
for every $z$:
either $d(y,z)\le  d(y',z)$, or for some $\bar z < z$ 
$d(y,\bar z) < d(y',\bar z)$.

\begin{conj}  \label{cuniquemindegleft} {\bf (Minimum degree)}
The fibre $\psi_T^{-1}(b)$ contains a unique element $y_b$ of 
minimum degree complexity. Minimum  means  $d(y_b)\le d(y)$, as per
the ordering on the degree tuples above,
for any $y \in \psi_T^{-1}(b)$. 
\end{conj} 

We call $y_b$ the {\em canonical basis element} 
associated with $b$, and the set
$\{y_b\}$  the {\em canonical basis} $CB({\cal A})$  of ${\cal A}$. 
The canonical basis  $CB(S_r)=\{x_b\}$ of $S_r$ is defined to be its dual.
The canonical basis $CB(S)$ of $S$ is $\cup_r CB(S_r)$.

\noindent {\bf Remark 1:}
The reader may wonder why we defined the canonical basis of ${\cal A}$ first,
and that of $S_r$ later, as its dual. Can we define the canonical
basis of $S_r$ directly? The algorithm for $S_r$ would be similar
as above. The main problem is to define the  degree  complexity
of an element  $x\in L(S_r)\cap S_\Q$. 
We have a natural projection from $Z=\mbox{End}(X_q^{\otimes d})$
to $S_r$, but not a natural injection
that injects $S_{r,\Q}=S_r \cap S_\Q$ into $Z_\Q$.
So the analogue of Proposition~\ref{platticeinjection} does not hold. 

\ignore{
\noindent {\bf Remark 2:} 
The definition of degree complexity above depends on the injection
of ${\cal A}$ into $Z$. Can we define the degree complexity purely internally
within ${\cal A}$, without having to embed ${\cal A}$ in $Z$?
In Section~\ref{sinternal}, we shall give an alternative,
conjectural internal definition of
degree complexity for $S_r$ using which we then get a direct 
algorithm for constructing the canonical basis of $S_r$ and hence $S$,
without having to go through ${\cal A}$ first.
But that definition  is not as efficient or satisfactory
as the definition of degree complexity above. 
}

\noindent {\bf Remark 2:} 
In the standard settting, Kashiwara and Lusztig 
give an efficient scheme for constructing each canonical basis 
element of the standard matrix coordinate ring ${\cal O}(M_q(V))$.
We do not have here an analogous efficient algorithm 
for constructing the canonical basis element $y_b$ in 
Conjecture~\ref{cuniquemindegleft}.
In the standard setting an efficient
cosntruction was possible  because the standard Drinfeld-Jimbo universal
enveloping algebra has an explicit presentation in terms of generators
and  defining relations.  This explicit presentation
is   crucially used in the  construction
of the standard canonical basis and also in proving correctness of
the construction.

The nonstandard universal algebra does not have an analogous explicit 
presentation as yet; cf. \cite{GCT7}. 
For this we need  explicit formulae   for the 
coefficients of the Laplace relation in \cite{GCT7}
among  the simplest nonstandard canonical basis elements in $S$, namely
nonstandard minors, since  as discussed in \cite{GCT7}, it is 
the mother relation in the representation theory of the nonstandard 
quantum group (just as in the standard setting).
That is, we need explicit interpretation for these coefficients
in the spirit of the explicit interpretation 
for the coefficients of the Kazhdan-Lusztig polynomials in terms perverse
sheaves.  This is the basic core problem 
that needs to be solved to prove that the preceding
algorithm for constructing the nonstandard canonical basis  is correct
and to give an explicit, efficient construction of $y_b$.
Furthermore, if  explicit presentation of the nonstandard universal
algebra is  so  nonelementary, as against the
elementary explicit presentation of the standard (Drinfled-Jimbo)
enveloping algebra,
then the task of proving correctness may be  formidable.


\subsection{Properties of the nonstandard canonical basis} \label{sconjp}
It may be conjectured that the nonstandard canonical bases $CB(S)$ and
$CB({\cal A})$
have properties
akin to the standard canonical basis of ${\cal O}(M_q(V))$:

\subsubsection{Cellular decomposition}

\begin{conj}  {\bf (Cell decomposition)} \label{ccelldecompleft}
The refined Peter-Weyl theorem, akin to the one proved by 
Lusztig \cite{lusztigbook} in the standard setting, holds
for $CB(S)$ and $CB({\cal A})$.
\end{conj} 

This  means the left, right and two-sided cells in ${\cal O}(M_q^H(X))$ 
with respect to  $CB(S)$ 
yield irreducible left, right, and two-sided (polynomial)
representations of $G_q^H$.
And furthermore, the left sub-cells of each left cell with respect
to the restricted $H_q$-action yield irreducible $H_q$-representations.
The left cell of ${\cal O}(M_q^H(X))$ is defined as follows.
Given $b\in CB(S)$, 
let $\Delta(b)= \sum_{b',b''}  c_b^{b',b''} 
b' \otimes b''$, where $\Delta$ denotes
comultiplication. Then we say that $b'' \leftarrow_L b$
if $c_b^{b',b''}$ is nonzero
for some $b'$. Let $<_L$ denote the transitive closure of $\leftarrow_L$.
Using $<_L$ we  define left cells in a natural way.
The right and two-sided cells are defined similarly. 
The left, right and two-sided subcells with respect to the $H_q$-action
are defined similarly. The definitions for $CB({\cal A})$ are similar.

By restricting 
the canonical basis to any left cell corresponding
to an irreducible polynomial representation $W_{q,\alpha}$ of $G^H_q$,
we get the canonical basis of $W_{q,\alpha}$; here the choice of the
left cell would conjecturally not matter up to scaling.

\subsubsection{Positivity in the Kronecker case}
\begin{conj}  {\bf (Positivity)} \label{cposkroneckerleft}
In the Kronecker case--i.e. when $H=GL(V)\times GL(W)$, 
$X=V\otimes W$ with the natural $H$-action--each
coefficient $g(q)$ of any canonical basis element in $CB({\cal A}_r^H)$
in the basis $CB(Z)$ (cf. eq.(\ref{eqyincanZ}))
is a positive polynomial in $q$.

Similarly, 
each  multiplicative or comultiplicative structural constant of
$CB({\cal A}_r^H)$  is  of the form
$\stackrel{+}{-} (q-\f{1}{q})^a f(q)$, where $a$ is a nonnegative
integer and $f(q)$ is a $-$-invariant
positive and unimodal polynomial in $q$ and $q^{-1}$.

The same also for $CB(S)$.
\end{conj}

Here by  a positive polynomial  we mean
a polynomial  with nonnegative rational coefficients.
By  unimodality of the  ($-$-invariant) polynomial $f(q)$, we mean 
its coefficients 
$f_{-k}, \ldots, f_k$
satisfy the condition
\[ f_{-k} \le f_{-k+1} \le \cdots \le f_{-1} \le 
f_{0} \ge f_{1}
\ge \cdots f_{k}.\]
By multiplicative structural constants, we mean the coefficients
$m^{b''}_{b,b'}$ 
in the expansion 
\[ 
b b'= \sum_{b''} m^{b''}_{b,b'} b'',\]
for $b,b' \in CB(S)$, and with $b''$ ranging over the elements in $CB(S)$.
Comultiplicative structural constants are defined similarly. 

For experimental evidence for  the dual nonstandard algebra
${\cal B}^H_r(q)$, see Section~\ref{sexpkronecker}.

Presumably, the nonnegative coefficients of $g(q)$ and $f(q)$ may 
have a topological interpretation in the spirit of that
for the coefficients of the Kazhdan-Lusztig polynomials, and
unimodality of $f(q)$ may be a consequence of some result
akin to the Hard Lefschetz theorem
in the spirit of the results on unimodal sequences in \cite{stanley}.

The Kronecker case is  fundamental
because ${\cal O}(M_q^H(X))$ therein is a nonflat deformation of
${\cal O}(M(X))$, though $\C_q^H[X]$ is a flat deformation of $\C[X]$.
Thus to prove 
the  positivity Conjecture~\ref{cposkroneckerleft}, some nonstandard 
extension of the theory of perverse sheaves and the work surrounding
the Riemann hypothesis over finite fields \cite{beilinson,weil2} that can
deal with nonflat, noncommutative varieties like ${\cal O}(M_q^H(X))$
may be neeeded.

\subsubsection{Nonstandard positivity and saturation} \label{ssaturationleft}
In the Kronecker case,
the braided symmetric algebra $\C^H_q[X]$ is a flat deformation 
of $\C[X]$; i.e. $\dim(C_q^{H,r}[X])=\dim(\C^r[X])$. 
But when  $\C^H_q[X]$ is a nonflat deformation 
the situation is  much more complex. 
To see this, 
let $c$ be a coefficient of any canonical basis element in $CB({\cal A}_r^H)$
in the basis $CB(Z)$.
As observed after  eq.(\ref{eqyincanZ}), it is integral over 
$\Q[q]$. Hence every coefficient of its minimal polynomial $f_c$
is a polynomial in $q$ with rational coefficients.
In the spirit of Conjecture~\ref{cposkroneckerleft},
 one may ask if the coefficients of this
polynomial are always nonnegative.
Unfortunately, this need not be so; cf. Section~\ref{sexphsl2gsl4} 
for  counterexamples in the dual setting of ${\cal B}_r^H(q)$.
The following is a relaxed version of Conjecture~\ref{cposkroneckerleft}
for the general  case. 

\begin{conj} {\bf (Saturation)} \label{cnonstdposleft0}
Each coefficient $a(q)$ of $f_c$ is a {\em saturated} polynomial
(in the terminology of \cite{GCT6}); 
this means $a(1)$ is a positive rational if 
$a(q)$ is not an identically zero polynomial.

Similarly, each coefficient $b(q)$ of the minimal polynomial 
of a multipliciative or comultiplicative structural constant 
of $CB({\cal A}_r^H)$ can be expressed in 
the form 
\begin{equation} \label{eqbexpli2}
 b(q) = (-1)^{e}  (q-\f{1}{q})^{e'}
c(q)
\end{equation}
for some nonnegative integers $e,e'$, where 
\begin{enumerate} 
\item $e$ is chosen so that 
the middle term of $c(q)$ is positive; here by the middle term we mean
the coefficient of $q^i$ for the smallest $i$ such that this coefficient is
nonzero,  and 
\item  $c(q)$ is a saturated
polynomial in $q$ and $q^{-1}$--this again means 
$c(1)$ is a positive rational if $c(q)$ is not an identically zero polynomial.
\end{enumerate}
\end{conj}

In the context of the plethysm problem, one is finally interested
in the behaviour of $b(q)$ at $q=1$ (cf. Section~\ref{scomplexity}),
so this relaxed saturation form
of positivity should be sufficient; see also 
\cite{GCT6} for the importance of  saturation
in the context of the  flip in GCT.
A stronger positivity
 conjecture that would specialize to Conjecture~\ref{cposkroneckerleft}
in the Kronecker case is:

\begin{conj} {\bf (Nonstandard Positivity)} \label{cnonstdposleft1}
Each polynomial $c(q)$ in Conjecture~\ref{cnonstdposleft0} is  almost
positive and unimodal. That is, it is of the form
$c^0(q)+c^1(q)$, where,  if $c(q)$ is not identically zero, 
\begin{enumerate}
\item $c^0(1) >> |c^1(1)|$, where $>>$ means much greater as $r \rightarrow
\infty$, 
and more generally,
\item $c^0(q)$ is a dominant positive unimodal  polynomial,
and $c^1(q)$ is a very small  error-correction polynomial.
Specifically, 
\[||c^1(q)||/||c^0(q)|| \le 1/\poly(\bitlength{\mu},\bitlength{\pi},\bitlength{r}),\]
 where $\mu$ and $\pi$ are as in the plethysm problem 
(cf. begining of Section~\ref{sintro}), $\bitlength{\quad}$ denotes 
the bitlength-of-specification function, and $\poly(\quad)$ means polynomial
of a fixed (constant) degree in the specified bitlengths, and $||\quad ||$
denotes the $L_2$-norm of the coefficient vector of the polynomial.
\end{enumerate}
\end{conj}

See Section~\ref{sexphsl2gsl4} for experimental evidence for 
Conjectures~\ref{cnonstdposleft0}-\ref{cnonstdposleft1}
in the dual setting of ${\cal B}_r^H(q)$.

Presumably, the nonnegative coefficients of such $c^0(q)$  may again
have a topological interpretation in the spirit of that
for the coefficients of the Kazhdan-Lusztig polynomials, and
unimodality of $c^0(q)$ may again be a consequence of some result
akin to the Hard Lefschetz theorem.
The  correction polynomial $c^1(q)$  may also have a
topological  interpretation that  depends on a cohomological
measure of nonflatness of $C_q^H[X]$. 
In the Kronecker case, when $\C_q^H[X]$ is a
flat deformation of $\C[X]$, this correction would then vanish, and
Conjecture~\ref{cnonstdposleft1} would reduce to 
Conjecture~\ref{cposkroneckerleft}.
Furthermore,  the conjectural   nonnegative value of $c(1)$ 
may also have an interpretation akin to the representation-theoretic
interpretation for the values of the Kazhdan-Lusztig polynomials at $q=1$.

If nonstandard extension of the work surronding the Riemann hypothesis
over finite fields as needed to prove the positivity 
Conjecture~ref{cposkroneckerleft} in the Kronecker case can be found,
that may open the 
way for investigating the more complicated nonstandard form of positivity 
in Conjecture~\ref{cnonstdposleft1}.

\section{Nonstandard canonical basis of ${\cal B}_r^H$} \label{scanonicalB}
Let ${\cal B}^H_r={\cal B}^H_r(q)$ be the nonstandard quantization of 
the symmetric group ring $\C[S_r]$ in \cite{GCT7}.
In this section,  we describe  an analogous conjecturally correct 
algorithm  for constructing a 
nonstandard canonical basis $E(r)$ of  ${\cal B}^H_r$.
In the standard setting--i.e., when $H=G$--this
basis would conjecturally specialize to the
Kazhdan-Lusztig basis of the Hecke-algebra, though the 
specialized algorithm here is 
different from the algorithm in \cite{kazhdan}.

Since ${\cal B}_r^H(q)$ is semisimple \cite{GCT4,GCT7}, 
by the Wedderburn structure theorem 
\begin{equation} \label{eqweder}
 {\cal B}_r^H(q) = \bigoplus_\alpha T_{q,\alpha}^* \otimes T_{q,\alpha},
\end{equation}
where $T_{q,\alpha}$ ranges over the 
irreducible representations of ${\cal B}^H_r$,
assuming that the underlying base field  is a suitable
algebraic extension of $\Q(q^{1/2})$. We shall denote this base field
by $K$--it is the same as the base field 
$K$ in Section~\ref{striplen}, except that the role of $q$ there is played by
$q^{1/2}$ here. Let $A_K, \bar A_K, K_\Q, \hat Q$ be as in 
Section~\ref{striplen},
with the role of $q$ played by $q^{1/2}$.

We assume that $H$ is the general linear group $GL(V)$ or a product of
general linear groups.
In this case  (cf. \cite{GCT7}), ${\cal B}^H_r$ is a $-$-invariant
subalgebra of a 
suitable Hecke-algebra or a product of Hecke algebras,
where $-$ denotes the usual 
bar-automorphism on the Hecke algebra \cite{kazhdan} 
(analogue of $-$ in Section~\ref{scanonical}).
Let ${\cal P}_i$'s and ${\cal Q}_i$'s denote the rescaled
 positive and negative
generators  of ${\cal B}^H_r$ as defined in \cite{GCT7} (denoted by
$p^{+,H}_{X,i}$ and $q^{+,H}_{X,i}$ therein) 
so that they belong to the usual $\Z[q^{1/2},q^{-1/2}]$-form on 
the ambient Hecke algebra (or the ambient product of Hecke algebras).
Let ${\cal B}^H_{r,\Q}$  denote
the  $K_\Q$-form 
of ${\cal B}^H_r$ generated ring-theoretically by ${\cal P}_i$'s,
or equivalently ${\cal Q}_i$'s. 

The goal is to construct an $A_K$-lattice $R(r) \subseteq {\cal B}^H_r$
so that the canonical basis $E(r)$ of ${\cal B}_r^H$ can then
be constructed by a nonstandard globalization procedure on 
the triple $( {\cal B}^H_{r,\Q},R(r), \bar R(r))$ analogous
to the one Section~\ref{sglobalgl}. Just as in Section~\ref{striplen},
 it will turn
out that this triple need not be balanced.
We will  resolve the ambiguity caused by  lack of
balance  using the notion of  minimum degree complexity
very much as in Section~\ref{sglobalgl}. 

\subsection{Nonstandard Gelfand-Tsetlin basis of ${\cal B}^H_r(q)$}
In the construction of Kazhdan-Lusztig polynomials as described in 
\cite{kazhdan,soergel}, the lattice in the Hecke algebra is constructed 
using its standard monomial basis. 
But in general ${\cal B}^H_r(q)$  does not have a naturally defined
monomial basis; see \cite{GCT4} for an example.
So we need a different way to
construct  the lattice $R(r)$.
The construction here will be analogous to 
the construction of the lattice 
in the standard matrix coordinate ring ${\cal O}(M_q(V))$ based on
its Gelfand-Tsetlin basis. This  construction in ${\cal O}(M_q(V))$
is different from
the  one  defined in Section~\ref{sstandard}. We shall recall it
using the same notation as in Section~\ref{sstandard}.
It is based on
the observation
that  the Gelfand-Tsetlin 
basis  of  an irreducible $H_q$-representation $V_{q,\mu}$ (after 
rescaling as described in section  7.3.3  of \cite{klimyk})
is  its local crystal basis: i.e.,  it  is an
$A$-basis of the lattice $L_\mu \subseteq V_{q,\mu}$, and that its projection 
in $L_\mu/q L_\mu$ is equal to the  basis $B_\mu$ (whose elements 
have disjoint monomial supports as described before). This observation
was in fact the starting point for the theory of local crystal basis 
\cite{kashiwara1}.
Let us denote
the (rescaled) Gelfand-Tsetlin basis of $V_{q,\mu}$  by $GT_{q,\mu}$. 
The Gelfand-Tsetlin basis of ${\cal O}(M_q(V))$ is
defined as per the standard Peter-Weyl theorem (~\ref{eqstdpeterweyl}):
\begin{equation} \label{eqpeterGTmu}
GT({\cal O}(M_q(V)))= \bigcup_{\mu} 
GT_{q,\mu}^* \otimes GT_{q,\mu}.
\end{equation}
Let $L_{GT}$ be the lattice generated by this Gelfand-Tsetlin basis,
and $B_{GT}$ the projection of the Gelfand-Tsetlin basis on
$L_{GT}/q L_{GT}$. Then $(L_{GT},B_{GT})$ coincides with
the standard crystal base $(L,B)$ of ${\cal O}(M_q(V))$ 
in Section~\ref{sstandard}.

The algebra ${\cal B}^H_r(q)$ has a 
natural analgoue of the Gelfand-Tseltin basis,
which  can then be used to construct the lattice $R(r)$. 
We begin by describing this basis 
for  an irreducible representation 
$T_{q,\alpha}$ of ${\cal B}_r^H(q)$.

We proceed by induction on $r$, the case
$r=1$ being easy. The following is a conjectural analogue of 
the standard Pieri's rule  in this setting:

\noindent {\bf C1':} $T_{q,\alpha}$ has a multiplicity-free
decomposition as a ${\cal B}^H_{r-1}$-module.

(This conjecture is  not really necessary as long as there is a natural
way to resolve the ambiguity caused by multiplicity). 
By induction, we have defined a basis for each irreducible 
${\cal B}^H_{r-1}(q)$-submodule
of $T_{q,\alpha}$. Putting these bases together, we get 
the sought {\em nonstandard Gelfand-Tsetlin basis}
$C_\alpha$ of $T_{q,\alpha}$. 

Assuming multiplicity-free decomposition, such a basis is unique,
up to scaling factors, which will be fixed in the course of the
algorithm below. 
Each element  $x\in C_\alpha$
can be indexed by a {\em nonstandard Gelfand-Tsetlin tableau}, which is an 
analogue of the standard 
Gelfand-Tsetlin tableau \cite{klimyk} in this setting. It is defined
to be the  tuple $(\alpha_r,\alpha_{r-1},\ldots)$, with $\alpha_r=\alpha$,
of the classifying labels--which we shall call {\em types}--of the
irreducible ${\cal B}^H_{i}$-submodules $T_{q,\alpha_i}$
containing $x$, where $T_{q,\alpha_i} \subset T_{q,\alpha_{i+1}}$.

We define the {\em nonstandard Gelfand-Tsetlin  basis}
$C(r)$ of ${\cal B}^H_q(r)$ as per the
decomposition (\ref{eqweder}):
\[ 
C(r)= \bigcup_\alpha C_\alpha^* \otimes C_\alpha.
\]
Each element of $C(r)$ is
indexed by a {\em nonstandard Gelfand-Tsetlin bi-tableau}
as per this decomposition. We shall denote the element of
$\C(r)$ indexed by a nonstandard Gelfand-Tsetlin bitableau $T$ by 
$c_T$.

\subsection{Local crystal base}
The sought lattice $R(r) \subseteq {\cal B}^H_r(q)$ 
will be  generated by  the elements of $C(r)$ after  scaling
them appropriately  in the course of the algorithm below. 
Let us assume at the moment that this  scaling has already been 
given to us, and thus $R(r)$ is fixed.
Let $b_T$ denote the image of $c_T$ under the projection
$\psi: R(r) \rightarrow R(r)/q^{1/2} R(r)$. Let $B(r)=\{b_T\}$ be the
basis of $R(r)/q^{1/2} R(r)$. Then $(R(r), B(r))$ is the 
analgoue of the local crystal base in the standard setting.

\subsection{Nonstandard globalization via minimization of degree complexity}
\label{snonstdfglobright}
The elements of $C(r)$ need not be $-$-invariant. Next we globalize 
$C(r)$ to get a $-$-invariant canonical basis 
$E(r)$ of ${\cal B}_r^H$  in the spirit of Kazhdan and 
Lusztig \cite{kazhdan}, with the role of the standard basis in 
\cite{kazhdan,soergel} played by the nonstandard Gelfand-Tsetlin basis 
of ${\cal B}_r^H(q)$ here.
As already mentioned, the main difference from the standard setting 
of Hecke algebra  is that $( {\cal B}^H_r(q), R(r),\bar R(r))$ 
need not be balanced. This is the main
problem that needs to be addresssed.
The nonstandard globalization procedure here is analogous
to the one in Section~\ref{sglobalgl}. It goes as follows.

\noindent (1) 
In Section~\ref{sglobalgl}, we described a partial  order $\le $
on the types (classifying
labels) of the irreducible modules $W_{q,\alpha}$ of ${\cal A}^H_r(q)$.
By the nonstadard duality conjecture \cite{GCT7}, this
induces a partial order $\le$ on the types (classifying labels)
of the (paired) irreducible modules $T_{q,\alpha}$ of ${\cal B}^H_r(q)$.
(In the standard setting, this
procedure would yield a partial  order on the partitions of 
size $r$, with the partition containing a single row of size $r$ at 
the top of the order and the partition containing a single 
column of size $r$ at the bottom of the order.)

\noindent (2) This induces  a lexicographic  partial
order $\le $ on the nonstandard 
Gelfand-Tsetlin tableaux, since they are  just tuples of types, and also on 
nonstandard Gelfand-Tsetlin bitableau which index the  basis elements $C(r)$. 

\noindent (3) 
Let ${\cal B}^{\le T}$ be the span of the basis elements 
$c_{T'} \in C(r)$ such that $T'\le T$. 
Let $R^{\le T}=R(r)\cap {\cal B}^{\le T}$,
$\bar R^{\le T}=\bar R(r)\cap {\cal B}^{\le T}$,  and 
${\cal B}^{\le T}_\Q = {\cal B}^H_{r,\Q} \cap {\cal B}^{\le T}$. Then
the triple $({\cal B}^{\le T}_\Q,R^{\le T},\bar R^{\le T})$ need 
not be balanced. To define a canonical basis element $e_T$ associated
with $T$, we  associate a
{\em degree complexity} with  each element $y \in {\cal B}^H_{r,\Q}$
in the spirit of Section~\ref{sglobalgl}.

This is done as follows.
Since we are assuming that $H$ is $GL(V)$ or 
a product of general linear groups, ${\cal B}^H_r$ is a subalgebra 
of a product of Hecke algebras, say 
$Z={\cal H}_{k_1}(q) \otimes \cdots \otimes 
 {\cal H}_{k_l}(q)$, where ${\cal H}_j(q)$ denotes
 the Hecke algebra with rank $j$.
Furthermore, 
\[{\cal B}^H_{r,\Q} \subseteq Z_\Q={\cal H}_{k_1,\Q} \times \cdots \times  {\cal H}_{k_l,\Q},\]
where ${\cal H}_{j,\Q}$ denote the $K_Q$-form of
${\cal H}_j(q)$ obtained by tensoring its usual $Q[q^{1/2},q^{-1/2}]$-form
with $K_\Q$. 
Consider the Kazhdan-Lusztig basis $KL(Z)$ of $Z$ formed by taking 
the product of the Kazhdan-Lusztig bases of its  Hecke algebra  factors.
Express $y \in {\cal B}^H_{r,\Q}$ in terms of $KL(Z)$:
\begin{equation}  \label{eqexpincanZright}
 y= \sum_z a(y,z) z, 
\end{equation}
where $z$ ranges over the elements in $KL(Z)$. Then each
coefficient $a(y,z) \in K_\Q$. Let $d(y,z)$ denote the degree of
$a(y,z)$; i.e., the order of its pole at $q=\infty$. If $a(y,z)=0$,
we define $d(y,z)=-\infty$.
We define {\em  the degree complexity} $d(y)$ of $y$ to be the tuple
$\langle\ldots,  d(y,z), \ldots  \rangle$
of these degrees. We put a partial order on degree complexities as follows.
Put a partial order $\le$ on the Kazhdan-Lusztig basis  of the Hecke algebra
${\cal H}_j(q)$ as per the reverse order on the (reduced) lengths of
the permutation indices of the basis elements--so $1$ is the highest
element as per this order. This also puts a partial order on 
$KL(Z)$. 
Let $<$ denote the strict less than relation
as per this partial order.
Given $y$ and $y'$, 
we say that $d(y) \le d(y')$ if
for every $z$: either $d(y,z)\le  d(y',z)$, or for some $\bar z < z$,
$d(y,\bar z) < d(y',\bar z)$. 

Consider the natural projection 
\[\psi_T: R^{\le T} \cap \bar R^{\le T} \cap {\cal B}^{\le T}_\Q
\rightarrow R^{\le T} /q^{1/2} R^{\le T}.\] 
Let $\psi_T^{-1}(b_T)$ be the fibre of $b_T \in B(r)$.
The following  is the analogue of Conjecture~\ref{cuniquemindegleft}
 in this context
(with different interpretation for $b,y,\psi$ etc. from there):

\begin{conj} {\bf (Minimum degree)}  \label{cunimindegright}
The fibre $\psi_T^{-1}(b_T)$ contains a unique element $e_T$ of 
minimum degree complexity. Minimum  means  $d(e_T)\le d(y)$, as per
the ordering on the degree tuples above,
for any $y \in \psi_T^{-1}(b_T)$. 
\end{conj} 

We  call $e_T$  the canonical basis element associated
with $T$, and $E(r)=\{e_T\}$ the canonical basis ${\cal B}^H_r(q)$. 

So far we have not discussed how to scale the nonstandard Gelfand-Tsetlin basis
of ${\cal B}^H_r(q)$ to get the lattice $R(r)$.
To complete the algorithm, it remains to fix this scaling.

Let $\{c'_T\}$ denote the nonstandard Gelfand-Tsetlin basis
of ${\cal B}^H_r(q)$ before scaling. The scaled $c_T$ will be
of the form $q^{a_T} c'_T$ for some rational $a_T$. We have to 
determine all $a_T$'s. Assume that $a_{T'}$, $T'< T$, have been 
fixed. For any rational $a$, let $c_{a,T}=q^a c'_T$. 
Let $R^{a,\le T}$ be the lattice generated by $c_{a,T}$ and
$c_{T'}$, $T<T$, and $\bar R^{a,\le T}$ obtained by applying $-$ to it.
Consider the projection 
\[\psi_{a,T}: R^{a,\le T} \cap \bar R^{a,\le T} \cap {\cal B}^{\le T}_\Q
\rightarrow R^{a,\le T} /q^{1/2} R^{a,\le T}.\] 
Let $b_{a,T}$ be the image of $c_{a,T}$ under the projection
$R^{a,\le T}/q^{1/2} R^{a,\le T}$. 
Let $\psi_{a,T}^{-1}(b_{a,T})$ be its  fibre.
The following  is the strengthened 
form  of Conjecture~\ref{cuniquemindegleft}.

\begin{conj} {\bf (Minimum degree)} \label{cunimindegright2}
There exists a unique $a_T$ and $e_{T} \in \psi_{a_T,T}^{-1}(b_{a_T,T})$
such that for any $a$ and any  $y \in \psi_{a,T}^{-1}(b_{a,T})$
$d(e_{T})\le d(y)$. That is, $e_{T}$ is the unique  element of 
minimum degree complexity over all choices of $a$.
\end{conj} 

This fixes $a_T$. Furthermore, 
$\psi_T=\psi_{a_T,T}$, $b_T=b_{a_T,T}$,
and $e_T$ in Conjecture~\ref{cunimindegright} is the same as  here.

If instead of the order $\le$ among the classifying labels $\alpha$'s 
of $T_{q,\alpha}$, we use its reverse order, and in the definition
of  degree complexity 
use the opposite of  the Kazhdan-Lusztig basis (obtained by replacing
${\cal Q}_i$ by ${\cal P}_i$),  we get another 
canonical basis  $E^{opp}(r)$ of ${\cal B}^H_r(q)$, which we shall call
its {\em opposite canonical basis}.

To prove Conjectures~\ref{cunimindegright}-\ref{cunimindegright2}, we need to
know  relations among the generators ${\cal P}_i$'s of 
${\cal B}^H_r$ explicitly, just as we know the relations among the
generators of the Hecke algebra explicitly. This is not known at
present. See \cite{GCT7} for the problems that arise in this context.

Each element $c$ of the  Kazhdan-Lusztig basis of the Hecke algebra can be 
expressed in the form 
\[ c= c_0 + \sum_{j>0} a(j) c_j, \] 
where each $c_j$ is a monomial in the generators of the Hecke algebra,
$a(j) \in \Q[q^{1/2},q^{-1/2}]$ and the length of each $c_j$, $j>0$, is
smaller than that of $c_0$. This need not be so in the nonstandard setting: 
there can be several monomials of maximum length with nontrivial coefficients
in any monomial representation of a nonstandard canonical basis element;
cf. Section~\ref{sexphsl2gsl4} and Figure~\ref{fv02} therein for an example.

\subsection{Conjectural properties} \label{sconjb}
It may be conjectured that the canonical bases
$E(r)$ and $E^{opp}(r)$ have properties akin to those of 
the Kazhdan-Lusztig basis of the Hecke algebra.

\subsubsection{Cellular decomposition}

\begin{conj} {\bf (Cell decomposition)}  \label{ccelldecompright}
Analogue of the cell decomposition property of the Kazhdan-Lusztig basis
also holds for $E(r)$ and $E^{opp}(r)$.
\end{conj} 

Specifically this means the following. Let us
define the left, right and two-sided cells of ${\cal B}_r^H$ with
respect to the canonical basis $E(r)$ very much as in Section~\ref{sconjp}.
Then it may be conjectured that they yield irreducible left, right
and two-sided representations of ${\cal B}_r^H$. The conjecture
 for $E^{opp}(r)$ is similar.

By restricting   $E(r)$ to any left cell corresponding to
an  irreducible ${\cal B}^H_r$-module $W_{q,\alpha}$, we get 
the canonical basis of $W_{q,\alpha}$; here the choice of the left 
cell would conjecturally not matter (up to scaling).

\subsection{Positivity in the Kronecker case}
The following is an analogue of Conjecture~\ref{cposkroneckerleft} here.

\begin{conj} {\bf (Positivity)} \label{cposkroneckerright}
Let  $c\in K_\Q$ be a 
multiplicative or comultiplicative structural constant of $E(r)$ or a
structural coefficient of a canonical basis element
in $E(r)$--i.e., a coefficient of its expression in terms of
the canonical basis of $Z$ as in eq.(\ref{eqexpincanZright}).

In the Kronecker case, 
$c$ is  of the form
$\stackrel{+}{-} (q^{1/2}-q^{-1/2})^a f(q)$, where $a$ is a nonnegative
integer and $f(q)$ is a $-$-invariant positive and unimodal 
polynomial in $q^{1/2}$ and $q^{-1/2}$.

The same for $E^{opp}(r)$.
\end{conj}

For experimental evidence see Sections~\ref{skroneckerexppos1} 
and \ref{skronrequla4}. 

\subsubsection{Nonstandard positivity and saturation}
The general case is much more complex as in Section~\ref{ssaturationleft}. 
The following is an analogoue of Conjecture~\ref{cnonstdposleft0}.

\begin{conj} {\bf (Saturation)} \label{cnonstdposright}
Let  $c\in K_\Q$ be a  multiplicative or comultiplicative structural
 constant of $E(r)$ or a structural coefficient of a canonical basis element
in $E(r)$--i.e., a coefficient of its expression in terms of
the canonical basis of $Z$ as in eq.(\ref{eqexpincanZright}).
Let $f_c$ be its minimal polynomial with coefficients in $\Q(q^{1/2})$.

Then  each coefficient $s(q)$ of $f_c$
can be expressed in 
the form 

\begin{equation} \label{eqbexpli3}
s(q) = (-1)^{e}  (q^{1/2}-q^{-1/2})^{e'}
g(q)
\end{equation}
for some nonnegative integers $e,e'$, where 
\begin{enumerate} 
\item $e$ is chosen so that 
the middle term of $g(q)$ is positive; and
\item   $g(q)$ is a saturated
polynomial in $q^{1/2}$ and $q^{-1/2}$.
\end{enumerate}
\end{conj} 

Analogue of the stronger Conjecture~\ref{cnonstdposleft1} in this case is:

\begin{conj} {\bf (Nonstandard Positivity, informal)} \label{cnonstdposright1}
Each polynomial  $g(q)$ above is almost
positive and unimodal; i.e. 
of the form
$g^0(q)+g^1(q)$, where, if $g(q)$ is not identically zero, 
\begin{enumerate} 
\item $g^0(1) >> |g^1(1)|$,
and more generally,
\item $g^0(q)$ is a dominant positive unimodal  polynomial,
and $g^1(q)$ is a very small  error-correction polynomial.
Specifically, 
\[||g^1(q)||/||g^0(q)|| \le 1/\poly(\bitlength{\mu},\bitlength{\pi},\bitlength{r}),\] with the terminology as in Conjecture~\ref{cnonstdposleft1}.

\end{enumerate}
\end{conj} 

For  experimental evidence,  see Section~\ref{snonstdposexp}.

Here, $g(1)$ and 
the conjectural nonnegative coefficients of  $g^0(q)$ 
may have a representation-theoretic/topological/cohomological
interpretation akin
to that sought for the analogous quantities in 
Section~\ref{ssaturationleft}.

\subsubsection{Quasi-cellular decomposition}

\begin{conj} \label{coppquasicell} 
The opposite canonical basis $E^{opp}$ also has the following 
quasi-cellular decomposition property.
\end{conj}

For this we define a quasi-subcellular decomposition of each 
left or right cell with respect to $E^{opp}(r)$. 
Specifically, given $e',e''$ belonging to the same left cell, 
express
\begin{equation}\label{eqmult1}
 e e'= \sum_{e''} \epsilon(e,e',e'') (q^{1/2}-q^{-1/2})^{\delta(e,e',e'')}
 d_{e,e'}^{e''} e'',
\end{equation}
where $e'' \in E(r)$, the sign $\epsilon(e,e',e'')$ is either $1$ or
$-1$, $\delta(e,e',e'')$ is a nonnegative integer,
  and $d_{e,e'}^{e''}$ is 
a $-$-invariant saturated  polynomial in $q^{1/2}$ and $q^{-1/2}$ 
as per the saturation Conjecture~\ref{cnonstdposright}.
We say that $e''\propto_L e'$ if, for some $e$, $d_{e,e'}^{e''}$ in
(\ref{eqmult1}) is nonzero and $\delta(e,e',e'')$ is zero;
 i.e., if, for some $e$,
$e''$ occurs with nonzero coefficient in the expansion of $e e'$ specialized 
at $q=1$.
Let $\prec_L$ denote the transitive closure of $\propto_L$. Using
$\prec_L$ we  define left quasi-subcells of a left-cell of $E^{opp}(r)$. 
It may be conjectured that each left quasi-subcell of $E^{opp}(r)$ yields 
an irreducible representation of the symmetric group $S_r$ at $q=1$. 
That is, when the ${\cal B}_r^H$-representation $Y$ corresponding this 
left cell
is specialized at $q=1$, so as to become a representation $Y_{q=1}$ of 
$\C[S_r]$, the partial order on its left quasi-sub-cells
induces a composition series of $Y_{q=1}$ whose factors are irreducible
representations of $\C[S_r]$. 

Fix one such quasi-subcell $C$ of $E^{opp}(r)$. Let $S_{\lambda(C)}$
be the irreducible representation (Specht module)
of $\C[S_r]$ that is isomorphic to the factor in correspondence with $C$
in this composition series of $Y_{q=1}$, where $\lambda(C)$ is a partition
depending on $C$.
The canonical basis elements in $C$,
after specialization at $q=1$ and projection, yield  a basis of
$S_{\lambda(C)}$.
It may be conjectured that this basis coincides with the Kazhdan-Lusztig
basis of $S_{\lambda(C)}$ (up to rescaling). By the Kazhdan-Lusztig basis 
of $S_{\lambda(C)}$, we mean specialization at $q=1$ of the Kazhdan-Lusztig
basis of the quantized Specht module $S_{q,\lambda(C)}$ of the Hecke 
algebra ${\cal H}_r(q)$.

But Conjecture~\ref{coppquasicell}
 need not hold for $E(r)$; cf. Section~\ref{sexphsl2gsl4} 
for a counterexample. This is analgous to the fact that 
the refined Peter-Weyl theorem in \cite{lusztigbook} for the coordinate
ring of the standard quantum group $H_q$ holds only for the ordering 
$\le$ (as defined in Section~\ref{sglobalgl}) among the labels
(highest weights)
of irreducible $H_q$-modules--there is no canonical basis of the standard
coordinate ring which
admits refined Peter-Weyl theorem for the opposite of the order $\le$. 

\section{Internal definition of degree complexity} \label{sinternal}
We give here an {\em internal} definition of degree complexity 
which may be used in place of the definition in Section~\ref{snonstdfglobright}
during the construction of the canonical basis.
By internal, we mean it is  based only on the structure of ${\cal B}^H_r(q)$
and does not depend on its embedding in the external ambient  algebra
$Z$ there. This notion of degree complexity does not coincide
with  the
one in Section~\ref{snonstdfglobright}, 
but the canonical basis constructed using
this definition may be conjectured to be the same as the one
constructed therein. 


Let ${\cal B}[i] \subseteq {\cal B}^H_r(q)$ be the span of the
monomials in ${\cal Q}_j$'s of length $j$, and ${\cal B}[< i]$ of
length $< i$. We say that a given set of monomials in ${\cal Q}_j$'s 
of length $i$ is {\em independent} if the images of these monomials
in ${\cal B}[i]/{\cal B}[< i]$ are linearly independent. 
An expression 
\begin{equation} \label{eqinternal1}
  a= \sum_m a_m m,
\end{equation}
where $m$ ranges over monomials in ${\cal Q}_j$'s and $a_m \in K_\Q$,
is called {\em valid} if, for each $i$, the  monomials $m$ of length $i$ with
$a_m\not = 0$ in this expression are independent. Assume that
$a$ is $-$-invariant, so that each $a_m$ is $-$-invariant.
The degree complexity $\hat d(a)$ of $a$ is defined to be 
the tuple $\langle \hat d_l(a), \ldots, \hat d_i(a),\ldots,  \hat d_0(a) 
\rangle$ 
where $\hat d_i(a)$ denotes the maximum degree (at $q=0$)  of $a_m$ 
for any $m$ of
length $i$, and $l$ is the maximum length of $m$ with $a_m\not = 0$ 
in the expression
 (\ref{eqinternal1}); by definition, $\hat d_i(a)=-\infty$ if there is
no $m$ in (\ref{eqinternal1}) of length $i$ with $a_m\not = 0$.
We order these degree complexities lexicographically. 
The degree complexity $\hat d(b)$ of an element $b \in {\cal B}_{r,\Q}^H(q)$
is defined to be the minimum degree complexity of its any valid expression.
It may be conjectured that if this definition of degree complexity,
with the lexicographic ordering as above, is used in place of
the definition of degree complexity in Section~\ref{snonstdfglobright}, 
the algorithm therein still works correctly and constructs the
same canonical basis $E(r)$.

For the opposite canonical basis $E^{opp}(r)$, one can similarly use
the internal definition as above with ${\cal P}_i$ in place of ${\cal Q}_i$.

The definition of degree complexity in
this section is not as satisfactory as in Section~\ref{snonstdfglobright}
because
${\cal B}^H_q(r)$ does not a natural monomial basis \cite{GCT4}. Hence
to find the  degree complexity of an element, 
one has to consider all its monomial
expressions, finite but huge in number.
It will be interesting to know if
there is a more efficient  internal definition.

\section{Experimental evidence for ${\cal B}^H_r(q)$} \label{sexpevid}
In this section we shall verify the conjectures in this paper
for two nontrivial special cases of 
the nonstandard algebra ${\cal B}={\cal B}^H_r(q)$.
The nonstandard canonical bases of ${\cal B}^H_r(q)$ in these cases
were computed with the help of a computer
using the algorithm in Section~\ref{scanonicalB}
and the notion of degree complexity
as in Section~\ref{snonstdfglobright}.

First, some notation.
Given a string $\sigma=i_1 \cdots i_k$ of positive integers,
we  let  ${\cal P}_{\sigma}$ denote the
monomial ${\cal P}_{i_1} \cdots {\cal P}_{i_k}$;
${\cal Q}_{\sigma}$ is defined similarly.
Given a $-$-invariant polynomial $g(q) \in \Q[q,q^{-1}]$,
we define the {\em vector} $a_g$ associated with $g(q)$ as follows.
Express $g(q)$ in
the form $\stackrel{+}{-}(q-1/q)^e h(q)$, where $e$ is the maximum possible.
Let $h_{-l},\ldots, h_0,\ldots, h_l$ be the coefficients of $h(q)$.
Then $a_g$ is defined to be $[h_0,\ldots,h_l]$. In particular if
$h(q)$ is (positive) unimodal, then $a_q$ is a (positive) 
nonincreasing sequence.
The vector associated with a $-$-invariant polynomial in $q^{1/2}$ and
$q^{-1/2}$ is defined similarly.

\subsection{Kronecker problem: $n=2, r=3$}\label{sexpkronecker}
Consider ${\cal B}={\cal B}^H_3(q)$ in the special case of 
the Kronecker problem for $n=2$ and $r=3$. Thus $H=Gl_2 \times Gl_2$, and
$G=Gl_4$ with $H$ embedded diagonally.
Let ${\cal P}_i$, $i=1,2$, be
as in Section~\ref{scanonicalB} and \cite{GCT4}.
The nonstandard 
canonical basis of ${\cal B}$ was computed in \cite{GCT4}
by an ad hoc method for $r=3$, but it coincides with the
one computed by the algorithm here.
It is as follows. 
Let
\[   c_1 = \frac{q^6 +2q^5 +3q^4 +4q^3 +3q^2 +2q +1}{q^3 }, \quad
\   c_2 = \frac{q^4 +q^3 +4q^2 +q +1}{q^2 }, \]
\[ b_1=-(q^2+1)^2/q^2, \ \mbox{and}\ b_2= (q+1)^2/q. \]

Then the opposite 
canonical basis $E^{opp}(3)$ of ${\cal B}$
consists of the following ten elements:
\begin{equation} 
\begin{array}{l}
\Sigma= c_1 {\cal P}_1 -c_2 {\cal P}_{121} + {\cal P}_{1 2 1 2 1}, \\
\gamma^i_1= b_1 {\cal P}_1 + {\cal P}_{1 2 1}, \quad i=1,2, \\
\gamma^i_{12}= b_1 {\cal P}_{1 2} + {\cal P}_{1 2 1 2}, \quad i=1,2, \\
\gamma^i_2= b_1 {\cal P}_2 + {\cal P}_{2 1 2}, \quad i=1,2, \\
\gamma^i_{21}= b_1 {\cal P}_{2 1} + {\cal P}_{2 1 2 1}, \quad i=1,2, \\
\mu=1.
\end{array}
\end{equation}
The canonical basis $E(3)$ is obtained by susbstituting ${\cal Q}_i$ 
for ${\cal P}_i$. In what follows, we shall only consider $E^{opp}(3)$.

\subsubsection{Cellular and quasi-cellular decomposition}
The basis $E^{opp}(3)$ has a cellular decomposition, in accordance
with Conjecture~\ref{ccelldecompright},  with the following  right cells:

\[ 
\begin{array}{lcl}
U_\sigma&=&\{  \Sigma\} \\
V_1&=& \{  \gamma_1^1,  \gamma_{12}^1 \}  \\
V_2&=& \{  \gamma_1^2,  \gamma_{12}^2 \}  \\
W_1&=& \{  \gamma_2^1,  \gamma_{21}^1 \}  \\
W_2&=& \{  \gamma_2^2,  \gamma_{21}^2 \} \\
U_\mu=&=& \{  \mu\}.
\end{array}
\]
The left cell decomposition is similar.
The representation of ${\cal B}$ supported by  $U_\sigma$
is the trivial one dimensional representation. 
The representation supported by $V_1$ or $W_1$ is isomorphic; let us
call it $\chi^1$. Similarly, 
the representation supported by $V_2$ or $W_2$ is isomorphic; let us
call it $\chi^2$. Then $\chi^1$ and $\chi^2$ are two nonisomorphic
two-dimensional representations of ${\cal B}$ which specialize at
$q=1$ to the two-dimensional Specht module of the symmetric group $S_3$
corresponding to the partition $(2,1)$. 
Thus quasi-cellular decomposition (Conjecture~\ref{coppquasicell}) holds
trivially here.

\subsubsection{Positivity} \label{skroneckerexppos1}
Coefficients of the elements of $W_1$ and $W_2$ in the 
 Kazhdan-Lusztig basis of 
${\cal H}_3(q)  \otimes {\cal H}_3(q) 
 \supseteq {\cal B}^H_3(q)$ are shown in Figure~\ref{fw1w2}
(with the Kazhdan-Lusztig basis  symmetrized and
appropriately ordered as described in \cite{GCT4}); the first 
column shows the coefficients of $\gamma^1_1$, the second of $\gamma^1_{12}$,
and so on.
It can be observed that all coefficients are positive, and unimodal
polynomials in $\Q[q,q^{-1}]$. 
The cofficients of other canonical basis elements can be found in
\cite{GCT4}; they too are positive, unimodal polynomials. 
This verifies the positivity 
Conjecture~\ref{cposkroneckerright} for the structural coefficients of the
canonical basis.
A few typical 
nonzero  multiplicative structural constants of the canonical basis
are shown in Figure~\ref{fmultkron2}, where the coefficient 
of $b b'$ with respect to  the basis element $b''$
is denoted by  $c(b,b';b'')$.
It can be seen that each
constant is a polynomial of the form 
\[ (-1)^a (q^{1/2}-q^{-1/2})^b f(q^{1/2},q^{-1/2}),\]
where $f$ is a positive unimodal polynomial. 
It was verified with computer that all multiplicative structural 
constants are of this form. This verifies 
the  positivity Conjecture~\ref{cposkroneckerright}
for the multiplicative structural constants as well.

\begin{figure}[p!] 
\[
\left[ \begin {array}{cccc} {\frac { \left( {q}^{2}+q+1 \right)  \left( 1+q \right) ^{4}}{{q}^{3}}}&{\frac { \left( {q}^{2}+q+1 \right)  \left( 1+q \right) ^{6}}{{q}^{4}}}&{\frac { \left( {q}^{2}+q+1 \right)  \left( 1+q \right) ^{4}}{{q}^{3}}}&{\frac { \left( {q}^{2}+q+1 \right)  \left( 1+q \right) ^{6}}{{q}^{4}}}\\\noalign{\medskip}{\frac { \left( 1+q \right) ^{5}}{{q}^{5/2}}}&{\frac { \left( 1+q \right) ^{7}}{{q}^{7/2}}}&{\frac { \left( 1+q \right) ^{5}}{{q}^{5/2}}}&{\frac { \left( 1+q \right) ^{7}}{{q}^{7/2}}}\\\noalign{\medskip}{\frac { \left( 1+q \right) ^{4}}{{q}^{2}}}&{\frac { \left( 1+q \right) ^{6}}{{q}^{3}}}&{\frac { \left( 1+q \right) ^{4}}{{q}^{2}}}&{\frac { \left( 1+q \right) ^{6}}{{q}^{3}}}\\\noalign{\medskip}{\frac { \left( 1+q \right) ^{3}}{{q}^{3/2}}}&{\frac { \left( 1+q \right) ^{5}}{{q}^{5/2}}}&{\frac { \left( 1+q \right) ^{3}}{{q}^{3/2}}}&{\frac { \left( 1+q \right) ^{5}}{{q}^{5/2}}}\\\noalign{\medskip}{\frac { \left( 1+q \right) ^{5}}{{q}^{5/2}}}&{\frac { \left( 1+q \right) ^{7}}{{q}^{7/2}}}&{\frac { \left( 1+q \right) ^{5}}{{q}^{5/2}}}&{\frac { \left( 1+q \right) ^{7}}{{q}^{7/2}}}\\\noalign{\medskip}{\frac { \left( 1+q \right) ^{4}}{{q}^{2}}}&{\frac { \left( 1+q \right) ^{6}}{{q}^{3}}}&{\frac { \left( 1+q \right) ^{4}}{{q}^{2}}}&{\frac { \left( 1+q \right) ^{6}}{{q}^{3}}}\\\noalign{\medskip}2\,{\frac {3\,{q}^{3}+1+3\,q+8\,{q}^{2}+{q}^{4}}{{q}^{2}}}&2\,{\frac { \left( 3\,{q}^{3}+1+3\,q+8\,{q}^{2}+{q}^{4} \right)  \left( 1+q \right) ^{2}}{{q}^{3}}}&2\,{\frac { \left( 1+q \right) ^{4}}{{q}^{2}}}&2\,{\frac { \left( 3\,{q}^{3}+1+3\,q+8\,{q}^{2}+{q}^{4} \right)  \left( 1+q \right) ^{2}}{{q}^{3}}}\\\noalign{\medskip}{\frac { \left( 1+q \right)  \left( {q}^{2}+6\,q+1 \right) }{{q}^{3/2}}}&2\,{\frac { \left( 1+q \right)  \left( 3\,{q}^{3}+1+3\,q+8\,{q}^{2}+{q}^{4} \right) }{{q}^{5/2}}}&2\,{\frac { \left( 1+q \right) ^{3}}{{q}^{3/2}}}&{\frac { \left( {q}^{2}+6\,q+1 \right)  \left( 1+q \right) ^{3}}{{q}^{5/2}}}\\\noalign{\medskip}8&{\frac { \left( {q}^{2}+6\,q+1 \right)  \left( 1+q \right) ^{2}}{{q}^{2}}}&2\,{\frac { \left( 1+q \right) ^{2}}{q}}&{\frac { \left( {q}^{2}+6\,q+1 \right)  \left( 1+q \right) ^{2}}{{q}^{2}}}\\\noalign{\medskip}{\frac { \left( {q}^{2}+4\,q+1 \right)  \left( 1+q \right) ^{2}}{{q}^{2}}}&{\frac { \left( {q}^{4}+5\,{q}^{3}+12\,{q}^{2}+5\,q+1 \right)  \left( 1+q \right) ^{2}}{{q}^{3}}}&{\frac { \left( {q}^{2}+4\,q+1 \right)  \left( 1+q \right) ^{2}}{{q}^{2}}}&{\frac { \left( {q}^{4}+5\,{q}^{3}+12\,{q}^{2}+5\,q+1 \right)  \left( 1+q \right) ^{2}}{{q}^{3}}}\\\noalign{\medskip}{\frac { \left( 1+q \right)  \left( {q}^{2}+6\,q+1 \right) }{{q}^{3/2}}}&{\frac { \left( {q}^{2}+6\,q+1 \right)  \left( 1+q \right) ^{3}}{{q}^{5/2}}}&2\,{\frac { \left( 1+q \right) ^{3}}{{q}^{3/2}}}&2\,{\frac { \left( 1+q \right)  \left( 3\,{q}^{3}+1+3\,q+8\,{q}^{2}+{q}^{4} \right) }{{q}^{5/2}}}\\\noalign{\medskip}4\,{\frac { \left( 1+q \right) ^{2}}{q}}&2\,{\frac {3\,{q}^{4}+6\,{q}^{3}+14\,{q}^{2}+6\,q+3}{{q}^{2}}}&4\,{\frac { \left( 1+q \right) ^{2}}{q}}&4\,{\frac { \left( 1+q \right) ^{4}}{{q}^{2}}}\\\noalign{\medskip}4\,{\frac {1+q}{\sqrt {q}}}&4\,{\frac { \left( 1+q \right) ^{3}}{{q}^{3/2}}}&4\,{\frac {1+q}{\sqrt {q}}}&4\,{\frac { \left( 1+q \right) ^{3}}{{q}^{3/2}}}\\\noalign{\medskip}2\,{\frac { \left( 1+q \right) ^{3}}{{q}^{3/2}}}&2\,{\frac { \left( 1+q \right)  \left( 3\,{q}^{3}+1+3\,q+8\,{q}^{2}+{q}^{4} \right) }{{q}^{5/2}}}&{\frac { \left( 1+q \right)  \left( {q}^{2}+6\,q+1 \right) }{{q}^{3/2}}}&{\frac { \left( {q}^{2}+6\,q+1 \right)  \left( 1+q \right) ^{3}}{{q}^{5/2}}}\\\noalign{\medskip}2\,{\frac { \left( 1+q \right) ^{2}}{q}}&{\frac { \left( {q}^{2}+6\,q+1 \right)  \left( 1+q \right) ^{2}}{{q}^{2}}}&2\,{\frac { \left( 1+q \right) ^{2}}{q}}&{\frac { \left( {q}^{2}+6\,q+1 \right)  \left( 1+q \right) ^{2}}{{q}^{2}}}\\\noalign{\medskip}8&8\,{\frac { \left( 1+q \right) ^{2}}{q}}&8&8\,{\frac { \left( 1+q \right) ^{2}}{q}}\\\noalign{\medskip}2\,{\frac { \left( 1+q \right) ^{2}}{q}}&{\frac { \left( {q}^{2}+6\,q+1 \right)  \left( 1+q \right) ^{2}}{{q}^{2}}}&8&{\frac { \left( {q}^{2}+6\,q+1 \right)  \left( 1+q \right) ^{2}}{{q}^{2}}}\\\noalign{\medskip}4\,{\frac {1+q}{\sqrt {q}}}&4\,{\frac { \left( 1+q \right) ^{3}}{{q}^{3/2}}}&4\,{\frac {1+q}{\sqrt {q}}}&4\,{\frac { \left( 1+q \right) ^{3}}{{q}^{3/2}}}\\\noalign{\medskip}2\,{\frac { \left( 1+q \right) ^{4}}{{q}^{2}}}&2\,{\frac { \left( 3\,{q}^{3}+1+3\,q+8\,{q}^{2}+{q}^{4} \right)  \left( 1+q \right) ^{2}}{{q}^{3}}}&2\,{\frac {3\,{q}^{3}+1+3\,q+8\,{q}^{2}+{q}^{4}}{{q}^{2}}}&2\,{\frac { \left( 3\,{q}^{3}+1+3\,q+8\,{q}^{2}+{q}^{4} \right)  \left( 1+q \right) ^{2}}{{q}^{3}}}\\\noalign{\medskip}2\,{\frac { \left( 1+q \right) ^{3}}{{q}^{3/2}}}&{\frac { \left( {q}^{2}+6\,q+1 \right)  \left( 1+q \right) ^{3}}{{q}^{5/2}}}&{\frac { \left( 1+q \right)  \left( {q}^{2}+6\,q+1 \right) }{{q}^{3/2}}}&2\,{\frac { \left( 1+q \right)  \left( 3\,{q}^{3}+1+3\,q+8\,{q}^{2}+{q}^{4} \right) }{{q}^{5/2}}}\\\noalign{\medskip}4\,{\frac { \left( 1+q \right) ^{2}}{q}}&4\,{\frac { \left( 1+q \right) ^{4}}{{q}^{2}}}&4\,{\frac { \left( 1+q \right) ^{2}}{q}}&2\,{\frac {3\,{q}^{4}+6\,{q}^{3}+14\,{q}^{2}+6\,q+3}{{q}^{2}}}\end {array} \right] 
\]
\caption{Coefficients of the elements of  W1 and W2 in the symmetrized 
 Kazhdan-Lusztig basis, as computed in \cite{GCT4}}
\label{fw1w2}
\end{figure}

\begin{figure}[p!] 
\[
\begin{array}{l}
c(\gamma^1_{1}; \gamma^1_{1}; \gamma^1_{1})=
c(\gamma^{12}_{1}; \gamma^1_{1}; \gamma^1_{1})
=-(1+q)^2*(q^2+q+1)*(q-1)^2/q^3; \\
c(\gamma^1_{2}; \gamma^1_{1}; \gamma^1_{21})=-(q^2+q+1)*(q-1)^2/q^2; \\
c(\gamma^1_{21}; \gamma^1_{1}; \gamma^1_{21})=-(1+q)^2*(q^2+q+1)*(q-1)^2/q^3;\\
c(\gamma^1_{1}; \gamma^1_{1}; \Sigma)=
c(\gamma^2_{1}; \gamma^1_{1}; \Sigma)=
c(\gamma^1_{2}; \gamma^1_{1}; \Sigma)=
c(\gamma^2_{2}; \gamma^1_{1}; \Sigma)=
1/q*(1+q)^2; \\
c(\gamma^1_{12}; \gamma^1_{12}; \Sigma)=
c(\gamma^1_{21}; \gamma^1_{12}; \Sigma)=
(1+q)^2*(6*q+5*q^2+5)/q^2; \\
c(\gamma^2_{21};\gamma^2_{21};\gamma^2_{21})=(1+q^2)^2*(q^2+q+1)*(q-1)^2/q^4;\\
c(\gamma^2_{21}; \gamma^2_{21}; \Sigma)=
c(\gamma^2_{12}; \gamma^2_{21}; \Sigma)=(1+q)^2*(2*q^2+q+1)*(q^2+q+2)/q^3.
\end{array}
\]
\caption{Multiplicative structural constants of the canonical basis
of ${\cal B}^H_3(q)$ in the Kronecker case, $n=2$, $r=3$}
\label{fmultkron2}
\end{figure}

\subsection{Kronecker case, $H=SL_2$,  $r=4$} \label{skronrequla4}
For the Kronecker case, $H=Gl_2\times GL_2$, $G=GL_4$, and $r=4$, we could
compute just one canonical basis element $\Sigma$ (akin to $\Sigma$ in
Section~\ref{sexpkronecker})  corresponding to the
trivial one dimensional representation of ${\cal B}^H_4(q)$.
Symbolic computations needed to compute other canonical basis elements
turned out to be  beyond the scope of MATLAB/Maple on an ordinary
workstation. 
The coefficients of $\Sigma$ in the Kazhdan-Luztig basis 
of ${\cal H}_4(q) \otimes {\cal H}_4(q) \supset {\cal B}^H_4(q)$ 
were computed in MATLAB/Maple. There are  $576$ coefficients in total. 
Figures~\ref{fcoefsi41}-\ref{fcoefsi43}  show the vectors associated with 
distinct nonzero coefficients among these. They can be seen to be positive
and nonincreasing in accordance with Conjecture~\ref{cposkroneckerright}.

\begin{figure}[p!] 
\[
\begin{array}{cccccccc}  \hline
10&4&3&&&&\\ \hline \noalign{\medskip}44&21&7&&&&\\ \hline \noalign{\medskip}22&17&7&1&&&\\ \hline \noalign{\medskip}50&30&15&2&&&\\ \hline \noalign{\medskip}20&12&4&&&&\\ \hline \noalign{\medskip}14&7&3&&&&\\ \hline \noalign{\medskip}44&31&14&5&&&\\ \hline \noalign{\medskip}19&12&4&1&&&\\ \hline \noalign{\medskip}6&5&3&1&&&\\ \hline \noalign{\medskip}94&64&29&4&&&\\ \hline \noalign{\medskip}88&65&28&7&&&\\ \hline \noalign{\medskip}39&24&8&1&&&\\ \hline \noalign{\medskip}80&45&17&2&&&\\ \hline \noalign{\medskip}40&32&16&4&&&\\ \hline \noalign{\medskip}28&21&10&3&&&\\ \hline \noalign{\medskip}75&45&19&5&&&\\ \hline \noalign{\medskip}38&31&16&5&1&&\\ \hline \noalign{\medskip}11&8&4&1&&&\\ \hline \noalign{\medskip}122&69&23&2&&&\\ \hline \noalign{\medskip}62&49&23&5&&& \\ \hline \end {array}  
\]
\caption{The vectors associated with  distinct nonzero coefficients 
of $\Sigma \in {\cal B}^H_4(q)$ (ignoring a positive, unimodal factor)}
\label{fcoefsi41}
\end{figure}

\begin{figure}[p!] 
\[
\begin{array}{cccccccc}  \hline
126&92&47&12&2&&\\ \hline \noalign{\medskip}158&93&33&4&&&\\ \hline \noalign{\medskip}153&93&35&7&&&\\ \hline \noalign{\medskip}78&63&32&9&1&&\\ \hline \noalign{\medskip}160&125&62&19&2&&\\ \hline \noalign{\medskip}72&48&20&4&&&\\ \hline \noalign{\medskip}150&120&64&24&5&&\\ \hline \noalign{\medskip}69&47&21&6&1&&\\ \hline \noalign{\medskip}22&19&12&5&1&&\\ \hline \noalign{\medskip}212&163&74&17&&&\\ \hline \noalign{\medskip}244&191&92&25&2&&\\ \hline \noalign{\medskip}111&72&28&5&&&\\ \hline \noalign{\medskip}102&71&33&9&1&&\\ \hline \noalign{\medskip}104&88&52&20&4&&\\ \hline \noalign{\medskip}100&85&52&22&6&1&\\ \hline \noalign{\medskip}30&23&13&5&1&&\\ \hline \noalign{\medskip}128&106&59&20&3&&\\ \hline \noalign{\medskip}316&251&126&37&4&&\\ \hline \noalign{\medskip}306&246&128&42&7&&\\ \hline \noalign{\medskip}141&95&41&10&1&& \\ \hline \end {array}  
\]
\caption{The vectors associated with  distinct nonzero coefficients 
of $\Sigma \in {\cal B}^H_4(q)$,
 continued}
\label{fcoefsi42}
\end{figure}

\begin{figure}[p!] 
\[
\begin{array}{cccccccc}  \hline
144&120&68&24&4&&\\ \hline \noalign{\medskip}41&31&17&6&1&&\\ \hline \noalign{\medskip}344&213&79&12&&&\\ \hline \noalign{\medskip}375&237&91&17&&&\\ \hline \noalign{\medskip}162&117&59&19&3&&\\ \hline \noalign{\medskip}222&183&100&33&5&&\\ \hline \noalign{\medskip}204&173&104&42&10&1&\\ \hline \noalign{\medskip}192&140&72&24&4&&\\ \hline \noalign{\medskip}60&53&36&18&6&1&\\ \hline \noalign{\medskip}220&191&124&58&18&3&\\ \hline \noalign{\medskip}264&188&92&28&4&&\\ \hline \noalign{\medskip}82&72&48&23&7&1&\\ \hline \noalign{\medskip}280&244&160&76&24&4&\\ \hline \noalign{\medskip}83&66&41&19&6&1&\\ \hline \noalign{\medskip}750&612&328&108&17&&\\ \hline \noalign{\medskip}345&237&107&28&3&&\\ \hline \noalign{\medskip}324&279&176&78&22&3&\\ \hline \noalign{\medskip}113&89&54&24&7&1&\\ \hline \noalign{\medskip}106&96&71&42&19&6&1\\ \hline \noalign{\medskip}528&452&280&120&32&4& \\ \hline \end {array}  
\]
\caption{The vectors associated with  distinct nonzero coefficients 
of $\Sigma \in {\cal B}^H_4(q)$ (continued)}
\label{fcoefsi43}
\end{figure}

\subsection{$H=sl_2$, $G=sl_4$} \label{sexphsl2gsl4}
Now we study the nonstandard algebra ${\cal B}={\cal B}^H_3(q)$, when
$H=Gl_2$, $X$ is its four dimensional irreducible 
representation, and $G=GL(X)=Gl_4$. 
It is a $21$-dimensional algebra  whose 
explicit presentation is given in Section 7.1 of \cite{GCT7}.
We  follow the notation as therein. Let ${\cal P}_i$ and ${\cal Q}_i$ be
as defined in the begining of  that section.
The monomials ${\cal P}_\sigma$, where $\sigma$ ranges over strings in
$1$ and $2$  of length $k\le 10$ with no consecutive $1$'s or $2$'s,
form a  basis of ${\cal B}$. This algebra  has one trivial one-dimensinal
representation, and five nonisomorphic two-dimensional representations,
so that
\[ 21= 1+ 2^2 +2^2+2^2 +2^2+2^2.\]

Let 
\begin{equation}
 disc=\left( 5\,{q}^{16}+8\,{q}^{12}-4\,{q}^{10}+18\,{q}^{8}-4\,{q}^{6}+8\,{q}^{4}+5 \right)  \left( {q}^{8}+1 \right) ^{2}{q}^{24},
\end{equation}
and
\[ x=disc^{1/2}.\] 
Since $disc$ is not a square, $x$ does not belong to $\Q(q)$.
Let $K=\Q(q)[x]$ be the algebraic extension of $\Q(q)$ obtained
by adjoining $x$. It is shown in \cite{GCT7} that 
${\cal B}$ admits 
a complete Wederburn-structure decomposition over $K$, but not $\Q(q)$.
In what follows,
we assume that ${\cal B}$ is defined over this base
field $K$.

The nonstandard  canonical bases $E(3)$ and $E^{opp}(3)$ of ${\cal B}$ 
were
computed in MATLAB/Maple using the algorithm in Section~\ref{scanonicalB}. 
They are as follows.

Let 
$U_i$, $1 \le i \le 5$, be the
$K$-span of the entries $u_i^1,u_i^{12},u_i^{21},u_i^2 \in {\cal B}$ 
of the matrix 
\[ u_i= \left[ \begin{array}{cc} u_i^1 &  u_i^{1 2} \\ u_i^{2 1} & u_i^2 \end{array} \right], \] 
where $u_1^1$ is as specified in 
Figure~\ref{fu1},
$u_2^1$ the element obtained from $u_1^1$ by substituting $-x$ for $x$,
and $u_3^1,u_4^1,u_5^1$ as specified in Figures~\ref{fu3}-\ref{fu5}.
Elements are specified in these figures by giving their nonzero
coefficients in 
the  $\{{\cal Q}_\sigma\}$ basis; the coefficient for ${\cal Q}_\sigma$
is shown in front of $\sigma$.
Let $u_i^2$, $1\le i \le 5$, be the element obtained from $u_i^1$ 
by interchanging ${\cal Q}_1$ and ${\cal Q}_2$. 
Let $u_i^{12}=u_i^1 {\cal Q}_2$, and $u_i^{2 1}={\cal Q}_2 u_i^1$, for 
$1 \le i \le 5$. 
Let $u_0=1$ (this definition of $u_0$ is different from that in \cite{GCT7}). 
Then $u_0$ and the entries of $u_i$ form the canonical basis $E(3)$
of ${\cal B}^H_3(q)$. 
The left cells of $E(e)$ are $\{u_0\}$ and the columns of $u_i$. 
The right  cells are $\{u_0\}$ and the rows  of $u_i$. 
The representation supported by $\{u_0\}$ is the trivial
one-dimensional representation; let us denote it by $\Sigma$.
The representations supported 
by the columns or rows of $u_i$ are two-dimensional representations
of ${\cal B}$, distinct for each $i$; let us denote them by $\chi^i$.
The left cell $\{u_0\}$ is at the top of the $\le_L$ partial order
and the left cells corresponding to the  columns  $U_i$ are at 
depth $1$ from the top in this partial order (and mutually incomparable). 
The situation for the right cells is similar.

Let 
\[ v_i= \left[ \begin{array}{cc} v_i^1 &  v_i^{1 2} \\ v_i^{2 1} & v_i^2 \end{array} \right], \] 
where  $v_i^\alpha$ is obtained from $u_i^\alpha$ by substituting 
${\cal P}_\sigma$ for ${\cal Q}_\sigma$ in the expression of $u_i^\alpha$ in 
the $\{Q_\sigma\}$ basis.
Let $v_0$ be the element whose coefficients in  the $\{{\cal P}_\sigma\}$ basis
are   as shown in Figures~\ref{fv01}-\ref{fv02}.
Then $v_0$ and the elements of $v_i$ form the opposite canonical
basis $E^{opp}(3)$ of ${\cal B}^H_3(q)$. 
The left cells are $\{v_0\}$ and the columns of $v_i$. 
The right  cells are $\{v_0\}$ and the rows  of $v_i$. 
The left cell $\{v_0\}$ is at the bottom of the $\le_L$ partial order
and the left cells corresponding to the  columns of  $v_i$ at 
height $1$ from the bottom (and mutually incomparable). The situation
for the right cells is similar.

Let $\Sigma'$ be the trivial one-dimensional representation of 
the subalgebra ${\cal B}^H_2(q) \subset {\cal B}={\cal B}^H_3(q)$
generated  by ${\cal P}_2$.
Let $\mu'$ be the other  one-dimensional representation of
${\cal B}^H_2(q)$  that specializes to the signed one-dimensional
representation of the symmetric group $S_2$ at $q=1$. Then 
the nonstandard Gelfand-Tsetlin tableau $T(b)$'s associated with the basis
elements $b$'s of $E(3)$  are as follows.

If $b=u_0$, then $T(b)=[\Sigma,\Sigma']$. 
If $b=u^1_i$, then $T(b)=[\chi^i,\Sigma']$.
If $b=u^{2 1}_i$, then $T(b)=[\chi^i,\mu']$.
If $b=u^2_i$, then $T(b)=[\chi^i,\Sigma']$.
If $b=u^{1 2}_i$, then $T(b)=[\chi^i,\mu']$.

The nonstandard Gelfand-Tsetlin tableau associated with the basis
elements in $E^{opp}(r)$ are similar.

\subsubsection{Violation of standard balance}
Let $R(3)$  be the $A_K$-lattice generated by $E(3)$,
$\bar R(3)=(R(3))^{-}$. Let $R^{opp}(3)$ and $\bar R^{opp}(3)$ be
defined similarly.
Then it turns out that the triple $({\cal B}^H_{3,\Q}, R(3),\bar R(3))$ 
associated
with the canonical basis $E(3)$  is balanced, but the triple
$({\cal B}^H_{3,\Q},R^{opp}(3),\bar R^{opp}(3))$ 
associated with the opposite canonical basis $E^{opp}(3)$ is not
balanced. Specifically, the fibre 
$\psi^{-1}_T(0) \not = \{0\}$
when $T=T(b)$ is the nonstandard tableau associated with
$b=v^2_i$, for any $i$ (it is zero for all other $b$'s). 

For example, with the help  of computer 
it was found that the $\Q$-module $\psi^{-1}_T(0)$, for $b=v^2_5, T=T(b)$,
is generated 
by the two elements $w$ and $x$ specified in Figures~\ref{fw1}-\ref{fx2},
 which 
give their nonzero cofficients in the $\{{\cal P}_\sigma\}$ basis.
Clearly $w$ belongs
to the $K_\Q$ form ${\cal B}^H_{3,\Q}$ since the coefficients
belong to $\Q[q,q^{-1}]$. It is 
$-$-invariant, since the coefficients are $-$-invariant.
It can be verified that $w \in q R^{opp}(3)$.
Specifically, 
it can be shown that 
\[w=a_0 v_0 + c^1 v^1_5 + c^{12} v^{12}_5 + 
c^2 v^2_5 + c^{21} v^{21}_5,\] 
where the coefficient vector $[a_0, c^1, c^{12}, c^2, c^{21}]$ is the following
\[
 \left[ \begin {array}{c} -{\frac {{q}^{6}}{ \left( 2\,{q}^{8}-2\,{q}^{6}+3\,{q}^{4}-2\,{q}^{2}+2 \right)  \left( {q}^{2}+1 \right) ^{2}}}\\\noalign{\medskip}{\frac { \left( {q}^{4}-{q}^{2}+1 \right)  \left( {q}^{4}+1 \right) q}{ \left( {q}^{2}+1 \right)  \left( 2\,{q}^{8}-2\,{q}^{6}+3\,{q}^{4}-2\,{q}^{2}+2 \right) }}\\\noalign{\medskip}-{\frac {{q}^{6}}{ \left( 2\,{q}^{8}-2\,{q}^{6}+3\,{q}^{4}-2\,{q}^{2}+2 \right)  \left( {q}^{2}+1 \right) ^{2}}}\\\noalign{\medskip}-{\frac {{q}^{6}}{ \left( 2\,{q}^{8}-2\,{q}^{6}+3\,{q}^{4}-2\,{q}^{2}+2 \right)  \left( {q}^{2}+1 \right) ^{2}}}\\\noalign{\medskip}{\frac { \left( {q}^{4}-{q}^{2}+1 \right)  \left( {q}^{4}+1 \right) q}{ \left( {q}^{2}+1 \right)  \left( 2\,{q}^{8}-2\,{q}^{6}+3\,{q}^{4}-2\,{q}^{2}+2 \right) }}\end {array} \right] 
\]
This means $\psi_T(w)=0$. It can be verified that $w$ belongs to
${\cal B}^{\le T}$. 
Similarly, $x$ is a $-$-invariant element of
 ${\cal B}^H_{3,\Q} \cap {\cal B}^{\le T}$ that
belongs to $q R^{opp}(3)$.

\subsection{Nonstandard globalization via minimization of degree complexity}
Thus each element in $\psi^{-1}_T(\bar b)$, $b=v^2_5$, $\bar b =\psi(b)$, 
$T=T(b)$, 
is a linear combination of $b, w$ and $x$. 
It easy to see from the explicit formulae in Figures~\ref{fw1}-\ref{fx2}
and Figure~\ref{fu5} that
  $b=v^2_5$ is the element of minimum degree complexity 
in $\psi^{-1}_T(\bar b)$, where the degree complexity is defined internally as
in Section~\ref{sinternal}. 
(Remember that $v^2_5$ is obtained from $u_5^1$ in Figure~\ref{fu5}
by substituting ${\cal P}_i$ for ${\cal Q}_i$ and then interchanging
${\cal P}_1$ and ${\cal P}_2$.) The same is also true for all $b$'s. 
This verifies the minimum degree conjecture (Conjecture~\ref{cunimindegright});
Conjecture~\ref{cunimindegright2} can also be verified similarly.

We could not  use the external definition of degree complexity 
as in Section~\ref{snonstdfglobright} here, since the smallest 
product of Hecke algebras containing
${\cal B}^H_3(q)$ is $Z={\cal H}_3(q)^{\otimes 9}$ with dimension
$10077696=6^9$. It is impossible to carry out symbolic computations
in an algebra of this size in MATLAB/Maple.

\begin{sidewaysfigure}[p!]
\[\begin {array}{|c|c|}\hline
\sigma & Coefficient \\ \hline 
 1&1/2\,( {q}^{4}-{q}^{2}+1 ) ^{2} ( {q}^{8}-{q}^{6}+{q}^{4}-{q}^{2}+1 ) ^{2} ( {q}^{4}+1 ) ^{2} ( {q}^{2}+1 ) ^{4} \\ & \quad \times  ( x+3\,{q}^{28}+4\,{q}^{24}-2\,{q}^{22}+10\,{q}^{20}-2\,{q}^{18}+4\,{q}^{16}+3\,{q}^{12} )/{{q}^{40}}\\ \hline\noalign{\medskip}121&-1/2\,( {q}^{2}+1 ) ^{2} ( 2\,{q}^{18}-295\,{q}^{28}-516\,{q}^{36}+x+210\,{q}^{26}+3\,{q}^{56}-3\,{q}^{54}+47\,{q}^{46}+9\,{q}^{52}-{q}^{48} \\ & \quad +2\,{q}^{50}-84\,{q}^{24}-295\,{q}^{40}+604\,{q}^{34}+462\,{q}^{30}-x{q}^{2}+47\,{q}^{22}-9\,{q}^{20}x+19\,{q}^{10}x-{q}^{26}x+{q}^{28}x-3\,{q}^{14}\\ & \quad +{q}^{24}x+4\,{q}^{22}x  +30\,{q}^{14}x+462\,{q}^{38}-516\,{q}^{32}+19\,{q}^{18}x+210\,{q}^{42}-{q}^{20}+9\,{q}^{16}-24\,{q}^{16}x-24\,{q}^{12}x\\ & \quad -9\,{q}^{8}x+4\,{q}^{6}x+{q}^{4}x-84\,{q}^{44}+3\,{q}^{12} )/{{q}^{36}}\\ \hline\noalign{\medskip}12121&1/2\,({q}^{18}-2\,{q}^{28}+22\,{q}^{36}+x+45\,{q}^{26}+2\,{q}^{46}+3\,{q}^{48}+22\,{q}^{24}+24\,{q}^{40}+45\,{q}^{34}+92\,{q}^{30}+18\,{q}^{22}+{q}^{20}x \\ & \quad +6\,{q}^{10}x+2\,{q}^{14}+{q}^{14}x+18\,{q}^{38}-2\,{q}^{32}+{q}^{42}+24\,{q}^{20}+9\,{q}^{16}+{q}^{16}x+{q}^{6}x+{q}^{4}x+9\,{q}^{44}+3\,{q}^{12})/{{q}^{30}}\\ \hline\noalign{\medskip}1212121&-1/2\,(22\,{q}^{20}+6\,{q}^{16}+6\,{q}^{24}+2\,{q}^{26}+2\,{q}^{14}+2\,{q}^{30}+2\,{q}^{10}+3\,{q}^{28}-2\,{q}^{22}-2\,{q}^{18}+3\,{q}^{12}+x)/{{q}^{20}}\\ \hline\noalign{\medskip}121212121&1
\\ \hline \end {array}
\]
\caption{Coefficients of $u_1^1$}
\label{fu1}
\end{sidewaysfigure}

\begin{sidewaysfigure}[p!]
\[
 \begin {array}{|c|c|}\hline
\sigma & Coefficient \\ \hline  1&( {q}^{8}-{q}^{6}+{q}^{4}-{q}^{2}+1 ) ^{2} ( {q}^{12}+{q}^{10}+2\,{q}^{8}+2\,{q}^{4}+{q}^{2}+1 ) ^{2} ( {q}^{2}+1 ) ^{4} ( {q}^{4}-{q}^{2}+1 ) ^{4}/{{q}^{32}}\\ \hline\noalign{\medskip}121&-( 1-4\,{q}^{10}+14\,{q}^{8}-30\,{q}^{14}+44\,{q}^{28}+73\,{q}^{16}+3\,{q}^{2}+14\,{q}^{32}-30\,{q}^{26}+73\,{q}^{24}+3\,{q}^{38} \\ & \quad + 102\,{q}^{20}-53\,{q}^{18}+{q}^{40}+44\,{q}^{12}-53\,{q}^{22}-4\,{q}^{30}+5\,{q}^{4}+5\,{q}^{36} )  ( {q}^{2}+1 ) ^{2} ( {q}^{4}-{q}^{2}+1 ) ^{2}/{{q}^{26}}\\ \hline\noalign{\medskip}12121&(3+72\,{q}^{18}+14\,{q}^{28}+3\,{q}^{36}+20\,{q}^{26}+10\,{q}^{24}+2\,{q}^{34}+2\,{q}^{30}+36\,{q}^{22}+14\,{q}^{8}+7\,{q}^{4} \\ & \quad +2\,{q}^{2}+7\,{q}^{32}+2\,{q}^{6}-10\,{q}^{20}-10\,{q}^{16}+10\,{q}^{12}+20\,{q}^{10}+36\,{q}^{14})/{{q}^{18}}\\ \hline\noalign{\medskip}1212121&-(1-2\,{q}^{12}+14\,{q}^{10}-2\,{q}^{8}+{q}^{4}+3\,{q}^{2}+4\,{q}^{6}+{q}^{20}+4\,{q}^{14}+3\,{q}^{18}+{q}^{16})/{{q}^{10}}\\ \hline\noalign{\medskip}121212121&1\\ \hline \end{array}  
\]
\caption{Coefficients of $u_3^1$}
\label{fu3}
\end{sidewaysfigure}

\begin{sidewaysfigure}[p!]
\[
 \begin {array}{|c|c|}\hline
\sigma & Coefficient \\ \hline  1&( {q}^{2}+1 ) ^{2} ( {q}^{4}-{q}^{2}+1 ) ^{2} ( {q}^{4}+1 ) ^{2} ( {q}^{8}-{q}^{6}+{q}^{4}-{q}^{2}+1 ) ^{2} ( {q}^{12}+{q}^{10}+2\,{q}^{8}+2\,{q}^{4}+{q}^{2}+1 ) ^{2}/{{q}^{30}}\\ \hline\noalign{\medskip}121&-(1+75\,{q}^{18}-49\,{q}^{28}+42\,{q}^{36}+206\,{q}^{26}-{q}^{46}+{q}^{52}+7\,{q}^{48}-49\,{q}^{24}+40\,{q}^{40}+75\,{q}^{34}+158\,{q}^{30}+158\,{q}^{22} \\ & \quad +22\,{q}^{8}+7\,{q}^{4}+17\,{q}^{14}+17\,{q}^{38}+{q}^{32}-{q}^{6}+{q}^{20}+42\,{q}^{16}+22\,{q}^{44}+40\,{q}^{12}){{q}^{26}}\\ \hline\noalign{\medskip}12121&(3+80\,{q}^{18}+10\,{q}^{28}+3\,{q}^{36}+26\,{q}^{26}-3\,{q}^{24}+{q}^{34}+5\,{q}^{30}+52\,{q}^{22}+10\,{q}^{8}+5\,{q}^{4}+52\,{q}^{14}+{q}^{2}+5\,{q}^{32} \\ & \quad +5\,{q}^{6}-19\,{q}^{20}-19\,{q}^{16}-3\,{q}^{12}+26\,{q}^{10})/{{q}^{18}}\\ \hline\noalign{\medskip}1212121&-(1-2\,{q}^{12}+14\,{q}^{10}-2\,{q}^{8}+3\,{q}^{2}+5\,{q}^{6}+{q}^{20}+5\,{q}^{14}+3\,{q}^{18})/{{q}^{10}}\\ \hline\noalign{\medskip}121212121&1\\ \hline \end{array}  
\]
\caption{Coefficients of $u_4^1$}
\label{fu4}
\end{sidewaysfigure}

\begin{sidewaysfigure}[p!]
\[
 \begin {array}{|c|c|}\hline
\sigma & Coefficient \\ \hline  1&( {q}^{2}+1 ) ^{2} ( {q}^{4}+1 ) ^{2} ( {q}^{12}+{q}^{10}+2\,{q}^{8}+2\,{q}^{4}+{q}^{2}+1 ) ^{2} ( {q}^{4}-{q}^{2}+1 ) ^{4}/{{q}^{26}}\\ \hline\noalign{\medskip}121&-( {q}^{36}+3\,{q}^{34}+10\,{q}^{32}+19\,{q}^{30}+33\,{q}^{28}+53\,{q}^{26}+64\,{q}^{24}+91\,{q}^{22}+84\,{q}^{20}+116\,{q}^{18}+84\,{q}^{16}+91\,{q}^{14} \\ & \quad +64\,{q}^{12}+53\,{q}^{10}+33\,{q}^{8}+19\,{q}^{6}+10\,{q}^{4}+3\,{q}^{2}+1 )  ( {q}^{4}-{q}^{2}+1 ) ^{2}/{{q}^{22}}\\ \hline\noalign{\medskip}12121&(80\,{q}^{16}+3\,{q}^{26}+26\,{q}^{24}+4\,{q}^{22}+{q}^{32}+7\,{q}^{28}+50\,{q}^{20}+3\,{q}^{6}+3\,{q}^{2}+50\,{q}^{12}+1+3\,{q}^{30}+7\,{q}^{4}-14\,{q}^{18} \\ & \quad -14\,{q}^{14}+4\,{q}^{10}+26\,{q}^{8})/{{q}^{16}}\\ \hline\noalign{\medskip}1212121&-(3+5\,{q}^{12}-2\,{q}^{10}+14\,{q}^{8}+5\,{q}^{4}+{q}^{2}-2\,{q}^{6}+{q}^{14}+3\,{q}^{16})/{{q}^{8}}\\ \hline\noalign{\medskip}121212121&1\\ \hline \end{array}  
\]
\caption{Coefficients of $u_5^1$}
\label{fu5}
\end{sidewaysfigure}

\begin{sidewaysfigure}[p!]
\[
\begin {array}{|c|c|}\hline
\sigma & Coefficient \\ \hline 
\emptyset &( {q}^{4}+{q}^{3}+{q}^{2}+q+1 )  ( {q}^{4}-{q}^{3}+{q}^{2}-q+1 )  ( {q}^{6}-{q}^{3}+1 )  ( {q}^{6}+{q}^{3}+1 )  ( {q}^{6}-{q}^{5}+{q}^{4}-{q}^{3}+{q}^{2}-q+1 ) \\ & \quad \times  ( {q}^{6}+{q}^{5}+{q}^{4}+{q}^{3}+{q}^{2}+q+1 )  ( 2\,{q}^{8}-2\,{q}^{6}+3\,{q}^{4}-2\,{q}^{2}+2 )  ( {q}^{2}+q+1 ) ^{2} ( {q}^{2}-q+1 ) ^{2} \\ & \quad \times   ( {q}^{4}+1 ) ^{2} ( q-1 ) ^{4} ( q+1 ) ^{4} ( {q}^{2}+1 ) ^{4} ( {q}^{4}-{q}^{2}+1 ) ^{5}/{{q}^{46}}\\ \hline\noalign{\medskip}2&-( {q}^{2}+1 ) ^{3} ( {q}^{4}+1 )  ( {q}^{4}-{q}^{2}+1 ) ^{3} ( 2+13\,{q}^{12}-12\,{q}^{10}+11\,{q}^{8}+7\,{q}^{4}-4\,{q}^{2}-9\,{q}^{6}+{q}^{36}-10\,{q}^{42} \\ & \quad +12\,{q}^{20}-12\,{q}^{50}-11\,{q}^{32}-12\,{q}^{14}-10\,{q}^{18}+13\,{q}^{16}+16\,{q}^{26}-11\,{q}^{28}+12\,{q}^{40}+16\,{q}^{34}+{q}^{38}+{q}^{24}+28\,{q}^{30} \\ & \quad +11\,{q}^{52}+{q}^{22}+13\,{q}^{48}-12\,{q}^{46}+13\,{q}^{44}-4\,{q}^{58}+7\,{q}^{56}-9\,{q}^{54}+2\,{q}^{60} )/{{q}^{41}}\\ \hline\noalign{\medskip}1&-( {q}^{2}+1 ) ^{3} ( {q}^{4}+1 )  ( {q}^{4}-{q}^{2}+1 ) ^{3} ( 2+13\,{q}^{12}-12\,{q}^{10}+11\,{q}^{8}+7\,{q}^{4}-4\,{q}^{2}-9\,{q}^{6}+{q}^{36}-10\,{q}^{42} \\ & \quad +12\,{q}^{20}-12\,{q}^{50}-11\,{q}^{32}-12\,{q}^{14}-10\,{q}^{18}+13\,{q}^{16}+16\,{q}^{26}-11\,{q}^{28}+12\,{q}^{40}+16\,{q}^{34}+{q}^{38}+{q}^{24}+28\,{q}^{30} \\ & \quad +11\,{q}^{52}+{q}^{22}+13\,{q}^{48}-12\,{q}^{46}+13\,{q}^{44}-4\,{q}^{58}+7\,{q}^{56}-9\,{q}^{54}+2\,{q}^{60} )/{{q}^{41}}\\ \hline\noalign{\medskip}12&( 2+13\,{q}^{12}-12\,{q}^{10}+11\,{q}^{8}+7\,{q}^{4}-4\,{q}^{2}-9\,{q}^{6}+{q}^{36}-10\,{q}^{42}+12\,{q}^{20}-12\,{q}^{50}-11\,{q}^{32}-12\,{q}^{14}-10\,{q}^{18} \\ & \quad +13\,{q}^{16}+16\,{q}^{26}-11\,{q}^{28}+12\,{q}^{40}+16\,{q}^{34}+{q}^{38}+{q}^{24}+28\,{q}^{30}+11\,{q}^{52}+{q}^{22}+13\,{q}^{48}-12\,{q}^{46}+13\,{q}^{44} \\ & \quad -4\,{q}^{58}+7\,{q}^{56}-9\,{q}^{54}+2\,{q}^{60} )  ( {q}^{2}+1 ) ^{2} ( {q}^{4}-{q}^{2}+1 ) ^{2}/{{q}^{36}}\\ \hline\noalign{\medskip}21&( 2+13\,{q}^{12}-12\,{q}^{10}+11\,{q}^{8}+7\,{q}^{4}-4\,{q}^{2}-9\,{q}^{6}+{q}^{36}-10\,{q}^{42}+12\,{q}^{20}-12\,{q}^{50}-11\,{q}^{32}-12\,{q}^{14}-10\,{q}^{18} \\ & \quad +13\,{q}^{16}+16\,{q}^{26}-11\,{q}^{28}+12\,{q}^{40}+16\,{q}^{34}+{q}^{38}+{q}^{24}+28\,{q}^{30}+11\,{q}^{52}+{q}^{22}+13\,{q}^{48}-12\,{q}^{46}+13\,{q}^{44} \\ & \quad -4\,{q}^{58}+7\,{q}^{56}-9\,{q}^{54}+2\,{q}^{60} )  ( {q}^{2}+1 ) ^{2} ( {q}^{4}-{q}^{2}+1 ) ^{2}/{{q}^{36}}\\ \hline\noalign{\medskip}212&-( {q}^{2}+1 )  ( {q}^{4}+1 )  ( {q}^{4}-{q}^{2}+1 )  ( 2-3\,{q}^{12}-9\,{q}^{10}-2\,{q}^{8}+2\,{q}^{4}-4\,{q}^{2}-2\,{q}^{6}+{q}^{36}-9\,{q}^{42}+27\,{q}^{20} \\ & \quad -4\,{q}^{50}+27\,{q}^{32}-13\,{q}^{14}-48\,{q}^{18}+{q}^{16}-110\,{q}^{26}+53\,{q}^{28}-3\,{q}^{40}-48\,{q}^{34}-13\,{q}^{38}+53\,{q}^{24}-77\,{q}^{30} \\ & \quad +2\,{q}^{52}-77\,{q}^{22}+2\,{q}^{48}-2\,{q}^{46}-2\,{q}^{44} )/{{q}^{31}}\\ \hline\noalign{\medskip}121&-( {q}^{2}+1 )  ( {q}^{4}+1 )  ( {q}^{4}-{q}^{2}+1 )  ( 2-3\,{q}^{12}-9\,{q}^{10}-2\,{q}^{8}+2\,{q}^{4}-4\,{q}^{2}-2\,{q}^{6}+{q}^{36}-9\,{q}^{42}+27\,{q}^{20} \\ & \quad -4\,{q}^{50}+27\,{q}^{32}-13\,{q}^{14}-48\,{q}^{18}+{q}^{16}-110\,{q}^{26}+53\,{q}^{28}-3\,{q}^{40}-48\,{q}^{34}-13\,{q}^{38}+53\,{q}^{24}-77\,{q}^{30}+2\,{q}^{52} \\ & \quad -77\,{q}^{22}+2\,{q}^{48}-2\,{q}^{46}-2\,{q}^{44} )/{{q}^{31}}\\ \hline\noalign{\medskip}1212&(2-3\,{q}^{12}-9\,{q}^{10}-2\,{q}^{8}+2\,{q}^{4}-4\,{q}^{2}-2\,{q}^{6}+{q}^{36}-9\,{q}^{42}+27\,{q}^{20}-4\,{q}^{50}+27\,{q}^{32}-13\,{q}^{14}-48\,{q}^{18} \\ & \quad +{q}^{16}-110\,{q}^{26}+53\,{q}^{28}-3\,{q}^{40}-48\,{q}^{34}-13\,{q}^{38}+53\,{q}^{24}-77\,{q}^{30}+2\,{q}^{52}-77\,{q}^{22}+2\,{q}^{48}-2\,{q}^{46}-2\,{q}^{44})/{{q}^{26}}\\ \hline\noalign{\medskip}2121&(2-3\,{q}^{12}-9\,{q}^{10}-2\,{q}^{8}+2\,{q}^{4}-4\,{q}^{2}-2\,{q}^{6}+{q}^{36}-9\,{q}^{42}+27\,{q}^{20}-4\,{q}^{50}+27\,{q}^{32}-13\,{q}^{14}-48\,{q}^{18}+{q}^{16} \\ & \quad -110\,{q}^{26}+53\,{q}^{28}-3\,{q}^{40}-48\,{q}^{34}-13\,{q}^{38}+53\,{q}^{24}-77\,{q}^{30}+2\,{q}^{52}-77\,{q}^{22}+2\,{q}^{48}-2\,{q}^{46}-2\,{q}^{44})/{{q}^{26}}\\ \hline \end {array}  
\]
\caption{First nine coefficients of $v_0$ in $\{{\cal P}_\sigma\}$ basis}
\label{fv01}
\end{sidewaysfigure}

\begin{sidewaysfigure}[p!]
\[
\begin {array}{|c|c|}\hline
\sigma & Coefficient \\ \hline 
21212&-( {q}^{2}+1 )  ( {q}^{4}+1 )  ( {q}^{4}-{q}^{2}+1 )  ( 3\,{q}^{16}+2\,{q}^{14}+14\,{q}^{8}+2\,{q}^{2}+3 )  ( {q}^{8}-{q}^{6}+{q}^{4}+1 ) ( {q}^{8}+{q}^{4}-{q}^{2}+1 )/{{q}^{21}}\\ \hline \noalign{\medskip}
12121&-( {q}^{2}+1 )  ( {q}^{4}+1 )  ( {q}^{4}-{q}^{2}+1 )  ( 3\,{q}^{16}+2\,{q}^{14}+14\,{q}^{8}+2\,{q}^{2}+3 )  ( {q}^{8}-{q}^{6}+{q}^{4}+1 )  ( {q}^{8}+{q}^{4}-{q}^{2}+1 )/{{q}^{21}} \\ \hline 
\noalign{\medskip} 121212&( 3\,{q}^{16}+2\,{q}^{14}+14\,{q}^{8}+2\,{q}^{2}+3 )  ( {q}^{8}+{q}^{4}-{q}^{2}+1 )  ( {q}^{8}-{q}^{6}+{q}^{4}+1 )/{{q}^{16}}\\ \hline\noalign{\medskip}212121&( 3\,{q}^{16}+2\,{q}^{14}+14\,{q}^{8}+2\,{q}^{2}+3 )  ( {q}^{8}+{q}^{4}-{q}^{2}+1 )  ( {q}^{8}-{q}^{6}+{q}^{4}+1 )/{{q}^{16}}\\ \hline\noalign{\medskip}2121212&( {q}^{2}+1 )  ( {q}^{4}-{q}^{2}+1 )  ( {q}^{4}+1 )  ( 3\,{q}^{16}-{q}^{14}+3\,{q}^{12}-3\,{q}^{10}+12\,{q}^{8}-3\,{q}^{6}+3\,{q}^{4}-{q}^{2}+3 )/{{q}^{13}}\\ \hline\noalign{\medskip}1212121&( {q}^{2}+1 )  ( {q}^{4}-{q}^{2}+1 )  ( {q}^{4}+1 )  ( 3\,{q}^{16}-{q}^{14}+3\,{q}^{12}-3\,{q}^{10}+12\,{q}^{8}-3\,{q}^{6}+3\,{q}^{4}-{q}^{2}+3 )/{{q}^{13}}\\ \hline\noalign{\medskip}12121212&-(3\,{q}^{16}-{q}^{14}+3\,{q}^{12}-3\,{q}^{10}+12\,{q}^{8}-3\,{q}^{6}+3\,{q}^{4}-{q}^{2}+3)/{{q}^{8}}\\ \hline\noalign{\medskip}21212121&-(3\,{q}^{16}-{q}^{14}+3\,{q}^{12}-3\,{q}^{10}+12\,{q}^{8}-3\,{q}^{6}+3\,{q}^{4}-{q}^{2}+3)/{{q}^{8}}\\ \hline\noalign{\medskip}212121212&-( {q}^{2}+1 )  ( {q}^{4}-{q}^{2}+1 )  ( {q}^{4}+1 )/{{q}^{5}}\\ \hline\noalign{\medskip}121212121&-( {q}^{2}+1 )  ( {q}^{4}-{q}^{2}+1 )  ( {q}^{4}+1 )/{{q}^{5}}\\ \hline\noalign{\medskip}1212121212&1\\ \hline\noalign{\medskip}2121212121&1 \\ \hline \end  {array}  
\]
\caption{Last twelve coefficients of $v_0$ in $\{{\cal P}_\sigma\}$ basis}
\label{fv02}
\end{sidewaysfigure}

\begin{sidewaysfigure}[p!]
\[\begin{array}{|c|c|}\hline
\sigma & Coefficient \\ \hline 
\emptyset& (( {q}^{4}-{q}^{3}+{q}^{2}-q+1 )  ( {q}^{4}+{q}^{3}+{q}^{2}+q+1 )  ( {q}^{6}-{q}^{3}+1 )  ( {q}^{6}+{q}^{3}+1 )  ( {q}^{6}-{q}^{5}+{q}^{4}-{q}^{3}+{q}^{2}-q+1 ) \\ & \quad  ( {q}^{6}+{q}^{5}+{q}^{4}+{q}^{3}+{q}^{2}+q+1 )  ( {q}^{4}+1 ) ^{2} ( {q}^{2}-q+1 ) ^{2} ( 1+{q}^{2}+q ) ^{2} ( q-1 ) ^{4} ( q+1 ) ^{4} ( {q}^{2}+1 ) ^{2} \\ & \quad  ( {q}^{4}-{q}^{2}+1 ) ^{5})/{{q}^{40}}\\\hline \noalign{\medskip}2&-(( {q}^{2}+1 )  ( {q}^{4}+1 )  ( {q}^{4}-{q}^{2}+1 ) ^{3} ( 1-{q}^{20}-3\,{q}^{38}+{q}^{44}-3\,{q}^{14}+{q}^{4}-{q}^{2}+{q}^{12}+{q}^{8}-{q}^{6}-2\,{q}^{10}+{q}^{40} \\ & \quad -14\,{q}^{26}-8\,{q}^{34}-{q}^{32}-3\,{q}^{24}-10\,{q}^{22}-3\,{q}^{28}+{q}^{48}-10\,{q}^{30}+{q}^{52}-8\,{q}^{18}-{q}^{46}-{q}^{50}-2\,{q}^{42} )) /{{q}^{35}}\\\hline \noalign{\medskip}1&-( ( {q}^{2}+1 )  ( {q}^{4}+1 )  ( {q}^{4}-{q}^{2}+1 ) ^{3} ( 1-{q}^{20}-3\,{q}^{38}+{q}^{44}-3\,{q}^{14}+{q}^{4}-{q}^{2}+{q}^{12}+{q}^{8}-{q}^{6}-2\,{q}^{10} \\ & \quad +{q}^{40}-14\,{q}^{26}-8\,{q}^{34}-{q}^{32}-3\,{q}^{24}-10\,{q}^{22}-3\,{q}^{28}+{q}^{48}-10\,{q}^{30}+{q}^{52}-8\,{q}^{18}-{q}^{46}-{q}^{50}-2\,{q}^{42} ))/{{q}^{35}}\\\hline \noalign{\medskip}12&(( {q}^{4}-{q}^{2}+1 ) ^{2} ( 1-{q}^{20}-3\,{q}^{38}+{q}^{44}-3\,{q}^{14}+{q}^{4}-{q}^{2}+{q}^{12}+{q}^{8}-{q}^{6}-2\,{q}^{10}+{q}^{40}-14\,{q}^{26}-8\,{q}^{34} \\ & \quad -{q}^{32}-3\,{q}^{24}-10\,{q}^{22}-3\,{q}^{28}+{q}^{48}-10\,{q}^{30}+{q}^{52}-8\,{q}^{18}-{q}^{46}-{q}^{50}-2\,{q}^{42} ))/ {{q}^{30}}\\\hline \noalign{\medskip}21&(( {q}^{4}-{q}^{2}+1 ) ^{2} ( 1-{q}^{20}-3\,{q}^{38}+{q}^{44}-3\,{q}^{14}+{q}^{4}-{q}^{2}+{q}^{12}+{q}^{8}-{q}^{6}-2\,{q}^{10}+{q}^{40}-14\,{q}^{26}-8\,{q}^{34}-{q}^{32} \\ & \quad -3\,{q}^{24}-10\,{q}^{22}-3\,{q}^{28}+{q}^{48}-10\,{q}^{30}+{q}^{52}-8\,{q}^{18}-{q}^{46}-{q}^{50}-2\,{q}^{42} )) / {{q}^{30}}\\\hline \noalign{\medskip}212&-(( {q}^{2}+1 )  ( {q}^{4}+1 )  ( {q}^{4}-{q}^{2}+1 )  ( {q}^{40}-3\,{q}^{38}+4\,{q}^{36}-4\,{q}^{34}+7\,{q}^{32}-10\,{q}^{30}+19\,{q}^{28}-13\,{q}^{26}+25\,{q}^{24} \\ & \quad -22\,{q}^{22}+40\,{q}^{20}-22\,{q}^{18}+25\,{q}^{16}-13\,{q}^{14}+19\,{q}^{12}-10\,{q}^{10}+7\,{q}^{8}-4\,{q}^{6}+4\,{q}^{4}-3\,{q}^{2}+1 )) /{{q}^{25}}\\\hline \noalign{\medskip}121&- (( {q}^{2}+1 )  ( {q}^{4}+1 )  ( {q}^{4}-{q}^{2}+1 )  ( {q}^{40}-3\,{q}^{38}+4\,{q}^{36}-4\,{q}^{34}+7\,{q}^{32}-10\,{q}^{30}+19\,{q}^{28}-13\,{q}^{26}+25\,{q}^{24} \\ & \quad -22\,{q}^{22}+40\,{q}^{20}-22\,{q}^{18}+25\,{q}^{16}-13\,{q}^{14}+19\,{q}^{12}-10\,{q}^{10}+7\,{q}^{8}-4\,{q}^{6}+4\,{q}^{4}-3\,{q}^{2}+1 ))/{{q}^{25}}\\\hline \noalign{\medskip}1212&(({q}^{40}-3\,{q}^{38}+4\,{q}^{36}-4\,{q}^{34}+7\,{q}^{32}-10\,{q}^{30}+19\,{q}^{28}-13\,{q}^{26}+25\,{q}^{24}-22\,{q}^{22}+40\,{q}^{20}-22\,{q}^{18}+25\,{q}^{16} \\ & \quad -13\,{q}^{14}+19\,{q}^{12}-10\,{q}^{10}+7\,{q}^{8}-4\,{q}^{6}+4\,{q}^{4}-3\,{q}^{2}+1))/{{q}^{20}}\end{array} \] 
\caption{A $-$-invariant  element $w$ in $q R^{opp}(3)$: the first eight coefficients}
\label{fw1}
\end{sidewaysfigure}

\begin{sidewaysfigure}[p!]
\[\begin{array}{|c|c|}\hline
\sigma & Coefficient \\ \hline 
2121&({q}^{40}-3\,{q}^{38}+4\,{q}^{36}-4\,{q}^{34}+7\,{q}^{32}-10\,{q}^{30}+19\,{q}^{28}-13\,{q}^{26}+25\,{q}^{24}-22\,{q}^{22}+40\,{q}^{20}-22\,{q}^{18}+25\,{q}^{16} \\ & \quad -13\,{q}^{14}+19\,{q}^{12}-10\,{q}^{10}+7\,{q}^{8}-4\,{q}^{6}+4\,{q}^{4}-3\,{q}^{2}+1)/{{q}^{20}}\\\hline \noalign{\medskip}21212&-(( {q}^{2}+1 )  ( {q}^{4}+1 )  ( {q}^{4}-{q}^{2}+1 )  ( {q}^{20}-3\,{q}^{18}+{q}^{16}-3\,{q}^{14}+3\,{q}^{12}-10\,{q}^{10}+3\,{q}^{8}-3\,{q}^{6}+{q}^{4}-3\,{q}^{2}+1))/ {{q}^{15}}\\\hline \noalign{\medskip}12121&-(( {q}^{2}+1 )  ( {q}^{4}+1 )  ( {q}^{4}-{q}^{2}+1 )  ( {q}^{20}-3\,{q}^{18}+{q}^{16}-3\,{q}^{14}+3\,{q}^{12}-10\,{q}^{10}+3\,{q}^{8}-3\,{q}^{6}+{q}^{4}-3\,{q}^{2}+1 ))/{{q}^{15}}\\\hline \noalign{\medskip}121212&
({q}^{20}-3\,{q}^{18}+{q}^{16}-3\,{q}^{14}+3\,{q}^{12}-10\,{q}^{10}+3\,{q}^{8}-3\,{q}^{6}+{q}^{4}-3\,{q}^{2}+1)/{q}^{10}\\\hline \noalign{\medskip}212121&({q}^{20}-3\,{q}^{18}+{q}^{16}-3\,{q}^{14}+3\,{q}^{12}-10\,{q}^{10}+3\,{q}^{8}-3\,{q}^{6}+{q}^{4}-3\,{q}^{2}+1)/{{q}^{10}}\\\hline \noalign{\medskip}2121212&-(( {q}^{2}+1 )  ( {q}^{4}+1 )  ( {q}^{4}-{q}^{2}+1 ) )/{{q}^{5}}\\\hline \noalign{\medskip}1212121&-(( {q}^{2}+1 )  ( {q}^{4}+1 )  ( {q}^{4}-{q}^{2}+1 ) )/{{q}^{5}}\\\hline \noalign{\medskip}12121212&1\\\hline \noalign{\medskip}21212121&1\end {array}  \]
\caption{A $-$-invariant  element $w$ in $q R^{opp}(3)$: the last  nine coefficients}
\label{fw2}
\end{sidewaysfigure}

\begin{sidewaysfigure}[p!]
\[\begin{array}{|c|c|}\hline
\sigma & Coefficient \\ \hline 
\emptyset&(( {q}^{4}-{q}^{3}+{q}^{2}-q+1 )  ( {q}^{4}+{q}^{3}+{q}^{2}+q+1 )  ( {q}^{6}-{q}^{3}+1 )  ( {q}^{6}+{q}^{3}+1 )  ( {q}^{6}-{q}^{5}+{q}^{4}-{q}^{3}+{q}^{2}-q+1 ) \\ & \quad  ( {q}^{6}+{q}^{5}+{q}^{4}+{q}^{3}+{q}^{2}+q+1 )  ( 2\,{q}^{8}-2\,{q}^{6}+3\,{q}^{4}-2\,{q}^{2}+2 )  ( {q}^{4}+1 ) ^{2} ( {q}^{2}-q+1 ) ^{2} ( {q}^{2}+q+1 ) ^{2} \\ & \quad  ( q-1 ) ^{4} ( q+1 ) ^{4} ( {q}^{2}+1 ) ^{3} ( {q}^{4}-{q}^{2}+1 ) ^{5})/{{q}^{45}}\\\hline \noalign{\medskip}2&-(( {q}^{2}+1 ) ^{2} ( {q}^{4}-{q}^{2}+1 ) ^{3} ( {q}^{4}+1 )  ( 2+7\,{q}^{4}-4\,{q}^{2}-9\,{q}^{6}+11\,{q}^{8}-12\,{q}^{46}+13\,{q}^{12}-12\,{q}^{10}+{q}^{24} \\ & \quad -10\,{q}^{18}+13\,{q}^{16}-12\,{q}^{14}+16\,{q}^{26}+13\,{q}^{48}+13\,{q}^{44}+28\,{q}^{30}-10\,{q}^{42}+11\,{q}^{52}+2\,{q}^{60}-9\,{q}^{54}+7\,{q}^{56} \\ & \quad -4\,{q}^{58}+12\,{q}^{20}-12\,{q}^{50}-11\,{q}^{28}+16\,{q}^{34}+{q}^{22}+{q}^{36}-11\,{q}^{32}+{q}^{38}+12\,{q}^{40} ) )/{{q}^{40}}\\\hline \noalign{\medskip}1&-(( {q}^{2}+1 ) ^{2} ( {q}^{4}-{q}^{2}+1 ) ^{3} ( {q}^{4}+1 )  ( 2+7\,{q}^{4}-4\,{q}^{2}-9\,{q}^{6}+11\,{q}^{8}-12\,{q}^{46}+13\,{q}^{12}-12\,{q}^{10}+{q}^{24} \\ & \quad -10\,{q}^{18}+13\,{q}^{16}-12\,{q}^{14}+16\,{q}^{26}+13\,{q}^{48}+13\,{q}^{44}+28\,{q}^{30}-10\,{q}^{42}+11\,{q}^{52}+2\,{q}^{60}-9\,{q}^{54}+7\,{q}^{56}-4\,{q}^{58} \\ & \quad +12\,{q}^{20}-12\,{q}^{50}-11\,{q}^{28}+16\,{q}^{34}+{q}^{22}+{q}^{36}-11\,{q}^{32}+{q}^{38}+12\,{q}^{40} ) )/{{q}^{40}}\\\hline \noalign{\medskip}12&(( {q}^{2}+1 )  ( {q}^{4}-{q}^{2}+1 ) ^{2} ( 2\,{q}^{8}-2\,{q}^{6}+3\,{q}^{4}-2\,{q}^{2}+2 )  ( 1+{q}^{4}-{q}^{2}-{q}^{6}+{q}^{8} \\ & \quad -{q}^{46}+{q}^{12}-2\,{q}^{10}-3\,{q}^{24}-8\,{q}^{18}-3\,{q}^{14}-14\,{q}^{26}+{q}^{48}+{q}^{44}-10\,{q}^{30}-2\,{q}^{42}+{q}^{52}-{q}^{20}-{q}^{50} \\ & \quad -3\,{q}^{28}-8\,{q}^{34}-10\,{q}^{22}-{q}^{32}-3\,{q}^{38}+{q}^{40} ))/ {{q}^{35}}\\\hline \noalign{\medskip}21&(( {q}^{2}+1 )  ( {q}^{4}-{q}^{2}+1 ) ^{2} ( 2\,{q}^{8}-2\,{q}^{6}+3\,{q}^{4}-2\,{q}^{2}+2 )  ( 1+{q}^{4}-{q}^{2}-{q}^{6}+{q}^{8}-{q}^{46}+{q}^{12}-2\,{q}^{10}\\ & \quad -3\,{q}^{24}-8\,{q}^{18}-3\,{q}^{14}-14\,{q}^{26}+{q}^{48}+{q}^{44}-10\,{q}^{30}-2\,{q}^{42}+{q}^{52}-{q}^{20}-{q}^{50}-3\,{q}^{28}-8\,{q}^{34}-10\,{q}^{22} \\ & \quad -{q}^{32}-3\,{q}^{38}+{q}^{40} ) )/{{q}^{35}}\\\hline \noalign{\medskip}212&-(( {q}^{4}-{q}^{2}+1 )  ( {q}^{4}+1 )  ( 2+2\,{q}^{4}-4\,{q}^{2}-2\,{q}^{6}-2\,{q}^{8}-2\,{q}^{46}-3\,{q}^{12}-9\,{q}^{10}+53\,{q}^{24}-48\,{q}^{18}+{q}^{16} \\ &\quad -13\,{q}^{14}-110\,{q}^{26}+2\,{q}^{48}-2\,{q}^{44}-77\,{q}^{30}-9\,{q}^{42}+2\,{q}^{52}+27\,{q}^{20}-4\,{q}^{50}+53\,{q}^{28}-48\,{q}^{34}-77\,{q}^{22} \\ & \quad +{q}^{36}+27\,{q}^{32}-13\,{q}^{38}-3\,{q}^{40} ))/{{q}^{30}}\\\hline \noalign{\medskip}121&-(( {q}^{4}-{q}^{2}+1 )  ( {q}^{4}+1 )  ( 2+2\,{q}^{4}-4\,{q}^{2}-2\,{q}^{6}-2\,{q}^{8}-2\,{q}^{46}-3\,{q}^{12}-9\,{q}^{10}+53\,{q}^{24}-48\,{q}^{18}+{q}^{16}\\ & \quad -13\,{q}^{14}-110\,{q}^{26}+2\,{q}^{48}-2\,{q}^{44}-77\,{q}^{30}-9\,{q}^{42}+2\,{q}^{52}+27\,{q}^{20}-4\,{q}^{50}+53\,{q}^{28}-48\,{q}^{34}-77\,{q}^{22} \\ & \quad +{q}^{36}+27\,{q}^{32}-13\,{q}^{38}-3\,{q}^{40} ) )/{{q}^{30}}\\\hline \noalign{\medskip}1212&(( 2\,{q}^{8}-2\,{q}^{6}+3\,{q}^{4}-2\,{q}^{2}+2 )  ( {q}^{40}-3\,{q}^{38}+4\,{q}^{36}-4\,{q}^{34}+7\,{q}^{32}-10\,{q}^{30}+19\,{q}^{28}-13\,{q}^{26} \\ & \quad +25\,{q}^{24}-22\,{q}^{22}+40\,{q}^{20}-22\,{q}^{18}+25\,{q}^{16}-13\,{q}^{14}+19\,{q}^{12}-10\,{q}^{10}+7\,{q}^{8}-4\,{q}^{6}+4\,{q}^{4}-3\,{q}^{2}+1 )  ( {q}^{2}+1 ))/{{q}^{25}} \\ \hline \end {array}\]  
\caption{A $-$-invariant  element $x$ in $q R^{opp}(3)$: the first eight coefficients}
\label{fx1}
\end{sidewaysfigure}

\begin{sidewaysfigure}[p!]
\[\begin{array}{|c|c|}\hline
\sigma & Coefficient \\ \hline 
2121&(( 2\,{q}^{8}-2\,{q}^{6}+3\,{q}^{4}-2\,{q}^{2}+2 )  ( {q}^{40}-3\,{q}^{38}+4\,{q}^{36}-4\,{q}^{34}+7\,{q}^{32}-10\,{q}^{30}+19\,{q}^{28}-13\,{q}^{26}+25\,{q}^{24} \\ & \quad -22\,{q}^{22}+40\,{q}^{20}-22\,{q}^{18}+25\,{q}^{16}-13\,{q}^{14}+19\,{q}^{12}-10\,{q}^{10}+7\,{q}^{8}-4\,{q}^{6}+4\,{q}^{4}-3\,{q}^{2}+1 )  ( {q}^{2}+1 ) )/{{q}^{25}}\\\hline \noalign{\medskip}21212&-(( {q}^{4}+1 )  ( {q}^{4}-{q}^{2}+1 )  ( 3\,{q}^{16}+2\,{q}^{14}+14\,{q}^{8}+2\,{q}^{2}+3 )  ( {q}^{8}+{q}^{4}-{q}^{2}+1 )  ( {q}^{8}-{q}^{6}+{q}^{4}+1 ) )/{{q}^{20}}\\\hline \noalign{\medskip}12121&-(( {q}^{4}+1 )  ( {q}^{4}-{q}^{2}+1 )  ( 3\,{q}^{16}+2\,{q}^{14}+14\,{q}^{8}+2\,{q}^{2}+3 )  ( {q}^{8}+{q}^{4}-{q}^{2}+1 )  ( {q}^{8}-{q}^{6}+{q}^{4}+1 ) )/{{q}^{20}}\\\hline \noalign{\medskip}121212&(( 2\,{q}^{8}-2\,{q}^{6}+3\,{q}^{4}-2\,{q}^{2}+2 )  ( {q}^{20}-3\,{q}^{18}+{q}^{16}-3\,{q}^{14}+3\,{q}^{12}-10\,{q}^{10}+3\,{q}^{8}-3\,{q}^{6}+{q}^{4}-3\,{q}^{2}+1 ) \\ & \quad   ( {q}^{2}+1 ) )/{{q}^{15}}\\\hline \noalign{\medskip}212121&(( 2\,{q}^{8}-2\,{q}^{6}+3\,{q}^{4}-2\,{q}^{2}+2 )  ( {q}^{20}-3\,{q}^{18}+{q}^{16}-3\,{q}^{14}+3\,{q}^{12}-10\,{q}^{10}+3\,{q}^{8}-3\,{q}^{6}+{q}^{4}-3\,{q}^{2}+1 ) \\ & \quad   ( {q}^{2}+1 ) )// {{q}^{15}}\\\hline \noalign{\medskip}2121212&(( {q}^{4}+1 )  ( {q}^{4}-{q}^{2}+1 )  ( -3\,{q}^{6}-3\,{q}^{10}+12\,{q}^{8}-{q}^{2}+3+3\,{q}^{4}+3\,{q}^{16}-{q}^{14}+3\,{q}^{12} ) )/{{q}^{12}}\\\hline \noalign{\medskip}1212121&(( {q}^{4}+1 )  ( {q}^{4}-{q}^{2}+1 )  ( -3\,{q}^{6}-3\,{q}^{10}+12\,{q}^{8}-{q}^{2}+3+3\,{q}^{4}+3\,{q}^{16}-{q}^{14}+3\,{q}^{12} ) )/{{q}^{12}}\\\hline \noalign{\medskip}12121212&(( 2\,{q}^{8}-2\,{q}^{6}+3\,{q}^{4}-2\,{q}^{2}+2 )  ( {q}^{2}+1 ) )/{{q}^{5}}\\\hline \noalign{\medskip}21212121&
(( 2\,{q}^{8}-2\,{q}^{6}+3\,{q}^{4}-2\,{q}^{2}+2 )  ( {q}^{2}+1 ) )/{{q}^{5}}\\\hline \noalign{\medskip}212121212&-(( {q}^{4}+1 )  ( {q}^{4}-{q}^{2}+1 ) )/{{q}^{4}}\\\hline \noalign{\medskip}121212121&-(( {q}^{4}+1 )  ( {q}^{4}-{q}^{2}+1 ) )/{{q}^{4}} \\ \hline \end {array} \] 
\caption{A $-$-invariant  element $x$ in $q R^{opp}(3)$: the last eleven  coefficients}
\label{fx2}
\end{sidewaysfigure}

\clearpage

\subsubsection{Nonstandard positivity} \label{snonstdposexp}
Now we shall describe evidence for Conjecture~\ref{cnonstdposright1}
in the case under consideration.

So far we are assuming  that ${\cal P}_i$'s 
are as defined in the begining of Section 7.1 in \cite{GCT7}.  
But as explained towards the end of that section, the actual ${\cal P}_i$'s 
differ from these (chosen for convenience and simplicity)
by a positive, unimodal  factor $\hat f_p(q) \in \Q[q,q^{-1}]$ given there.
As it turns this does not matter in the calculations so far,
but for rescaling of the picture, but it does  in
the study of the positivity  properties below. 
Rescaling of ${\cal P}_i$ by $\hat f_p(q)$ implies that each
structural coefficient or  constant $c(q)$ computed so far has to be
multiplied by an appropriate power of $\hat f_p(q)^{m(c)}$, where
$m(c)$ is a nonnegative integer depending on $c$.
In what follows, it is implicitly assumed that
each  $c$ computed so far has been rescaled  by a suitable
power $\hat f_p(q)^{n(c)}$, 
where $n(c) \le m(c)$ is the smallest nonnegative integer chosen 
 so that 
the (nonstandard)
positivity property of the coefficients of $c$ becomes apparent
The picture remains 
the same even if we were to  multiply by $\hat f_p(q)^{m(c)}$,
but we choose $n(c)$ as small as possible to keep the 
the degrees of the polynomials from blowing up.


Figures~\ref{fv0sigma11}-\ref{fv0sigma4}
show the vectors associated with the nonzero coefficients of
$v_0$ in the $\{{\cal P}_\sigma\}$ basis. The vector 
(as defined
in the begining of Section~\ref{sexpevid}) for
the coefficient  
corresponding to each  $\sigma$ is obtained by concatenating
the rows in front of that $\sigma$.
Figure~\ref{ftracev11} shows the vectors associated with the
traces of the  nonzero coefficients 
of $v^1_1$ and Figures~\ref{fnormv11}-\ref{fnormv11_2} show their norms. 
Here the trace and norm of an element in the algerbraic extension $K$ 
of $\Q(q)$ is defined in the usual fashion as the sum and product
of its images under the  Frobenius automorphisms of $K$
over $\Q(q)$; they are  coefficients of the minimal polynomial of the
element. It can be seen that all  vectors in 
Figures~\ref{fv0sigma11}-\ref{fnormv11_2}
are positive and 
nonincreasing, except for the  vectors associated with $v_0$ for
$\sigma=212,121,1212,2121$, which are almost positive and nonincreasing.
It was verified with the help of computer that the vectors associated 
with the  coefficients of other canonical basis elements are similarly either
positive and nonincreasing or almost positive and nonincreasing.
This is in accordance with 
Conjectures~\ref{cnonstdposright}-\ref{cnonstdposright1}.

Figures~\ref{fmultcon1}-\ref{fmult9_2} show vectors associated 
with a few multiplicative constants, taking  norms and traces whenever
necessary; the coefficient 
of $b b'$ with respect to  the basis element $b''$
is denoted by $c(b,b';b'')$. Again it can be seen that these
vectors are positive and nonincreasing, except  a few,
which are almost positive and nonincreasing. It was verified with 
the help of computer that the picture is the same for other multiplicative
constants as well.
This too is in accordance with 
Conjectures~\ref{cnonstdposright}-\ref{cnonstdposright1}.

\begin{figure} 
\[
\begin {array}{|c|ccccccccc|} \hline 
\sigma& \mbox{vector} \\ \hline
\emptyset&34116640&34028832&33766665&33333910&32736719&31983492&31084702\\\noalign{\medskip}&30052720&28901524&27646408&26303647&24890162&23423208&21920062\\\noalign{\medskip}&20397697&18872456&17359791&15874002&14428069&13033484&11700135\\\noalign{\medskip}&10436190&9248100&8140602&7116788&6178174&5324820&4555450\\\noalign{\medskip}&3867635&3257976&2722277&2255718&1853025&1508640&1216881\\\noalign{\medskip}&972102&768790&601662&465735&356396&269443&201126\\\noalign{\medskip}&148131&107564&76935&54142&37439&25404&16892\\\noalign{\medskip}&10988&6976&4308&2576&1484&821&434\\\noalign{\medskip}&217&100&40&12&2&&\\ \hline \noalign{\medskip}2&13180&13086&12992&12744&12496&12124&11752\\\noalign{\medskip}&11225&10698&10112&9526&8890&8254&7584\\\noalign{\medskip}&6914&6294&5674&5083&4492&3979&3466\\\noalign{\medskip}&3036&2606&2256&1906&1638&1370&1178\\\noalign{\medskip}&986&840&694&603&512&450&388\\\noalign{\medskip}&335&282&259&236&206&176&153\\\noalign{\medskip}&130&116&102&85&68&54&40\\\noalign{\medskip}&34&28&21&14&9&4&4\\\noalign{\medskip}&4&2&&&&& \\ \hline \end {array}  
\]
\caption{The vectors associated with the coefficients of $v_{0}$}
\label{fv0sigma11}
\end{figure}

\begin{figure} 
\[
\begin {array}{|c|ccccccccc|} \hline 
\sigma& \mbox{vector} \\ \hline
1&13180&13086&12992&12744&12496&12124&11752\\\noalign{\medskip}&11225&10698&10112&9526&8890&8254&7584\\\noalign{\medskip}&6914&6294&5674&5083&4492&3979&3466\\\noalign{\medskip}&3036&2606&2256&1906&1638&1370&1178\\\noalign{\medskip}&986&840&694&603&512&450&388\\\noalign{\medskip}&335&282&259&236&206&176&153\\\noalign{\medskip}&130&116&102&85&68&54&40\\\noalign{\medskip}&34&28&21&14&9&4&4\\\noalign{\medskip}&4&2&&&&&\\ \hline \noalign{\medskip}12&3432&3432&3379&3326&3242&3158&3033\\\noalign{\medskip}&2908&2744&2580&2417&2254&2069&1884\\\noalign{\medskip}&1709&1534&1371&1208&1062&916&797\\\noalign{\medskip}&678&581&484&411&338&287&236\\\noalign{\medskip}&202&168&143&118&108&98&84\\\noalign{\medskip}&70&65&60&56&52&43&34\\\noalign{\medskip}&30&26&23&20&15&10&7\\\noalign{\medskip}&4&4&4&2&&& \\ \hline \end {array}  
\]
\caption{The vectors associated with the coefficients of $v_{0}$ (continued)}
\label{fv0sigma12}
\end{figure}

\begin{figure} 
\[
\begin {array}{|c|ccccccccc|} \hline 
\sigma& \mbox{vector} \\ \hline
21&3432&3432&3379&3326&3242&3158&3033&2908&2744\\\noalign{\medskip}&2580&2417&2254&2069&1884&1709&1534&1371&1208\\\noalign{\medskip}&1062&916&797&678&581&484&411&338&287\\\noalign{\medskip}&236&202&168&143&118&108&98&84&70\\\noalign{\medskip}&65&60&56&52&43&34&30&26&23\\\noalign{\medskip}&20&15&10&7&4&4&4&2&\\ \hline \noalign{\medskip}212&13352&13285&13218&12957&12696&12341&11986&11462&10938\\\noalign{\medskip}&10358&9778&9131&8484&7812&7140&6498&5856&5240\\\noalign{\medskip}&4624&4088&3552&3079&2606&2236&1866&1552&1238\\\noalign{\medskip}&1026&814&646&478&373&268&198&128&93\\\noalign{\medskip}&58&35&12&4&-4&-4&-4&-4&-4\\\noalign{\medskip}&-4&-4&-2&&&&&&\\ \hline \noalign{\medskip}121&13352&13285&13218&12957&12696&12341&11986&11462&10938\\\noalign{\medskip}&10358&9778&9131&8484&7812&7140&6498&5856&5240\\\noalign{\medskip}&4624&4088&3552&3079&2606&2236&1866&1552&1238\\\noalign{\medskip}&1026&814&646&478&373&268&198&128&93\\\noalign{\medskip}&58&35&12&4&-4&-4&-4&-4&-4\\\noalign{\medskip}&-4&-4&-2&&&&&&\\ \hline \noalign{\medskip}1212&3472&3472&3427&3382&3293&3204&3093&2982&2810\\\noalign{\medskip}&2638&2483&2328&2132&1936&1772&1608&1423&1238\\\noalign{\medskip}&1098&958&820&682&587&492&398&304&249\\\noalign{\medskip}&194&153&112&85&58&37&16&10&4\\\noalign{\medskip}&2&0&-2&-4&-4&-4&-2&&\\ \hline \noalign{\medskip}2121&3472&3472&3427&3382&3293&3204&3093&2982&2810\\\noalign{\medskip}&2638&2483&2328&2132&1936&1772&1608&1423&1238\\\noalign{\medskip}&1098&958&820&682&587&492&398&304&249\\\noalign{\medskip}&194&153&112&85&58&37&16&10&4\\\noalign{\medskip}&2&0&-2&-4&-4&-4&-2&& \\ \hline \end{array}  
\]
\caption{The vectors associated with the coefficients of $v_{0}$ (continued)}
\label{fv0sigma2}
\end{figure}

\begin{figure} 
\[
\begin {array}{|c|ccccccccc|} \hline 
\sigma& \mbox{vector} \\ \hline
21212&5496&5453&5410&5286&5162&5008&4854&4600&4346\\\noalign{\medskip}&4068&3790&3497&3204&2894&2584&2295&2006&1749\\\noalign{\medskip}&1492&1280&1068&889&710&586&462&366&270\\\noalign{\medskip}&215&160&117&74&56&38&27&16&11\\\noalign{\medskip}&6&3&&&&&&&\\ \hline \noalign{\medskip}12121&5496&5453&5410&5286&5162&5008&4854&4600&4346\\\noalign{\medskip}&4068&3790&3497&3204&2894&2584&2295&2006&1749\\\noalign{\medskip}&1492&1280&1068&889&710&586&462&366&270\\\noalign{\medskip}&215&160&117&74&56&38&27&16&11\\\noalign{\medskip}&6&3&&&&&&&\\ \hline \noalign{\medskip}121212&1434&1434&1406&1378&1346&1314&1267&1220&1128\\\noalign{\medskip}&1036&961&886&799&712&633&554&470&386\\\noalign{\medskip}&334&282&231&180&146&112&82&52&42\\\noalign{\medskip}&32&24&16&11&6&3&&&\\ \hline \noalign{\medskip}212121&1434&1434&1406&1378&1346&1314&1267&1220&1128\\\noalign{\medskip}&1036&961&886&799&712&633&554&470&386\\\noalign{\medskip}&334&282&231&180&146&112&82&52&42\\\noalign{\medskip}&32&24&16&11&6&3&&&\\ \hline \noalign{\medskip}2121212&1004&992&980&957&934&908&882&829&776\\\noalign{\medskip}&716&656&597&538&472&406&346&286&240\\\noalign{\medskip}&194&158&122&94&66&51&36&26&16\\\noalign{\medskip}&11&6&3&&&&&& \\ \hline \end{array}  
\]
\caption{The vectors associated with the coefficients of $v_{0}$ (continued)}
\label{fv0sigma33}
\end{figure}

\begin{figure} 
\[
\begin {array}{|c|ccccccccc|} \hline 
\sigma& \mbox{vector} \\ \hline
1212121&1004&992&980&957&934&908&882&829&776\\\noalign{\medskip}&716&656&597&538&472&406&346&286&240\\\noalign{\medskip}&194&158&122&94&66&51&36&26&16\\\noalign{\medskip}&11&6&3&&&&&&\\ \hline \noalign{\medskip}12121212&258&258&252&246&245&244&237&230&208\\\noalign{\medskip}&186&168&150&132&114&95&76&60&44\\\noalign{\medskip}&37&30&23&16&11&6&3&&\\ \hline \noalign{\medskip}21212121&258&258&252&246&245&244&237&230&208\\\noalign{\medskip}&186&168&150&132&114&95&76&60&44\\\noalign{\medskip}&37&30&23&16&11&6&3&&\\ \hline \noalign{\medskip}212121212&68&67&66&65&64&63&62&58&54\\\noalign{\medskip}&49&44&39&34&28&22&17&12&9\\\noalign{\medskip}&6&4&2&1&&&&&\\ \hline \noalign{\medskip}121212121&68&67&66&65&64&63&62&58&54\\\noalign{\medskip}&49&44&39&34&28&22&17&12&9\\\noalign{\medskip}&6&4&2&1&&&&&  \\ \hline \end{array}  
\]
\caption{The vectors associated with the coefficients of $v_{0}$ (continued)}
\label{fv0sigma4}
\end{figure}

\begin{figure} 
\[
\begin {array}{|c|ccccccccc|} \hline 
\sigma& \mbox{vector} \\ \hline
1&15864&15864&15680&15496&15160&14824&14334&13844&13228\\\noalign{\medskip}&12612&11932&11252&10505&9758&9008&8258&7534&6810\\\noalign{\medskip}&6126&5442&4834&4226&3702&3178&2738&2298&1948\\\noalign{\medskip}&1598&1333&1068&868&668&534&400&310&220\\\noalign{\medskip}&165&110&80&50&34&18&12&6&3\\ \hline \noalign{\medskip}121&14690&14690&14484&14278&13930&13582&13065&12548&11880\\\noalign{\medskip}&11212&10509&9806&9032&8258&7498&6738&6031&5324\\\noalign{\medskip}&4684&4044&3498&2952&2515&2078&1729&1380&1124\\\noalign{\medskip}&868&698&528&405&282&212&142&101&60\\\noalign{\medskip}&42&24&15&6&3&&&&\\ \hline \noalign{\medskip}12121&4974&4974&4886&4798&4683&4568&4380&4192&3914\\\noalign{\medskip}&3636&3367&3098&2803&2508&2217&1926&1669&1412\\\noalign{\medskip}&1205&998&825&652&529&406&316&226&172\\\noalign{\medskip}&118&89&60&41&22&14&6&3&\\ \hline \noalign{\medskip}1212121&708&708&694&680&673&666&642&618&565\\\noalign{\medskip}&512&464&416&365&314&262&210&170&130\\\noalign{\medskip}&106&82&62&42&30&18&11&4&2 \\ \hline
\end {array}  
\]
\caption{The vectors associated with the traces of the 
coefficients of $v^1_1$}
\label{ftracev11}
\end{figure}

\begin{figure} 
\[
\begin {array}{|c|ccccccccc|} \hline 
\sigma& \mbox{vector} \\ \hline
1&942062408&940617136&936306916&929157344&919242745&906637444&891460752\\\noalign{\medskip}&873831980&853909410&831851324&807845688&782080468&754764061&726104864\\\noalign{\medskip}&696321516&665632656&634256120&602409744&570301452&538139168&506112609\\\noalign{\medskip}&474411492&443199968&412642188&382872865&354026712&326205988&299512952\\\noalign{\medskip}&274016717&249786396&226859750&205274540&185039618&166163836&148631230\\\noalign{\medskip}&132425836&117511844&103853444&91399832&80100204&69893842&60720028\\\noalign{\medskip}&52513000&45206996&38734986&33029940&28027078&23661620&19873490\\\noalign{\medskip}&16602612&13795174&11397364&9362691&7644664&6204292&5002584\\\noalign{\medskip}&4007841&3188364&2519160&1975236&1537471&1186744&908860\\\noalign{\medskip}&689624&518854&386368&285084&207920&150131&106972\\\noalign{\medskip}&75388&52324&35881&24160&16048&10432&6681\\\noalign{\medskip}&4164&2550&1508&874&484&262&132\\\noalign{\medskip}&64&28&12&4&1&&\\ \hline \noalign{\medskip}121&835471628&833946584&829404381&821877948&811459617&798241720&782371271\\\noalign{\medskip}&763995284&743308085&720504000&695809491&669451020&641674579&612726160\\\noalign{\medskip}&582858876&552325840&521370802&490237512&459149620&428330776&397976239\\\noalign{\medskip}&368281268&339402663&311497224&284682092&259074408&234750212&211785544\\\noalign{\medskip}&190216267&170078244&151372598&134100452&118233841&103744800&90582172\\\noalign{\medskip}&78694800&68016456&58480912&50013272&42538640&35978861&30255780\\\noalign{\medskip}&25293821&21017408&17356415&14240716&11608322&9397244&7554889\\\noalign{\medskip}&6028664&4775532&3752456&2925473&2260620&1732150&1314316\\\noalign{\medskip}&988221&734968&541214&393616&283165&200852&140775\\\noalign{\medskip}&97032&65989&44012&28926&18556&11701&7160\\\noalign{\medskip}&4299&2484&1405&752&391&188&88\\\noalign{\medskip}&36&14&4&1&&& \\ \hline \end {array}  
\]
\caption{The vectors associated with the norms of the coefficients of
$v^1_1$}
\label{fnormv11}
\end{figure}

\begin{figure} 
\[
\begin {array}{|c|ccccccccc|} \hline 
12121&100225178&100001236&99336179&98236776&96719406&94800448&92504960\\\noalign{\medskip}&89858000&86893481&83645316&80151751&76451032&72582611&68585940\\\noalign{\medskip}&64501436&60369516&56227388&52112260&48055958&44090308&40241933\\\noalign{\medskip}&36537456&32995876&29636192&26469934&23508632&20757113&18220204\\\noalign{\medskip}&15896382&13784124&11876674&10167276&8645972&7302804&6125315\\\noalign{\medskip}&5101048&4216788&3459320&2815838&2273536&1820593&1445188\\\noalign{\medskip}&1137127&886216&684162&522672&395267&295468&218449\\\noalign{\medskip}&159384&114887&81572&57144&39308&26622&17644\\\noalign{\medskip}&11492&7284&4521&2704&1573&868&461\\\noalign{\medskip}&224&102&40&15&4&1&\\ \hline \noalign{\medskip}1212121&2212002&2205476&2186350&2155076&2112378&2058980&1995675\\\noalign{\medskip}&1923256&1842975&1756084&1663849&1567536&1468102&1366504\\\noalign{\medskip}&1263855&1161268&1059834&960644&864459&772040&684050\\\noalign{\medskip}&601152&523831&452572&387494&328716&276117&229576\\\noalign{\medskip}&188839&153652&123592&98236&77108&59732&45640\\\noalign{\medskip}&34364&25488&18596&13346&9396&6493&4384\\\noalign{\medskip}&2894&1848&1142&672&377&196&95\\\noalign{\medskip}&40&15&4&1&&& \\ \hline \end {array}  
\]
\caption{The vectors associated with the norms of the coefficients of
$v^1_1$ (continued)}
\label{fnormv11_2}
\end{figure}

\begin{figure} 
\[
\begin {array}{cccccccccc}  
9846&9820&9750&9644&9501&9320&9093\\\noalign{\medskip}8812&8493&8152&7779&7364&6917&6448\\\noalign{\medskip}5966&5480&4989&4492&4001&3528&3080\\\noalign{\medskip}2664&2274&1904&1569&1284&1038&820\\\noalign{\medskip}631&472&346&256&186&120&69\\\noalign{\medskip}44&31&16&4&&&\end {array}  
\]
\caption{The vector for the multiplicative constant $c(v0,v^1_5;v0)$}
\label{fmultcon1}
\end{figure}

\begin{figure} 
\[
\begin{array}{cccccccccc} 
13026&12964&12777&12464&12048&11552&10964\\\noalign{\medskip}10272&9523&8764&7990&7196&6398&5612\\\noalign{\medskip}4867&4192&3569&2980&2444&1980&1583\\\noalign{\medskip}1248&967&732&539&384&267&188\\\noalign{\medskip}131&80&42&24&16&8&2\end {array}  
\]
\caption{The vector for the multiplicative constant $c(v0,v^1_4;v0)$}
\label{fmult2}
\end{figure}

\begin{figure} 
\[
\begin{array}{cccccccccc}  
14180&14088&13832&13432&12894&12224&11448\\\noalign{\medskip}10592&9682&8744&7800&6872&5976&5128\\\noalign{\medskip}4342&3632&2998&2440&1956&1544&1202\\\noalign{\medskip}928&706&520&374&272&198&136\\\noalign{\medskip}88&56&36&24&16&8&2\end {array}  
\]
\caption{The vector for the multiplicative constant  $c(v0,v^1_3;v0)$}
\label{fmult3}
\end{figure}

\begin{figure} 
\[
\begin{array}{cccccccccc}  
1356922&1356922&1341857&1326792&1297628&1268464\\\noalign{\medskip}1227083&1185702&1133960&1082218&1023821&965424\\\noalign{\medskip}902616&839808&776306&712804&650849&588894\\\noalign{\medskip}531320&473746&421861&369976&325182&280388\\\noalign{\medskip}243022&205656&175677&145698&122582&99466\\\noalign{\medskip}82260&65054&52971&40888&32575&24262\\\noalign{\medskip}19002&13742&10480&7218&5402&3586\\\noalign{\medskip}2573&1560&1107&654&431&208\\\noalign{\medskip}135&62&35&8&4&\end {array}  
\]
\caption{The vector for the trace of the multiplicative constant 
$c(v0,v^1_1;v0)$}
\label{fmult4}
\end{figure}

\begin{figure} 
\[
\begin{array}{ccc}  
128607887512140&128408739182416&127813071267725\\\noalign{\medskip}126826189536408&125456853005552&123717149426056\\\noalign{\medskip}121622324577366&119190568251368&116442761279740\\\noalign{\medskip}113402187959064&110094219510476&106545974210632\\\noalign{\medskip}102785960441146&98843708903656&94749400393129\\\noalign{\medskip}90533495522016&86226372383558&81857978142544\\\noalign{\medskip}77457499613617&73053057887224&68671430858636\\\noalign{\medskip}64337807515464&60075576278933&55906149694176\\\noalign{\medskip}51848826289840&47920690427296&44136549385454\\\noalign{\medskip}40508906927184&37047971392480&33761696363504\\\noalign{\medskip}30655850888930&27734116255008&24998205742188\\\noalign{\medskip}22448003806144&20081720818165&17896059499880\\\noalign{\medskip}15886389332178&14046925218184&12370907081114\\\noalign{\medskip}10850777077832&9478351722036&8244986211000\\\noalign{\medskip}7141729026165&6159464877872&5289044837698\\\noalign{\medskip}4521402501856&3847655805468&3259194107520\\\noalign{\medskip}2747750779948&2305461534304&1924909254250\end {array}  
\]
\caption{The vector for the norm of the multiplicative constant
$c(v0,v^1_1;v0)$}
\label{fmult5}
\end{figure}

\begin{figure} 
\[
\begin{array}{ccc}  
1599156102128&1321763961652&1086804278768\\\noalign{\medskip}888858554957&723010747256&584832826172\\\noalign{\medskip}470364742664&376089969502&298907782312\\\noalign{\medskip}236103258694&185315973800&144508153155\\\noalign{\medskip}111933043504&86104032184&65765045520\\\noalign{\medskip}49862540548&37519404368&28010898646\\\noalign{\medskip}20742786784&15231640340&11087321280\\\noalign{\medskip}7997549136&5714462144&4043023409\\\noalign{\medskip}2831123144&1961213430&1343311944\\\noalign{\medskip}909211769&607734400&400883734\\\noalign{\medskip}260758832&167107927&105406152\\\noalign{\medskip}65367982&39805384&23766532\\\noalign{\medskip}13889944&7930388&4412904\\\noalign{\medskip}2386644&1250232&631716\\\noalign{\medskip}306184&141372&61592\\\noalign{\medskip}25006&9272&3049\\\noalign{\medskip}848&190&32\\\noalign{\medskip}4&&\end {array}  
\]
\caption{The vector for the norm of the multiplicative constant
$c(v0,v^1_1;v0)$ (continued)}
\label{fmult6}
\end{figure}

\clearpage

\begin{figure} 
\[
\begin{array}{cccccccccc}  
612764&609940&601539&587774&568990&545654&518332\\\noalign{\medskip}487666&454355&419136&382753&345926&309336&273610\\\noalign{\medskip}239300&206862&176658&148958&123938&101678&82176\\\noalign{\medskip}65362&51106&39226&29503&21696&15554&10828\\\noalign{\medskip}7279&4686&2851&1604&800&316&52\\\noalign{\medskip}-68&-101&-90&-64&-38&-18&-6\\\noalign{\medskip}-1&&&&&&\end {array}  
\]
\caption{The vector for  the multiplicative constant 
$c(v^1_5, v^1_5;v^1_5)$}
\label{fmult7_3}
\end{figure}

\begin{figure} 
\[
\begin{array}{cccccccccc}  
9738&9694&9563&9348&9052&8676&8228\\\noalign{\medskip}7732&7209&6658&6077&5484&4900&4332\\\noalign{\medskip}3784&3268&2794&2360&1962&1604&1292\\\noalign{\medskip}1028&809&626&471&344&246&172\\\noalign{\medskip}116&76&50&32&17&6&1\end {array}  
\]
\caption{The vector for  the multiplicative constant 
 $c(v^1_4,v^1_4;v^1_4)$}
\label{fmult8}
\end{figure}

\begin{figure} 
\[
\begin{array}{cccccccccc}  
81298&80886&79660&77650&74908&71510&67546\\\noalign{\medskip}63110&58306&53254&48074&42870&37740&32786\\\noalign{\medskip}28097&23732&19734&16144&12987&10258&7938\\\noalign{\medskip}6010&4449&3212&2251&1526&999&628\\\noalign{\medskip}374&208&107&50&20&6&1\end {array}  
\]
\caption{The vector for  the multiplicative constant
 $c(v^1_3,v^1_3;v^1_3)$}
\label{fmult9}
\end{figure}

\begin{figure} 
\[
\begin{array}{ccccccc}  
976672&974152&971632&954184&936736&915476\\\noalign{\medskip}894216&860888&827560&791463&755366&713035\\\noalign{\medskip}670704&627743&584782&540877&496972&454258\\\noalign{\medskip}411544&373144&334744&298301&261858&232239\\\noalign{\medskip}202620&175584&148548&128938&109328&91580\\\noalign{\medskip}73832&62540&51248&41348&31448&25972\\\noalign{\medskip}20496&15650&10804&8712&6620&4729\\\noalign{\medskip}2838&2215&1592&986&380&312\\\noalign{\medskip}244&113&-18&-9&0&-15\\\noalign{\medskip}-30&-15&&&&\end {array}  
\]
\caption{The vector for the trace of the multiplicative constant 
 $c(v^1_1,v^1_1;v^1_1)$}
\label{fmult7_4}
\end{figure}

\begin{figure} 
\[
\begin{array}{ccc}  
80255448992640&80138274925167&79787740098774\\\noalign{\medskip}79206792680861&78400301248652&77374989989221\\\noalign{\medskip}76139349486502&74703523453335&73079174468928\\\noalign{\medskip}71279332267508&69318226392392&67211105097154\\\noalign{\medskip}64974044241996&62623750255631&60177360010954\\\noalign{\medskip}57652240221271&55065789702876&52435247936361\\\noalign{\medskip}49777512793214&47108969815061&44445335040340\\\noalign{\medskip}41801513256271&39191473712706&36628144773627\\\noalign{\medskip}34123327701708&31687629536553&29330415837966\\\noalign{\medskip}27059783595895&24882552858520&22804275396566\\\noalign{\medskip}20829260005316&18960613784310&17200296928344\\\noalign{\medskip}15549188413159&14007161465134&12573167695619\\\noalign{\medskip}11245327257264&10021022340390&8896992772108\\\noalign{\medskip}7869432671924&6934086054972&6086339325247\\\noalign{\medskip}5321309766266&4633929430149&4019023248184\\\noalign{\medskip}3471380269320&2985817677720&2557237529092\\\noalign{\medskip}2180675978000&1851344837246&1564665615708\end {array}  
\]
\caption{The vector for the norm of the multiplicative constant
 $c(v^1_1,v^1_1;v^1_1)$}
\label{fmult8_2}
\end{figure}

\begin{figure} 
\[
\begin{array}{ccc}  
1316296386448&1102151942468&918417720699\\\noalign{\medskip}761557915658&628318302423&515724544156\\\noalign{\medskip}421076738262&341940657408&276136144552\\\noalign{\medskip}221723424848&176988052223&140424814566\\\noalign{\medskip}110720869883&86738676456&67499233418\\\noalign{\medskip}52165764220&40027907728&30486730148\\\noalign{\medskip}23040841355&17273591498&12841266667\\\noalign{\medskip}9462395940&6908272529&4994579014\\\noalign{\medskip}3573971135&2529616304&1769689525\\\noalign{\medskip}1222700770&833511103&559989972\\\noalign{\medskip}370279871&240567010&153252303\\\noalign{\medskip}95475892&57962133&34120478\\\noalign{\medskip}19338377&10435196&5256321\\\noalign{\medskip}2374094&864025&140844\\\noalign{\medskip}-155240&-236220&-220838\\\noalign{\medskip}-171968&-120043&-77166\\\noalign{\medskip}-46045&-25460&-12974\\\noalign{\medskip}-6040&-2518&-900\\\noalign{\medskip}-255&-50&-5\end {array}  
\]
\caption{The vector for the norm of the multiplicative constant
 $c(v^1_1,v^1_1;v^1_1)$ (continued)}
\label{fmult9_2}
\end{figure}

\subsection{Experimental evidence for crystalization} \label{expcrystalevi}
Let  $(L^\rho_\alpha, B^\rho_\alpha)$ be an upper  crystal
basis of $W^\rho_{q,\alpha}$ as in Conjecture~\ref{clocalcrystalb}. 
Since it is also a local crystal basis of $W^\rho_{q,\alpha}$ 
as an $H_q$-module, there is a crystal graph over $B^{\rho}_\alpha$
whose connected components correspond to irreducible 
$H_q$-submodules of $W^\rho_{q,\alpha}$. The elements of $B^{\rho}_\alpha$
that correspond to the highest weight nodes  of these connected  components 
are  called the {\em highest weight crystal elements}  of
the upper crystal base $(L^\rho_\alpha, B^{\rho}_\alpha)$
with respect to $H_q$.

\subsection{Kronecker problem: $n=2, r=3$}\label{sexpkroneckercrystal}
Consider again the special case of 
the Kronecker problem for $n=2$ and $r=3$ as in Section~\ref{sexpkronecker}.
Thus $H=Gl_2 \times Gl_2$,  $G=Gl_4$ with $H$ embedded diagonally,
$X=X_q=V_q\otimes W_q$ is the standard four dimensional representation of $H_q$,
where $V_q\cong W_q$ is the standard two representation of $GL_q(2)$.
Let 
\[x_1=v_1\otimes w_1, x_2=v_1\otimes w_2, x_3=v_2\otimes w_1, 
x_4=v_2\otimes w_2,\] 
be the standard basis of $X_q$.  Let $b_1,\ldots,b_4$ be the 
corresponding standard crystal basis of $L(X_q)$. 
The irreducible representations of $G^H_q$ that occur in $X_{q}^{\otimes 3}$
are: 
\begin{enumerate} 
\item  $C^{H,3}_q(X)$, the $16$-dimensional 
the degree-three component of the braided symmetric algebra of $G^H_q$,
\item $\wedge^{H,3}_q(X)$, the four dimensional degree three component of
the braided exterior algebra of $G^H_q$,
\item two copies of the 16-dimensional $G^H_q$-representation 
$W_{q,(2,1),1}(X)$ defined in \cite{GCT4} (it is denoted by 
$V_{q,(2,1),1}(X)$ there), and 
\item two copies of 
the $4$-dimensional representation  $W_{q,(2,1),2}(X)$ of  $G^H_q$ as also
defined there (it is called $V_{q,(2,1),2}(X)$ there).
\end{enumerate} 
Embeddings of the braided symmetric and exterior algebra components
in $X_q^{\otimes 3}$ are uniquely defined. We denote their embedded images
by $C^{H,3}_q(X)$ and $\wedge^{H,3}_q(X)$ again.
We choose appropriate embeddings of $W_{q,(2,1),2}(X)$ and
$W_{q,(2,1),2}(X)$ in $X_q^{\otimes 3}$ and denote them by the same symbols
again.  
As 
$H_q=GL_q(V)\times GL_q(W)$-modules,

\[
\begin{array}{lcl}
\C^{H,3}_q(X)&=& V_{q,(3)}(V)\otimes V_{q,(3)}(W) \oplus 
W_{q,(2,1)}(V)\otimes V_{q,(2,1)}(W), \\ \\
\wedge^{H,3}_q(X)&=& V_{q,(2,1)}(V)\otimes V_{q,(2,1)}(W), \\ \\
W_{q,(2,1),1}(X)&=&
V_{q,(2,1)}(V)\otimes V_{q,(3)}(W) \oplus
V_{q,(3)}(V)\otimes V_{q,(2,1)}(W) \\ \\
W_{q,(2,1),2}(X) &=& V_{q,(2,1)}(V)\otimes V_{q,(2,1)}(W).
\end{array}
\]

It was verified by computer that they have upper crystal bases
as per Conjecture~\ref{clocalcrystalb}.
  The highest weight crystal elements with
respect to $H_q$ for these  upper crystal bases, as shown separately for
each module, are as follows; we denote the monomial
basis element $b_{i_1}\otimes b_{i_2} \otimes b_{i_3}$ of $B(X_q^{\otimes 3})$
by $b_{i_1 i_2 i_3}$.

\[\begin{array}{ll}
\C^{H,3}_q(X):& \{b_{1 1 4} + b_{1 4 1}, b_{1 1 1}\}. \\
\wedge^{H,3}_q(X): \{ b_{1 2 3} + b_{1 3 2} \}. &   \\
W_{q,(2,1),1}(X):&  \{b_{1 2 1} , b_{1 3 1} \}. \\
W_{q,(2,1),2}(X):&  \{b_{1 1 4} - b_{1 4 1}\}.  \\
\end{array}
\] 

The highest weight crystal elements whose monomial support have
size two correspond to the  four dimensional $H_q$-module
$V_{q,(2,1)}(V)\otimes V_{q,(2,1)}(W)$. The element $b_{1 1 1}$ 
corresponds to the $H_q$-module $V_{q,(3)}(V)\otimes V_{q,(3)}(W)$, 
the element $b_{1 2 1}$ to the $H_q$-module
$V_{q,(3)}(V)\otimes V_{q,(2,1)}(W)$, and $b_{1 3 1}$ to the $H_q$-module
$V_{q,(2,1)}(V)\otimes V_{q,(3)}(W)$.
Notice that not all highest weight crystal elements have monomial supports
of size one as in the standard setting.

\subsubsection{$H=sl_2$, $G=sl_4$} \label{sexphsl2gsl4crystal}
Now we consider the case when
$H_q=Gl_q(2)$, $X_q$ its four dimensional irreducible 
representation, and $G_q^H$  as in Section~\ref{sexphsl2gsl4}.
Let $W_0,\ldots,W_5$ be the irreducible representations of $G^H_q$ 
occuring in $X_q^{\otimes 3}$ as defined in Section 6.1.2 of \cite{GCT7},
with $W_0=\C_q^{H,3}[X]$.

As $H_q$-modules, 
\begin{equation} 
\begin{array}{lcl} 
W_0&\cong& V_{q,(9)}(2) \oplus V_{q,(7,2)}(2), \\
W_1&\cong& V_{q,(6,3)}(2), \\
W_2&\cong& V_{q,(6,3)}(2), \\
W_3&\cong& V_{q,(8,1)}(2), \\
W_4&\cong& V_{q,(5,4)}(2), \\
W_5&\cong& V_{q,(7,2)}(2),
\end{array}
\end{equation}
where $V_{q,\lambda}(n)$ denotes the $q$-Weyl module of $GL_q(n)$ corresponding
to the partition $\lambda$. 
Their dimensions are $16,4,4,8,2$ and $6$, respectively. Though 
$W_1$ and $W_2$ are isomorphic as $H_q$-modules,
they  are nonisomorphic as $G^H_q$-modules.

It was verified by computer that they--or rather their 
embeddings in $X_q^{\otimes 3}$--have upper crystal bases
as per Conjecture~\ref{clocalcrystalb}. The highest weight crystal elements of
of the embedding of $W_1,\ldots,W_5$ have monomial supports of size one.
Let $b_{i_1 i_2 i_3}=b_{i_1}\otimes b_{i_2} \otimes b_{i_3}$
 denote the monomial basis elements of
 $B(X_q^{\otimes 3})$.
Then the highest weight crystal elements of
the uniquely defined embedding of $W_0$ are $b_{1 1 1}$ and 
$b=b_{1 1 3} + b_{1 3 1}$. The latter element $b$ here has monomial 
support of size two, a phenomenon not seen in the 
standard setting. Nonzero coefficients
of the  element $x$ in the lattice $L(\C_q^{3,H}(X))$
whose crystalization is $b$  is shown
in Figure~\ref{fa1}, wherein $x_1,\ldots,x_4$ are the standard basis vectors
of $X_q$.

\begin{sidewaysfigure}[p]
\[
\begin {array}{|c|c|}\hline
Monomial & Coefficient \\ \hline  
x_1\otimes x_1 \otimes x_3&-( {q}^{4}+1 ) ^{2} ( {q}^{2}+1 ) ^{4} ( {q}^{4}-{q}^{2}+1 ) ^{5} ( {q}^{4}+{q}^{3}+{q}^{2}+q+1 )  ( {q}^{4}-{q}^{3}+{q}^{2}-q+1 ) 
\\ & \quad  ( {q}^{6}-{q}^{5}+{q}^{4}-{q}^{3}+{q}^{2}-q+1 )  ( {q}^{6}+{q}^{5}+{q}^{4}+{q}^{3}+{q}^{2}+q+1 )  ( 2\,{q}^{20}-2\,{q}^{18}+{q}^{16}+{q}^{10}-{q}^{8}+{q}^{6}+{q}^{4}-{q}^{2}+1 ) \\ & \quad  ( {q}^{2}+q+1 ) ^{2} ( {q}^{2}-q+1 ) ^{2} ( q-1 ) ^{4} ( q+1 ) ^{4}/{{q}^{46}}\\ \hline \noalign{\medskip}x_1\otimes x_2\otimes x_2&-( {q}^{2}+1 ) ^{5} ( 2\,{q}^{6}+1 )  ( {q}^{6}+{q}^{5}+{q}^{4}+{q}^{3}+{q}^{2}+q+1 )  ( {q}^{6}-{q}^{5}+{q}^{4}-{q}^{3}+{q}^{2}-q+1 ) \\ & \quad   ( {q}^{4}-{q}^{3}+{q}^{2}-q+1 )  ( {q}^{4}+{q}^{3}+{q}^{2}+q+1 )  ( {q}^{2}+q+1 ) ^{2} ( {q}^{2}-q+1 ) ^{2} ( {q}^{4}+1 ) ^{2} ( q-1 ) ^{5} ( q+1 ) ^{5} \\ & \quad  ( {q}^{4}-{q}^{2}+1 ) ^{5}/{{q}^{37}}\\ \hline \noalign{\medskip}x_1\otimes x_3 \otimes x_1&-( {q}^{2}+1 ) ^{4} ( {q}^{4}+{q}^{3}+{q}^{2}+q+1 )  ( {q}^{4}-{q}^{3}+{q}^{2}-q+1 )  ( {q}^{6}+{q}^{5}+{q}^{4}+{q}^{3}+{q}^{2}+q+1 ) \\ & \quad  ( {q}^{6}-{q}^{5}+{q}^{4}-{q}^{3}+{q}^{2}-q+1 )  ( {q}^{16}+{q}^{14}-{q}^{12}+{q}^{10}+{q}^{4}-{q}^{2}+1 )  ( {q}^{4}+1 ) ^{2} ( {q}^{2}-q+1 ) ^{2} ( {q}^{2}+q+1 ) ^{2} \\ & \quad ( q-1 ) ^{4} ( q+1 ) ^{4} ( {q}^{4}-{q}^{2}+1 ) ^{5}/{{q}^{46}}\\ \hline\noalign{\medskip}x_2\otimes x_1 \otimes x_2&-( {q}^{2}+1 ) ^{5} ( 2\,{q}^{6}+1 )  ( {q}^{6}+{q}^{5}+{q}^{4}+{q}^{3}+{q}^{2}+q+1 )  ( {q}^{6}-{q}^{5}+{q}^{4}-{q}^{3}+{q}^{2}-q+1 ) \\ & \quad  ( {q}^{4}-{q}^{3}+{q}^{2}-q+1 )  ( {q}^{4}+{q}^{3}+{q}^{2}+q+1 )  ( {q}^{2}+q+1 ) ^{2} ( {q}^{2}-q+1 ) ^{2} ( {q}^{4}+1 ) ^{2} ( q-1 ) ^{5} ( q+1 ) ^{5} \\ & \quad  ( {q}^{4}-{q}^{2}+1 ) ^{5}/{{q}^{40}}\\ \hline\noalign{\medskip}x_2\otimes x_2 \otimes x_1&-( {q}^{2}+1 ) ^{5} ( 2\,{q}^{6}+1 )  ( {q}^{6}+{q}^{5}+{q}^{4}+{q}^{3}+{q}^{2}+q+1 )  ( {q}^{6}-{q}^{5}+{q}^{4}-{q}^{3}+{q}^{2}-q+1 )  ( {q}^{4}-{q}^{3}+{q}^{2}-q+1 ) \\ & \quad  ( {q}^{4}+{q}^{3}+{q}^{2}+q+1 )  ( {q}^{2}+q+1 ) ^{2} ( {q}^{2}-q+1 ) ^{2} ( {q}^{4}+1 ) ^{2} ( q-1 ) ^{5} ( q+1 ) ^{5} ( {q}^{4}-{q}^{2}+1 ) ^{5}/{{q}^{43}}\\ \hline\noalign{\medskip}x_3\otimes x_1 \otimes x_1&-( {q}^{2}+1 ) ^{4} ( {q}^{4}+{q}^{3}+{q}^{2}+q+1 )  ( {q}^{4}-{q}^{3}+{q}^{2}-q+1 )  ( {q}^{6}-{q}^{5}+{q}^{4}-{q}^{3}+{q}^{2}-q+1 )  ( {q}^{6}+{q}^{5} \\ & \quad +{q}^{4}+{q}^{3}+{q}^{2}+q+1 )  ( {q}^{12}-{q}^{10}+{q}^{8}+{q}^{6}+{q}^{4}-{q}^{2}+1 )  ( {q}^{2}+q+1 ) ^{2} ( {q}^{2}-q+1 ) ^{2} ( {q}^{4}+1 ) ^{2} \\ & \quad ( q-1 ) ^{4} ( q+1 ) ^{4} ( {q}^{4}-{q}^{2}+1 )^{5}/{{q}^{42}} \\ \hline \end {array} 
\] 
\caption{Nonzero coefficients  of $x$ in the lattice $L(\C_q^{3,H}(X))$}
\label{fa1}
\end{sidewaysfigure}

\section{Complexity theoretic properties of the canonical basis} \label{scomplexity}
In the standard setting,
elements of the canonical basis of $V_{q,\lambda}$ are indexed (labelled)
by semistandard tableau, for $H=GL(V)$, and by LS-paths \cite{littelmann},
for general
semisimple $H$. And
combinatorial analogues of Kashiwara's crystal operators
\cite{littelmann} on these labels can be computed efficiently \cite{GCT6}.
This is enough to imply a $\#P$-formula for  the generalized
Littlewood-Richardson coefficient   though 
the canonical basis of $V_{q,\lambda}$ is hard to compute.

In the same spirit, it may be  conjectured that the canonical basis 
of ${\cal O}(M_q^H(X)$ (or rather the set of its  labels) has
additional complexity theoretic 
properties (to be described in the full version),
based on  its cellular and refined 
sub-cellular decomposition (Conjecture~\ref{ccelldecompleft}),
that imply 
a positive $\#P$-formula for the multiplicity  $n_\pi^\alpha$ of 
the irreducible $H_q$-module
$V_{q,\pi}$ in $W_{q,\alpha}$. 
This would solve
the  problem P1 in \cite{GCT7}. 

Similarly, let  $m^\alpha_\lambda$ denote
the multiplicity of the Specht module 
$S_\lambda$ of the symmetric group $S_r$ corresponding to the partition
$\lambda$ in 
$\limit_{q\rightarrow 1} T_{q,\alpha}$, considered as an $S_r$-module. 
It may be conjectured
that  the canonical basis ${\cal B}^H_r(q)$ (or rather the set
of its labels) has similar 
additional complexity theoretic properties (to be
described in the full version) based on its
cellular and quasi-subcellular decompositions (Conjecture~\ref{ccelldecompright}).
This would imply a positive $\#P$-formula 
for the multiplicity $m^\pi_\lambda$, as needed 
in  the problem P2 in \cite{GCT7}.


\begin{thebibliography}{[Welzl]}

\bibitem[BBD]{beilinson} A. Beilinson, J. Bernstein, P. Deligne, Faisceaux pervers, Ast\'erisque 
100, (1982), Soc. Math. France.

\bibitem[BZ]{berenstein} A. Berenstein, S. Zwicknagl, Braided symmetric and exterior 
algebras, arXiv:math/0504155v3, April, 2007.


\bibitem[DJM]{date} M. Date, M. Jimbo, T. Miwa, Representations of 
$U_q(\hat gl(n,\C))$ at $q=0$ and the Robinson-Schensted correspondence,
in Physics and Mathematics of Strings, World Scientific, Singapore, 
1990, pp. 185-211.


\bibitem[Dl2]{weil2} P. Deligne, La conjecture de Weil II, Publ. Math. Inst. Haut. \'Etud. Sci. 52,
(1980) 137-252. 


\bibitem[Dri]{drinfeld} V. Drinfeld, Quantum groups, Proc. Int. Congr. Math. Berkeley, 1986,
vol. 1, Amer. Math. Soc. 1988, 798-820.


\bibitem[GCTflip1]{GCTflip1} K. Mulmuley,
On P. vs. NP, geometric complexity theory, and the flip I: a high-level 
view, Technical Report TR-2007-13, Computer Science Department,
The University of Chicago, September 2007.
Available at: http://ramakrishnadas.cs.uchicago.edu


\bibitem[GCT4]{GCT4} K. Mulmuley, M. Sohoni, Geometric complexity theory IV: 
quantum group for the Kronecker problem, cs. ArXiv preprint cs. CC/0703110,
March, 2007. Available at:
http://ramakrishnadas.cs.uchicago.edu



\bibitem[GCT6]{GCT6} K. Mulmuley, Geometric complexity theory VI:
the flip via saturated  and 
positive integer programming in representation theory and algebraic 
geometry,  Technical report TR 2007-04, Comp. Sci. Dept., 
The University of Chicago, May, 2007.
Available at: http://ramakrishnadas.cs.uchicago.edu.



\bibitem[GCT7]{GCT7} K. Mulmuley, Geometric complexity theory VII: 
Nonstandard quantum group for the plethysm problem, 
technical report TR-2007-14,
computer science dept., The university of Chicago, Sept. 2007.
Available at: http://ramakrishnadas.cs.uchicago.edu.


\ignore{\bibitem[GCT10]{GCT10} K. Mulmuley, Geometric complexity theory X:
On  class varieties, and the natural proof barrier, under preparation.
}

\bibitem[Ji]{jimbo} M. Jimbo, A $q$-difference analogue of $U({\cal G})$
 and the Yang-Baxter equation,
Lett. Math. Phys. 10 (1985), 63-69.

\bibitem[Kas1]{kashiwara1} M. Kashiwara, Crystalizing the $q$-analogue of universal enveloping
algebras, Comm. Math. Phys. 133 (1990), 249-260.


\bibitem[Kas1]{kashiwaracrystal} M. Kashiwara, On crystal bases of the $q$-analogue of
 universal enveloping algebras, Duke Math. J. 63 (1991), 465-516.

\bibitem[Kas2]{kashiwaraglobal} M. Kashiwara, Global crystal bases of quantum groups, 
Duke Mathematical Journal, vol. 69, no.2, 455-485.

\bibitem[KL1]{kazhdan} D. Kazhdan, G. Lusztig, Representations of Coxeter groups
and Hecke algebras, Invent. Math. 53 (1979), 165-184.


\bibitem[KL2]{kazhdan1} D. Kazhdan, G. Lusztig, Schubert varieties and Poincare
duality, Proc. Symp. Pure Math., AMS, 36 (1980), 185-203. 

\bibitem[Kli]{klimyk} A. Klimyk, K. Schm\"udgen, Quantum groups and their representations, 
Springer, 1997.

\bibitem[Li]{littelmann} 
 P. Littelmann, Paths and root operators in representation theory,
Ann. of Math. 142 (1995), 499-525.


\bibitem[Lu1]{lusztigcanonical} G. Lusztig, Canonical bases arising from 
quantized enveloping algebras, J. Amer. Math. Soc. 3, (1990), 447-498.



\bibitem[Lu2]{lusztigbook} G. Lusztig,
Introduction to quantum groups, Birkh\"auser, 1993.


\bibitem[RTF]{rtf} N. Reshetikhin, L. Takhtajan, L. Faddeev, Quantization of Lie groups 
and Lie algebras, Leningrad Math. J., 1 (1990), 193-225. 

\bibitem[So]{soergel} W. Soergel, Kazhdan-Lusztig polynomials and
 a combinatoric for tilting modules, Representation theory 1, (1997)


\bibitem[St]{stanley}
 R. Stanley, Positivity problems and conjectures in algebraic combinatorics, In
Mathematics: frontiers and perspectives, 295-319, Amer. Math. Soc. Providence, RI (2000).


\bibitem[W]{wor1} S. Woronowicz: Compact matrix pseudogroups, Commun. Math. Phys. 111 (1987), 613-665.


\end{thebibliography}
\end{document}